\documentclass[12pt,twoside]{book}
\usepackage[dvips]{graphicx}
\usepackage{color}
\usepackage{fancyhdr}

\newcommand{\figurecaption}[1]{\refstepcounter{figure}

\vspace{3.0ex}{\bf Figure~\thefigure:} #1}

\newcommand{\tablecaption}[1]{\refstepcounter{table}\vspace{5ex}

{\bf Table~\thetable:} #1}

\definecolor{acblue}{rgb}{0,0,0.6}
\definecolor{acred}{rgb}{0.8,0,0}
\definecolor{acgreen}{rgb}{0,0.4,0}

\newcommand{\Floor}[1]{\left\lfloor #1 \right\rfloor}

\newcommand{\Intv}[2]{\left[\,{#1,\:#2}\,\right)}
\newcommand{\Range}[2]{\left|\,{#1,\:#2}\,\right\rangle}

\newcounter{nExample}
\newcommand{\theExample}{\arabic{nExample}}

\newenvironment{Example}{\refstepcounter{nExample}
\vspace{3ex} {\pagebreak[3] \noindent {\color{acgreen} \large \sc Example \theExample } }
\begin{list}{\color{acgreen} $\triangleleft$}{\setlength{\leftmargin}{0.25in}\setlength{\rightmargin}{0.20in}\setlength{\listparindent}{0.2in}}\item}
{{\color{acgreen} $\triangleright$}\end{list}}

\newcounter{nAlgorithm}
\newcommand{\theAlgorithm}{\arabic{nAlgorithm}}

\newcommand{\COM}[1]{\` \parbox{2.8in}{\color{acblue} $\star$ \textit{\footnotesize #1}} \\ }
\newcommand{\CON}[1]{\` \parbox{2.6in}{\color{acblue} $\star$ \textit{\footnotesize #1}} \\ }
\newcommand{\COO}[1]{\` \parbox{2.4in}{\color{acblue} $\star$ \textit{\footnotesize #1}} \\ }

\newenvironment{Algorithm}[1]{\begin{figure}[tp]\refstepcounter{nAlgorithm}\vspace*{3ex}
{\noindent \hrulefill \textsc{\color{acblue} \ Algorithm \theAlgorithm}} \\
\noindent \textsf{#1} \begin{small} \begin{tabbing}}
{\end{tabbing} \vspace*{-3ex} \hrulefill {\color{acblue} {\ } $\bullet$} \end{small}\end{figure}}

\newenvironment{AAlgorithm}[1]{\begin{figure}[tp]\refstepcounter{nAlgorithm}\vspace*{3ex}
{\noindent \hrulefill \textsc{\color{acblue} \ Algorithm \theAlgorithm}} \\
\noindent \textsf{#1} \begin{small} \begin{tabbing}}
{\end{tabbing} \vspace*{-3ex} \hrulefill {\color{acblue} {\ } $\bullet$} \end{small}\end{figure}}

\begin{document}

%
%

\begin{titlepage}

%
\begin{figure}[tp]
\setlength{\unitlength}{1mm}
\begin{picture}(160,50)
\put(140,20){\includegraphics[width=20mm]{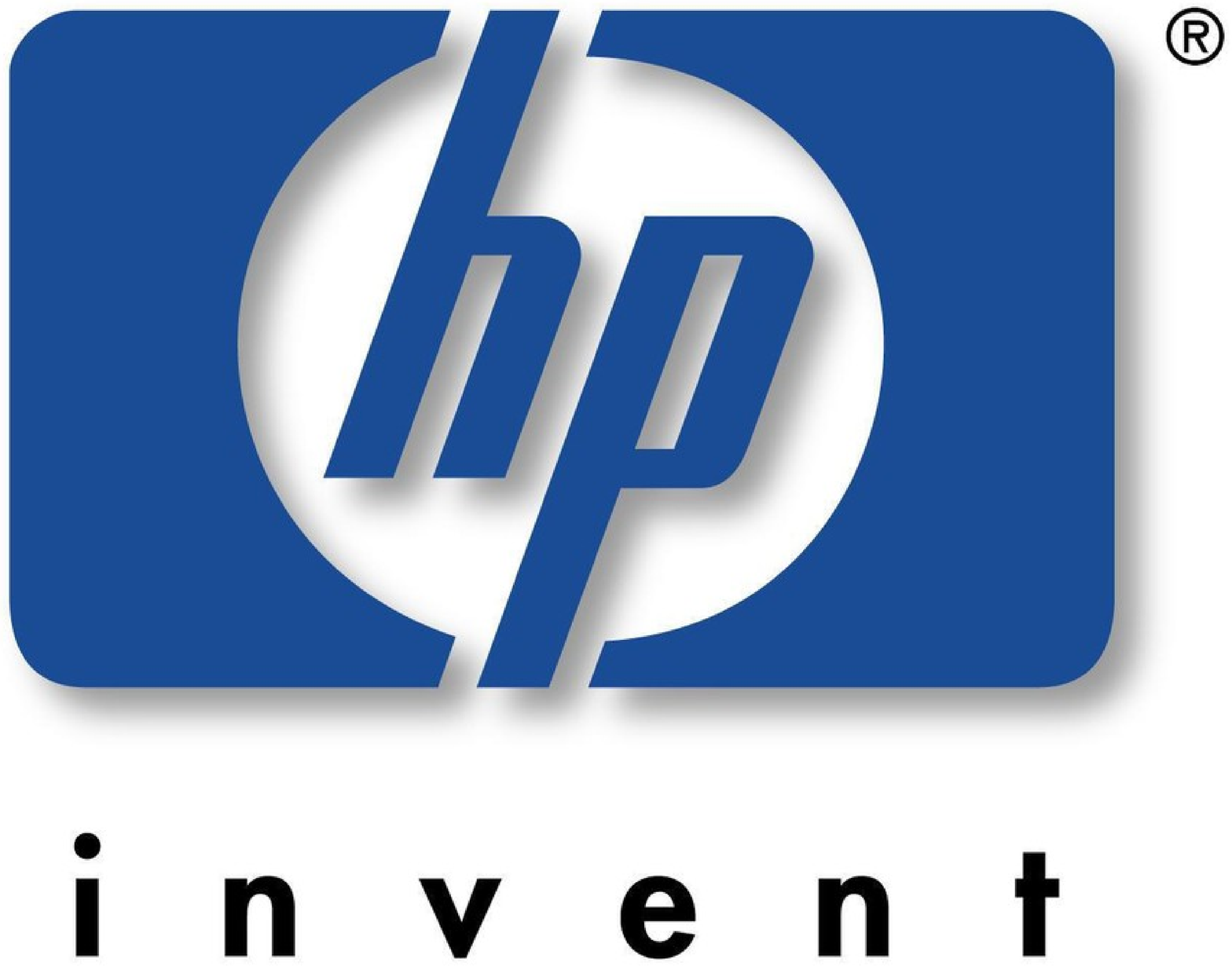}}
\end{picture}
\end{figure}

\noindent {\large \bf Introduction to Arithmetic Coding -- Theory and Practice}\footnote{Published as a chapter in {\em Lossless Compression Handbook}, ed. Khalid Sayood, \copyright Academic Press}\\

\vspace{2mm}
\noindent {\bf Amir Said}\\
Imaging Systems Laboratory\\
Hewlett-Packard Laboratories, Palo Alto, CA\\
\ \\
HPL-2004-76\\
April 21, 2004\footnote{Internal Accession Data Only, Approved for External Publication}

\vspace{10mm}

\noindent \parbox[t]{35mm}{\noindent Entropy coding,\\compression,\\complexity}
\parbox[t]{128mm}{This introduction to arithmetic coding is divided in two parts. The first 
explains how and why arithmetic coding works. We start presenting it in very 
general terms, so that its simplicity is not lost under layers of 
implementation details. Next, we show some of its basic properties, which 
are later used in the computational techniques required for a practical 
implementation.

\ 

In the second part, we cover the practical implementation aspects, including 
arithmetic operations with low precision, the subdivision of coding and 
modeling, and the realization of adaptive encoders. We also analyze the
arithmetic coding computational complexity, and techniques to reduce it.

\ 

We start some sections by first introducing the notation and most of the 
mathematical definitions. The reader should not be intimidated if at first
their motivation is not clear: these are always followed by examples and 
explanations.}

\end{titlepage}

\thispagestyle{empty}
\clearpage

\pagenumbering{roman}

\begin{small}
\tableofcontents
\end{small}

%
%
\chapter{Arithmetic Coding Principles}

\fancyhead[RE]{\rightmark}
\fancyhead[LO]{\leftmark}
\pagenumbering{arabic}

\section{Data Compression and Arithmetic Coding}

Compression applications employ a wide variety of techniques, have quite 
different degrees of complexity, but share some common processes. Figure~\ref{fgCompProc} 
shows a diagram with typical processes used for data compression. These 
processes depend on the data type, and the blocks in Figure~\ref{fgCompProc} may be in 
different order or combined. Numerical processing, like predictive coding 
and linear transforms, is normally used for waveform signals, like 
images and audio~\cite{Jayant84,Gersho92,Pennebaker92,Bhaskaran97,Sayood99}.
Logical processing consists of changing the data to a form more suited
for compression, like run-lengths, zero-trees, set-partitioning information,
and dictionary entries~\cite{Golomb66,Jayant84,ITU93,Shapiro93,Burrows94,Said95,Said96,Sayood99}.
The next stage, source modeling, is used to account for variations in the 
statistical properties of the data. It is responsible for gathering 
statistics and identifying data contexts that make the source models more 
accurate and
reliable~\cite{Rissanen81,Pennebaker88b,Bell90,Weinberger96,Wu96,LoPresto97,Witten99}.

What most compression systems have in common is the fact that the final process
is \emph{entropy coding}, which is the process of representing information in the
most compact form. It may be responsible for doing most of the compression work, 
or it may just complement what has been accomplished by previous stages.

When we consider all the different entropy-coding methods, and their possible
applications in compression applications, \textit{arithmetic coding} stands out in
terms of elegance, effectiveness and versatility, since it is able to work most
efficiently in the largest number of circumstances and purposes.
Among its most desirable features we have the following.
\begin{itemize}
 \item When applied to independent and identically distributed (i.i.d.) sources, 
  the compression of each symbol is provably optimal
  (Section~\ref{scOptimal}). 
 \item It is effective in a wide range of situations and compression ratios. The 
  same arithmetic coding implementation can effectively code all the diverse 
  data created by the different processes of Figure~\ref{fgCompProc}, such as modeling
  parameters, transform coefficients, signaling, etc. (Section~\ref{ssDynSrc}).
 \item It simplifies automatic modeling of complex sources, yielding near-optimal 
  or significantly improved compression for sources that are not i.i.d
  (Section~\ref{ssSepCodMod}).
 \item Its main process is arithmetic, which is supported with ever-increasing 
  efficiency by all general-purpose or digital signal processors (CPUs, DSPs)
  (Section~\ref{ssComplexA}).
 \item It is suited for use as a ``compression black-box'' by those that are not 
  coding experts or do not want to implement the coding algorithm themselves.
\end{itemize}

%
 \begin{figure}[tp]
 \setlength{\unitlength}{1.2mm}
 \begin{center}
 \begin{picture}(120,40)
%
%
  \thicklines
  \put(40,25){\framebox(25,10){\color{acblue} \shortstack{\textsf{Numerical} \\ \textsf{processing}}}}
  \put(80,25){\framebox(25,10){\color{acblue} \shortstack{\textsf{Logical} \\ \textsf{processing}}}}
  \put(40,5){\framebox(25,10){\color{acblue} \shortstack{\textsf{Entropy} \\ \textsf{coding}}}}
  \put(80,5){\framebox(25,10){\color{acblue} \shortstack{\textsf{Source} \\ \textsf{modeling}}}}
  \put(5,25){\makebox(25,10){\color{red} \shortstack{\textsf{Original} \\ \textsf{data}}}}
  \put(5,5){\makebox(25,10){\color{red} \shortstack{\textsf{Compressed} \\ \textsf{data}}}}
%
%
  \put(26,30){\vector(1,0){14}}
  \put(65,30){\vector(1,0){15}}
  \put(80,10){\vector(-1,0){15}}
  \put(40,10){\vector(-1,0){10}}
  \put(90,25){\vector(0,-1){10}}
%
%
  \thinlines
  \put(115,35){\Large \makebox(0,0){$\Omega$}}
  \put(72,0){\line(1,0){0.5}}
  \multiput(73.5,0)(2,0){18}{\line(1,0){1}}
  \put(110,0){\line(-1,0){0.5}}
  \put(4,20){\line(1,0){0.5}}
  \multiput(5.5,20)(2,0){33}{\line(1,0){1}}
  \put(72,20){\line(-1,0){0.5}}
  \put(4,40){\line(1,0){0.5}}
  \multiput(5.5,40)(2,0){52}{\line(1,0){1}}
  \put(110,40){\line(-1,0){0.5}}
  \put(4,20){\line(0,1){0.5}}
  \multiput(4,21.5)(0,2){9}{\line(0,1){1}}
  \put(4,40){\line(0,-1){0.5}}
  \put(72,0){\line(0,1){0.5}}
  \multiput(72,1.5)(0,2){9}{\line(0,1){1}}
  \put(72,20){\line(0,-1){0.5}}
  \put(110,0){\line(0,1){0.5}}
  \multiput(110,1.5)(0,2){19}{\line(0,1){1}}
  \put(110,40){\line(0,-1){0.5}}
 \end{picture}
\end{center}
 \figurecaption{System with typical processes for data compression.
  Arithmetic coding is normally the final stage, and the other stages
  can be modeled as a single data source $\Omega$.}
                                                      \label{fgCompProc}
\end{figure}

Even with all these advantages, arithmetic coding is not as popular and well 
understood as other methods. Certain practical problems held back its 
adoption.
\begin{itemize}
 \item The complexity of arithmetic operations was excessive for coding 
   applications. 
 \item Patents covered the most efficient implementations. Royalties and the fear 
   of patent infringement discouraged arithmetic coding in commercial products.
 \item Efficient implementations were difficult to understand.
\end{itemize}

However, these issues are now mostly overcome. First, the relative 
efficiency of computer arithmetic improved dramatically, and new techniques 
avoid the most expensive operations. Second, some of the patents have 
expired (e.g.,~\cite{Langdon78p,Langdon81p}), or became obsolete. Finally,
we do not need to 
worry so much about complexity-reduction details that obscure the inherent 
simplicity of the method. Current computational resources allow us to 
implement simple, efficient, and royalty-free arithmetic coding.

\section{Notation}                                           \label{ssNotation}

Let $\Omega$ be a data source that puts out symbols $s_k$ coded as integer 
numbers in the set $\{0,1,\ldots,M-1\}$, and let $S=\{s_1,s_2,\ldots,s_N\}$
be a sequence of $N$ random symbols put out by
$\Omega$~\cite{Shannon48,Gallager68,Jelinek68,McEliece84,Sayood99,Salomon00}.
For now, we assume that the source symbols are independent and identically
distributed~\cite{Papoulis84}, with probability
\begin{equation}
                                                           \label{eqNota1}
 p(m) = \mbox{Prob}\{s_k = m\}, \quad m = 0,1,2,\ldots,M-1,
  \quad k = 1,2,\ldots,N.
\end{equation}

We also assume that for all symbols we have $p(m) \ne 0$, and define $c(m)$ 
to be the cumulative distribution,
\begin{equation}
                                                           \label{eqNota2}
 c(m) = \sum\limits_{s=0}^{m-1} {\,p(s)}, \quad m = 0,1,\ldots,M.
\end{equation}

Note that $c(0) \equiv 0, \; c(M) \equiv 1$, and
\begin{equation}
                                                           \label{eqNota3}
 p(m) = c(m+1) - c(m).
\end{equation}

We use bold letters to represent the vectors with all $p(m)$ and $c(m)$ 
values, i.e.,
\begin{eqnarray*}
  {\bf p} & = & [\,p(0)\;p(1)\;\cdots\;p(M-1)\;], \\
  {\bf c} & = & [\,c(0)\;c(1)\;\cdots\;\;c(M-1)\;c(M)\;].
\end{eqnarray*}
We assume that the compressed data (output of the 
encoder) is saved in a vector (buffer) $\bf d$. The output alphabet has 
$D$ symbols, i.e., each element in $\bf d$ belongs to set 
$\{0,1,\ldots,D-1\}$.

Under the assumptions above, an optimal coding method~\cite{Shannon48}
codes each symbol $s$ from $\Omega$ with an average number of bits
equal to
\begin{equation}
                                                           \label{eqNota4}
 {\rm B}(s) = -\log_2 p(s)\quad\mbox{bits.}
\end{equation}

\begin{Example}                                               \label{exTextSrc}    
Data source $\Omega$ can be a file with English text: each symbol from this
source is a single byte representatinh a character. This data alphabet
contains $M=256$ symbols, and symbol numbers are defined by the ASCII
standard. The probabilities of the symbols can be estimated by gathering
statistics using a large number of English texts. Table~\ref{tbASCIITxt} shows
some characters, their ASCII symbol values, and their estimated probabilities.
It also shows the number of bits required to code symbol $s$ in an optimal
manner, $-\log_2 p(s)$. From these numbers we conclude that, if data symbols
in English text were i.i.d., then the best possible text compression ratio
would be about 2:1 (4~bits/symbol). Specialized text compression
methods~\cite{Ziv77,Ziv78,Bell90,Burrows94} can yield significantly better
compression ratios because they exploit the statistical dependence between
letters.
\end{Example}

%

\begin{table}[tbp]
 \begin{center} \begin{small}
  \begin{tabular}{|c|c|c|c|} \hline
    {\bf Character} & {\bf ASCII}  & {\bf Probability} & {\bf Optimal number} \\
                    & {\bf Symbol} &                   & {\bf of bits} \\
                    & $s$          &      $p(s)$       & $-\log_2p(s)$ \\ \hline \hline
    Space &  32 & 0.1524 &  2.714 \\ \hline
      ,   &  44 & 0.0136 &  6.205 \\ \hline
      .   &  46 & 0.0056 &  7.492 \\ \hline
      A   &  65 & 0.0017 &  9.223 \\ \hline
      B   &  66 & 0.0009 & 10.065 \\ \hline
      C   &  67 & 0.0013 &  9.548 \\ \hline
      a   &  97 & 0.0595 &  4.071 \\ \hline
      b   &  98 & 0.0119 &  6.391 \\ \hline
      c   &  99 & 0.0230 &  5.441 \\ \hline
      d   & 100 & 0.0338 &  4.887 \\ \hline
      e   & 101 & 0.1033 &  3.275 \\ \hline
      f   & 102 & 0.0227 &  5.463 \\ \hline
      t   & 116 & 0.0707 &  3.823 \\ \hline
      z   & 122 & 0.0005 & 11.069 \\ \hline
 \end{tabular}
 \end{small}
\end{center}
 \tablecaption{Estimated probabilities of some letters and punctuation marks
  in the English language. Symbols are numbered according to the ASCII
  standard.}
                                                             \label{tbASCIITxt}
\end{table}

This first example shows that our initial assumptions about data sources are 
rarely found in practical cases. More commonly, we have the following 
issues.

\begin{enumerate}
 \item The source symbols are not identically distributed.
 \item The symbols in the data sequence are not independent (even if
  uncorrelated)~\cite{Papoulis84}.
 \item We can only \emph{estimate} the probability values, the statistical
  dependence between symbols, and how they change in time.
\end{enumerate}

However, in the next sections we show that the generalization of
arithmetic coding to time-varying sources is straightforward, and
we explain how to address all these practical issues.

\section{Code Values}                                            \label{ssCoVa}

Arithmetic coding is different from other coding methods for which we know 
the exact relationship between the coded symbols and the actual bits that 
are written to a file. It codes one data symbol at a time, and assigns to 
each symbol a real-valued number of bits (see examples in the last column of
Table~\ref{tbASCIITxt}). To figure out how this is possible, we have to
understand the \emph{code value} representation: coded messages mapped to
real numbers in the interval~[0,~1).

The code value $v$ of a compressed data sequence is the real number with 
fractional digits equal to the sequence's symbols. We can convert sequences 
to code values by simply adding ``0.'' to the beginning of a coded sequence, 
and then interpreting the result as a number in base-$D$ notation, where $D$ is 
the number of symbols in the coded sequence alphabet. For example, if a 
coding method generates the sequence of bits \mbox{0011000101100,} then we have
\begin{equation}
 \begin{array}{lrrl}
  \mbox{Code sequence} & {\bf d} = & [\;\underbrace{0011000101100}\:] \, \\
  \mbox{Code value}    & v = & {0.\overbrace{0011000101100} {}_2 } & = \mbox{0.19287109375}
\end{array}
                                                         \label{eqCova1}
\end{equation}
where the ``2'' subscript denotes \mbox{base-2} notation. As usual, we omit the 
subscript for decimal notation.

This construction creates a convenient mapping between infinite sequences of symbols 
from a $D$-symbol alphabet and real numbers in the interval [0,~1), where any 
data sequence can be represented by a real number, and vice-versa. The code 
value representation can be used for any coding system and it provides a 
universal way to represent large amounts of information independently of the 
set of symbols used for coding (binary, ternary, decimal, etc.). For 
instance, in~(\ref{eqCova1}) we see the same code with base-2 and base-10 
representations.

We can evaluate the efficacy of any compression method by analyzing the 
distribution of the code values it produces. From Shannon's information 
theory~\cite{Shannon48} we know that, if a coding method is optimal, then the cumulative 
distribution~\cite{Papoulis84} of its code values has to be a straight line from point 
(0,~0) to point~(1,~1).

\begin{figure}[tp]
\begin{center}
  \includegraphics[scale=0.9]{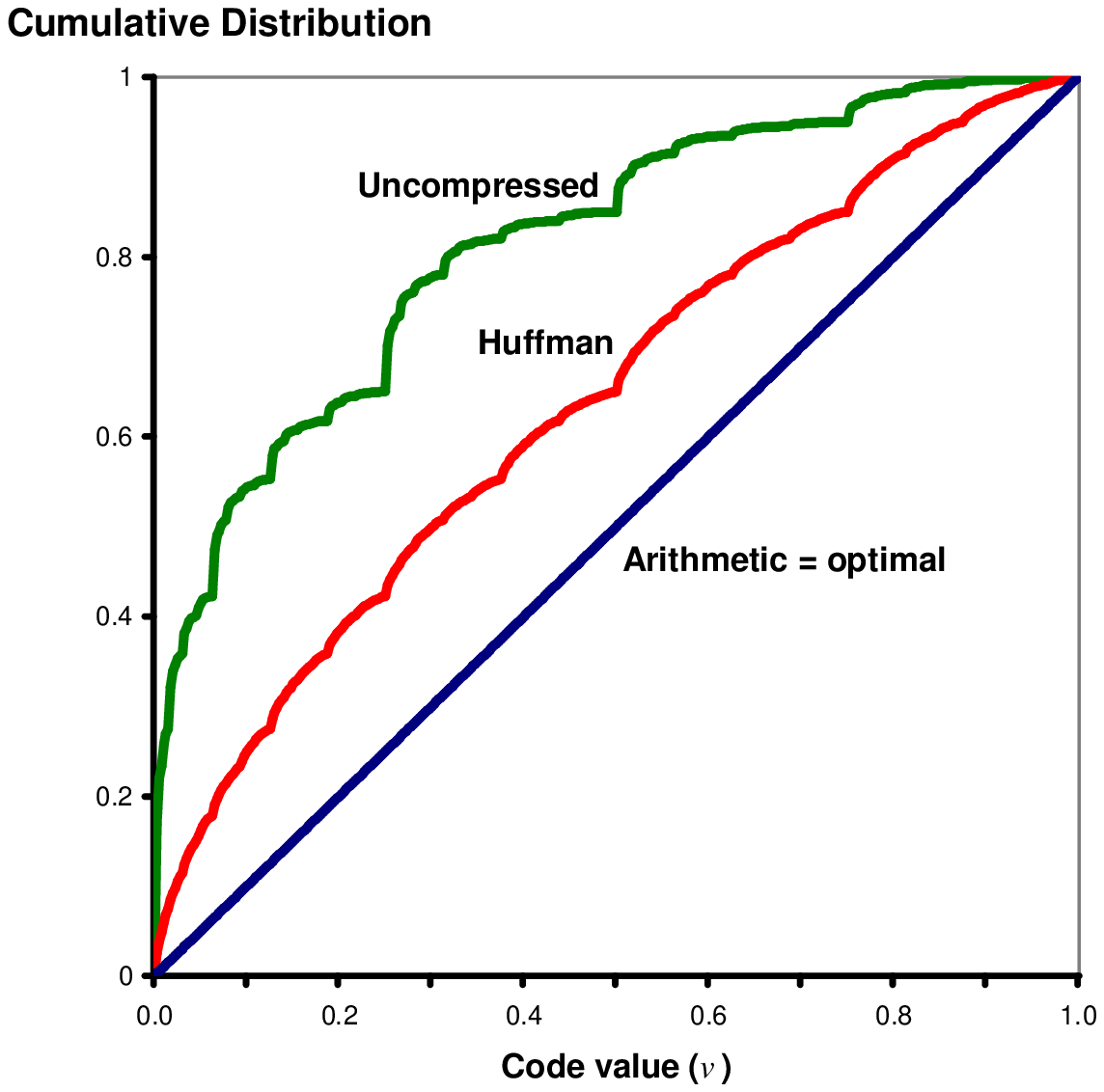}
\end{center}
 \figurecaption{Cumulative distribution of code values generated by different
 coding methods when applied to the source of Example~\ref{exCValGraph}.}
                                                           \label{fgCValGraph}
\end{figure}

\begin{Example}                                             \label{exCValGraph}    
Let us assume that the i.i.d.\ source $\Omega$ has four symbols, and the 
probabilities of the data symbols are ${\bf p}=[\,0.65\;\;0.2\;\;0.1\;\;0.05\,]$. 
If we code random data sequences from this source 
with two bits per symbols, the resulting code values produce a cumulative 
distribution as shown in Figure~\ref{fgCValGraph}, under the label ``uncompressed.''
Note how the distribution is skewed, indicating the possibility for
significant compression.

The same sequences can be coded with the Huffman code for
$\Omega$~\cite{Huffman52,Gallager68,McEliece84,Sayood99,Salomon00},
with one bit used for symbol ``0'', two bits for symbol ``1'', and three
bits for symbols ``2'' and ``3''. The corresponding code value cumulative
distribution in Figure~\ref{fgCValGraph} shows that there is substantial 
improvement over the uncompressed case, but this coding method 
is still clearly not optimal. The third line in Figure~\ref{fgCValGraph} shows that the 
sequences compressed with arithmetic coding simulation produce a code value 
distribution that is practically identical to the optimal.
\end{Example}

%

The straight-line distribution means that if a coding method is optimal then 
there is no statistical dependence or redundancy left in the compressed 
sequences, and consequently its code values are uniformly distributed on the 
interval [0,~1). This fact is essential for understanding of how arithmetic 
coding works. Moreover, code values are an integral part of the arithmetic 
encoding/decoding procedures, with arithmetic operations applied to real 
numbers that are directly related to code values.

One final comment about code values: two infinitely long different sequences 
can correspond to the same code value. This follows from the fact that for 
any $D>1$ we have

\begin{equation}
                                                         \label{eqCova2}
 \sum_{n=k}^\infty {(D-1)D^{-n}} = D^{1-k}.
\end{equation}

For example, if $D=10$ and $k=2$, then~(\ref{eqCova2}) is the equality
$0.09999999\ldots=0.1$.
This fact has no important practical significance for coding purposes, 
but we need to take it into account when studying some theoretical 
properties of arithmetic coding.

\section{Arithmetic Coding}                                  \label{ssAritCode}

\subsection{Encoding Process}                                 \label{ssEncProc}

In this section we first introduce the notation and equations that describe 
arithmetic encoding, followed by a detailed example. Fundamentally, the 
arithmetic encoding process consists of creating a sequence of nested 
intervals in the form $\Phi_k(S)=\Intv{\alpha_k}{\beta_k},\;k = 0,1,\ldots,N,$
where $S$ is the source data sequence, $\alpha_k$ and $\beta_k$ are real
numbers such that $0 \le \alpha_k \le \alpha_{k+1},$
and $\beta_{k+1} \le \beta_k \le 1$. For a simpler way to describe 
arithmetic coding we represent intervals in the form $\Range{b}{l}$, where
$b$ is called the \emph{base} or \emph{starting point} of the interval, and $l$ the 
\emph{length} of the interval. The relationship between the traditional
and the new interval notation is
\begin{equation}
                                                         \label{eqAritCode1}
  \Range{b}{l} = \Intv{\alpha}{\beta} \quad \mbox{if}\quad b = \alpha \quad
    \mbox{and}\quad l = \beta-\alpha.
\end{equation}

The intervals used during the arithmetic coding process are, in this new 
notation, defined by the set of recursive equations~\cite{Jelinek68,Rubin79}
\begin{eqnarray}
                                                         \label{eqAritCode2}
 \Phi_0(S) & = & \Range{b_0}{l_0} = \Range{0}{1}, \\
                                                         \label{eqAritCode3}
 \Phi_k(S) & = & \Range{b_k}{l_k} = \Range{b_{k-1}+c(s_k)\,l_{k-1}}
  {p(s_k)\,l_{k-1}},\quad k = 1,2,\,\ldots,N.
\end{eqnarray}

The properties of the intervals guarantee that $0 \le b_k \le b_{k+1} < 1$, 
and $0 < l_{k+1} < l_k \le 1$. Figure~\ref{fgDinSys} shows a dynamic system 
corresponding to the set of recursive equations~(\ref{eqAritCode3}). We later
explain how to choose, at the end of the coding process, a code value in the
final interval, i.e., $\hat{v}(S)\in\Phi_N(S)$.

%
 \setlength{\unitlength}{1.2mm}
 \begin{figure}[tp]
 \begin{center}
 \begin{picture}(110,55)
%
%
  \thicklines
  \put(5,45){\framebox(20,10){\color{acblue} \shortstack{\textsf{Data} \\ \textsf{source}}}}
  \put(35,45){\framebox(25,10){\color{acblue} \shortstack{\textsf{Source model} \\ \textsf{(tables)}}}}
  \put(10,16){\framebox(20,8){\color{acblue} \textsf{Delay}}}
  \put(65,16){\framebox(20,8){\color{acblue} \textsf{Delay}}}
%
%
  \put(42,33){\circle{6.2}}
  \put(39.88,30.88){\line(1,1){4.24}}
  \put(39.88,35.12){\line(1,-1){4.24}}
  \put(42,20){\circle{6.2}}
  \put(39,20){\line(1,0){6}}
  \put(42,17){\line(0,1){6}}
  \put(53,20){\circle{6.2}}
  \put(50.88,17.88){\line(1,1){4.24}}
  \put(50.88,22.12){\line(1,-1){4.24}} 
  \put(42,2){\circle*{1}}
  \put(90,8){\circle*{1}}
  \put(61,20){\circle*{1}}
%
%
  \put(25,50){\vector(1,0){10}}
  \put(5,20){\vector(1,0){5}}
  \put(5,2){\vector(1,0){95}}
  \put(30,20){\vector(1,0){9}}
  \put(65,20){\vector(-1,0){9}}
  \put(90,20){\vector(-1,0){5}}
  \put(61,33){\vector(-1,0){16}}
  \put(53,8){\vector(1,0){47}}
  \put(42,45){\vector(0,-1){9}}
  \put(42,30){\vector(0,-1){7}}
  \put(53,45){\vector(0,-1){22}}
  \put(61,20){\line(0,1){13}}
  \put(5,2){\line(0,1){18}}
  \put(42,2){\line(0,1){20}}
  \put(53,8){\line(0,1){9}}
  \put(90,8){\line(0,1){12}}
%
%
  \put(30,53){\large \makebox(0,0){$s_k$}}
  \put(36,15){\large \makebox(0,0){$b_{k-1}$}}
  \put(60,15){\large \makebox(0,0){$l_{k-1}$}}
  \put(105,8){\large \makebox(0,0){$l_k$}}
  \put(105,2){\large \makebox(0,0){$b_k$}}
  \put(35,41){\large \makebox(0,0){$c(s_k)$}}
  \put(60,41){\large \makebox(0,0){$p(s_k)$}}
%
%
  \put(80,35){\color{acblue} \shortstack[l]{
    $s$ -- data symbol \\
    $p$ -- symbol probability \\
    $c$ -- cumulative distribution \\
    $b$ -- interval base \\
    $l\,$ -- interval length}}
 \end{picture}
 \figurecaption{Dynamic system for updating arithmetic coding intervals.}
                                                              \label{fgDinSys}
\end{center}
\end{figure}

The coding process defined by (\ref{eqAritCode2}) and~(\ref{eqAritCode3}),
also called \emph{Elias coding,} was first described in~\cite{Jelinek68}.
Our convention of representing an interval using its 
base and length has been used since the first arithmetic coding
papers~\cite{Martin79,Rubin79}. Other authors have intervals represented by their extreme points, 
like [base,~base+length), but there is no mathematical difference between 
the two notations.

\begin{figure}[tp]
\begin{center}
  \includegraphics[scale=1.1]{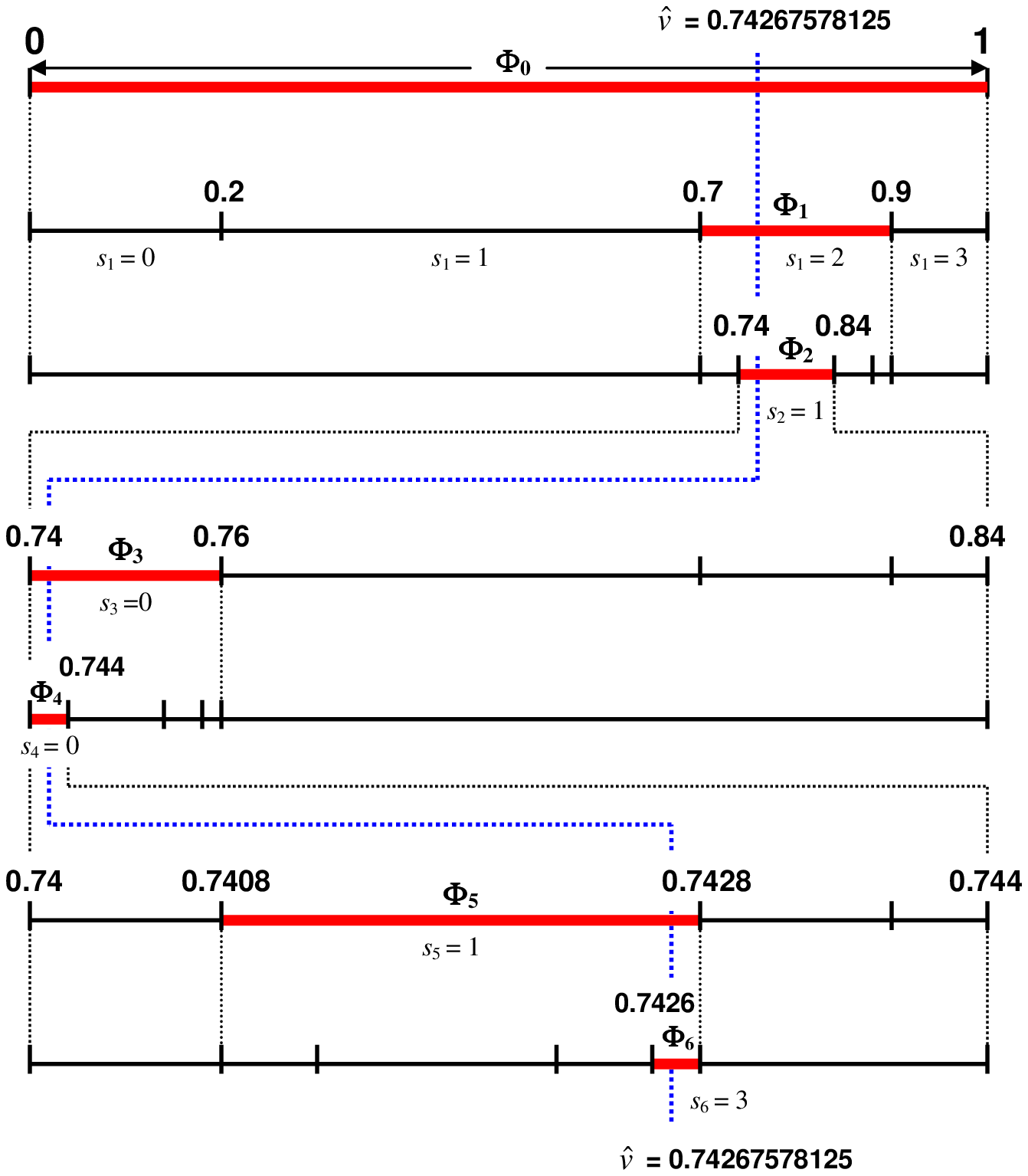}
\end{center}
 \figurecaption{Graphical representation of the arithmetic coding process of
  Example~\ref{exFirstEnc}: the interval $\Phi_0 = [0,1)$ is divided in
  nested intervals according to the probability of the data symbols.
  The selected intervals, corresponding to data sequence 
  $S=\{2,1,0,0,1,3\}$ are indicated by thicker lines.}
                                                             \label{fgCodeIntv}
\end{figure}

\begin{Example}
Let us assume that source $\Omega $ has four symbols ($M=4$), the 
probabilities and distribution of the symbols are
${\bf p} = [\,0.2\;\;0.5\;\;0.2\;\;0.1\,]$ and ${\bf c} = [\,0\;\;0.2\;\;0.7\;\;0.9\;\;1\,]$,
and the sequence of ($N=6$) symbols to be encoded is $S = \{2,1,0,0,1,3\}$.

Figure~\ref{fgCodeIntv} shows graphically how the encoding process corresponds to the 
selection of intervals in the line of real numbers. We start at the top of 
the figure, with the interval [0,~1), which is divided into four 
subintervals, each with length equal to the probability of the data symbols. 
Specifically, interval [0,~0.2) corresponds to $s_1=0$, interval 
[0.2,~0.7) corresponds to $s_1=1$, interval [0.7,~0.9) corresponds to
$s_1=2$, and finally interval [0.9,~1) corresponds to $s_1=3$. The next set 
of allowed nested subintervals also have length proportional to the 
probability of the symbols, but their lengths are also proportional to the 
length of the interval they belong to. Furthermore, they represent more than 
one symbol value. For example, interval [0,~0.04) corresponds to
$s_1=0,\;s_2=0$, interval [0.04,~0.14) corresponds to $s_1=0,\;s_2=1$, and 
so on.

The interval lengths are reduced by factors equal to symbol probabilities in 
order to obtain code values that are uniformly distributed in the interval 
[0,~1) (a necessary condition for optimality, as explained in Section~\ref{ssCoVa}). 
For example, if 20{\%} of the sequences start with symbol ``0'', then 20{\%} 
of the code values must be in the interval assigned to those sequences, 
which can only be achieved if we assign to the first symbol ``0'' an 
interval with length equal to its probability, 0.2. The same reasoning 
applies to the assignment of the subinterval lengths: every occurrence of 
symbol ``0'' must result in a reduction of the interval length to 20{\%} its 
current length. This way, after encoding several symbols the distribution of 
code values should be a very good approximation of a uniform distribution.

Equations (\ref{eqAritCode2}) and~(\ref{eqAritCode3}) provide the formulas
for the sequential computation of the intervals. Applying them to our example
we obtain:
\begin{eqnarray*}
 \Phi_0(S) & = & \Range{0}{1} = \Intv{0}{1}, \\
 \Phi_1(S) & = & \Range{b_0+c(2)l_0}{p(2)l_0} =
                 \Range{0+0.7 \times 1}{0.2 \times 1} = \Intv{0.7}{0.9}, \\
 \Phi_2(S) & = & \Range{b_1+c(1)l_1}{p(1)l_1} =
                 \Range{0.7 + 0.2 \times 0.2}{0.5 \times 0.2} = \Intv{0.74}{0.84}, \\
  & \vdots & \\
 \Phi_6(S) & = & \Range{b_5+c(3)l_5}{p(3)l_5} =
                 \Range{0.7426}{0.0002} = \Intv{0.7426}{0.7428},
\end{eqnarray*}

The list with all the encoder intervals is shown in the first four columns
of Table~\ref{tbFirsCod}. Since the intervals quickly become quite small,
in Figure~\ref{fgCodeIntv} we have to graphically magnify them (twice) so
that we can see how the coding process continues. Note that even though
the intervals are shown in different magnifications, the intervals values
do not change, and the process to subdivide intervals continues in
exactly the same manner.
                                                      \label{exFirstEnc}
\end{Example}

%

%

\begin{table}[tbp]
\begin{center}
  \begin{tabular}{|c|c|l|l|l|c|} \hline
  {\bf Iteration} & {\bf  Input} & {\bf Interval} & {\bf Interval} & \multicolumn{1}{|c|}{\bf Decoder} & {\bf Output} \\
  & {\bf Symbol} & \multicolumn{1}{|c|}{\bf base} & \multicolumn{1}{|c|}{\bf length} &
    {\bf updated value} & {\bf symbol} \\
  $k$ & $s_k $ & \multicolumn{1}{|c|}{$b_k$} & \multicolumn{1}{|c|}{$l_k$} & 
    \multicolumn{1}{|c|}{$\tilde{v}_k=\frac{\hat{v}-b_{k-1}}{l_{k-1}}$} & $\hat{s}_k$ \\ \hline \hline
  0 & --- & 0      & 1      & ---           & --- \\ \hline
  1 & 2  & 0.7    & 0.2    & 0.74267578125 &  2  \\ \hline
  2 & 1  & 0.74   & 0.1    & 0.21337890625 &  1  \\ \hline
  3 & 0  & 0.74   & 0.02   & 0.0267578125  &  0  \\ \hline
  4 & 0  & 0.74   & 0.004  & 0.1337890625  &  0  \\ \hline
  5 & 1  & 0.7408 & 0.002  & 0.6689453125  &  1  \\ \hline
  6 & 3  & 0.7426 & 0.0002 & 0.937890625   &  3  \\ \hline
  7 & --- &  ---   &  ---   & 0.37890625    &  1  \\ \hline
  8 & --- &  ---   &  ---   & 0.3578125     &  1  \\ \hline
\end{tabular}
\end{center}
 \tablecaption{Arithmetic encoding and decoding results for Examples \ref{exFirstEnc} and
  \ref{exFirstDec}. The last two rows show what happens when decoding continues past
  the last symbol.}
                                                      \label{tbFirsCod}
\end{table}

The final task in arithmetic encoding is to define a code value $\hat{v}(S)$
that will represent data sequence $S$. In the next section we show how 
the decoding process works correctly for \textit{any} code value $\hat{v} \in \Phi_N(S)$.
However, the code value cannot be provided to the decoder as a pure 
real number. It has to be stored or transmitted, using a conventional number 
representation. Since we have the freedom to choose any value in the final 
interval, we want to choose the values with the shortest representation. For 
instance, in Example~\ref{exFirstEnc}, the shortest decimal representation comes
from choosing $\hat{v}=0.7427,$ and the shortest binary representation is 
obtained with $\hat{v}=0.10111110001_2=0.74267578125$.

The process to find the best binary representation is quite simple and best 
shown by induction. The main idea is that for relatively large intervals we 
can find the optimal value by testing a few binary sequences, and as the 
interval lengths are halved, the number of sequences to be tested has to 
double, increasing the number of bits by one. Thus, according to the 
interval length $l_N$, we use the following rules:

\begin{itemize}
 \item If $l_N \in [0.5,\;1),$ then choose code value
  {\small $\hat{v} \in \{0,0.5\}=\{0.0_2,0.1_2\}$} 
  for a 1-bit representation.
 \item If $l_N \in [0.25,\;0.5),$ then choose value
  {\small $\hat{v} \in \{0,0.25,0.5,0.75\}=\{0.00_2,0.01_2,0.10_2,0.11_2\}$}
  for a 2-bit representation.
 \item If $l_N \in [0.125,\;0.25),$ then choose value
  {\small $\hat{v} \in \{0,0.125,0.25,0.375,0.5,0.625,0.75,0.875\}=$
  $\{0.000_2,0.001_2,0.010_2,0.011_2,0.100_2,0.101_2,0.110_2,0.111_2\}$}
  for a 3-bit representation.
\end{itemize}

By observing the pattern we conclude that the minimum number of bits 
required for representing $\hat{v} \in \Phi_N(S)$ is
\begin{equation}
                                                           \label{eqAritCode4}
 {\rm B}_{\min } = \left\lceil {-\log_2(l_N )} \right\rceil \quad \mbox{bits,}
\end{equation}
where $\left\lceil x \right\rceil$ represents the smallest integer greater 
than or equal to $x$.

We can test this conclusion observing the results for Example~\ref{exFirstEnc} in Table~\ref{tbFirsCod}. 
The final interval is $l_N = 0.0002$, and thus
${\rm B}_{\min } = \left\lceil {-\log_2 (0.0002)} \right\rceil =$ 13~bits.
However, in Example~\ref{exFirstEnc} we can choose $\hat{v}=0.10111110001_2$,
and it requires only 11~bits!

The origin of this inconsistency is the fact that we can choose binary 
representations with the number of bits given by~(10), and then remove the 
trailing zeros. However, with optimal coding the \emph{average} number of bits that can be 
saved with this process is only one~bit, and for that reason, it is rarely 
applied in practice.

\subsection{Decoding Process}                                 \label{ssDecProc}

In arithmetic coding, the decoded sequence is determined solely by the code 
value $\hat{v}$ of the compressed sequence. For that reason, we represent 
the decoded sequence as
\begin{equation}
                                                           \label{eqAritCode5}
 \hat{S}(\hat{v}) = \left\{\hat{s}_1(\hat{v}),\hat{s}_2(\hat{v}),\ldots,
  \hat{s}_N(\hat {v}) \right\}.
\end{equation}

We now show the decoding process by which any code value $\hat{v} \in \Phi_N(S)$
can be used for decoding the correct sequence (i.e., $\hat{S}(\hat{v})=S)$.
We present the set of recursive equations that implement 
decoding, followed by a practical example that provides an intuitive idea of 
how the decoding process works, and why it is correct.

The decoding process recovers the data symbols in the same sequence that 
they were coded. Formally, to find the numerical solution, we define a 
sequence of normalized code values
$\left\{ \tilde{v}_1,\tilde {v}_2,\ldots,\tilde {v}_N \right\}$.
Starting with $\tilde{v}_1 = \hat{v}$,
we sequentially find $\hat {s}_k$ from $\tilde {v}_k$, and then we 
compute $\tilde{v}_{k+1}$ from $\hat{s}_k$ and $\tilde{v}_k$.

The recursion formulas are
\begin{eqnarray}
                                                           \label{eqAritCode6}
 \tilde{v}_1 & = & \hat{v}, \\
                                                           \label{eqAritCode7}
 \hat{s}_k(\hat{v}) & = & \left\{\,s:\;c(s)\le\tilde{v}_k<c(s+1) \right\},\quad
  k = 1,2,\ldots,N, \\
                                                           \label{eqAritCode8}
 \tilde{v}_{k+1} & = & \frac{\tilde{v}_k - c(\hat{s}_k(\hat{v}))}
  {p(\hat{s}_k(\hat{v})) },\quad k = 1, 2,\,\ldots,N-1.
\end{eqnarray}
(In equation (\ref{eqAritCode7}) the colon means ``$s$ that satisfies the inequalities.'')

A mathematically equivalent decoding method---which later we show to be 
necessary when working with fixed-precision arithmetic---recovers the 
sequence of intervals created by the encoder, and searches for the correct 
value $\hat {s}_k (\hat {v})$ in each of these intervals. It is defined by
\begin{eqnarray}
                                                           \label{eqAritCode9}
 \Phi_0(\hat{S}) & = & \Range{b_0}{l_0} = \Range{0}{1}, \\
                                                           \label{eqAritCode10}
 \hat{s}_k(\hat{v}) & = & \left\{\,s:\;c(s)\le\frac{\hat{v}-b_{k-1}}{l_{k-1}}<c(s + 1) \right\},\quad
  k = 1,2,\ldots,N, \\
                                                           \label{eqAritCode11}
 \Phi_k(\hat{S}) & = & \Range{b_k}{l_k} = \Range{b_{k-1}+c(\hat{s}_k(\hat{v}))\,l_{k-1}}
  {p(\hat{s}_k(\hat{v}))\,l_{k-1}},\quad k = 1,2,\,\ldots,N.
\end{eqnarray}

The combination of recursion~(\ref{eqAritCode8}) with
recursion~(\ref{eqAritCode11}) yields
\begin{equation}
                                                           \label{eqAritCode12}
 \tilde {v}_k = \frac{\hat{v}-\sum\limits_{i=1}^{k-1} c(\hat{s}_i)
  \prod\limits_{j=1}^{i-1} p(\hat{s}_j)}{\prod\limits_{i=1}^{k-1}p(\hat{s}_i)} =
  \frac{\hat{v} - b_{k-1}}{l_{k-1}}.
\end{equation}
showing that~(\ref{eqAritCode7}) is equivalent to~(\ref{eqAritCode10}).

\begin{Example}
Let us apply the decoding process to the data obtained in Example~\ref{exFirstEnc}.
In Figure~\ref{fgCodeIntv}, we show graphically the meaning of $\hat{v}$:
it is a value that 
belongs to all nested intervals created during coding. The dotted line shows 
that its position moves as we magnify the graphs, but the value remains 
the same. From Figure~\ref{fgCodeIntv}, we can see that we can start decoding from the 
first interval $\Phi_0(S) = [0,\,1)$: we just have to compare $\hat{v}$ 
with the cumulative distribution \textbf{c} to find the only possible value 
of $\hat{s}_1$
\[
 \hat{s}_1(\hat{v}) = \left\{\, s:\; c(s)\le\hat{v}=0.74267578125<c(s+1) \right\} = 2.
\]

We can use the value of $\hat{s}_1$ to find out interval $\Phi_1(S)$, and use 
it for determining $\hat{s}_2 $. In fact, we can ``remove'' the 
effect of $\hat{s}_1 $ in $\hat{v}$ by defining the normalized code value
\[
 \tilde{v}_2 = \frac{\hat{v} - c(\hat{s}_1)}{p(\hat{s}_1)} = 0.21337890625.
\]

Note that, in general, $\tilde {v}_2 \in [0,\;1)$, i.e., it is a value 
normalized to the initial interval. In this interval we can use the same 
process to find
\[
 \hat{s}_2(\hat {v}) = \left\{\, s:\; c(s)\le\tilde{v}_2=0.21337890625<c(s+1) \right\} = 1.
\]

The last columns of Table~\ref{tbFirsCod} show how the process continues, and the updated 
values computed while decoding. We could say that the process continues 
until $\hat{s}_6$ is decoded. However, how can the decoder, having only 
the initial code value $\hat{v}$, know that it is time to stop decoding? 
The answer is simple: it can't. We added two extra rows to Table~\ref{tbFirsCod} to show 
that the decoding process can continue normally after the last symbol is 
encoded. Below we explain what happens.
                                                      \label{exFirstDec}
\end{Example}

It is important to understand that arithmetic encoding maps intervals to 
\emph{sets of sequences}. Each real number in an interval corresponds to
one infinite sequence. 
Thus, the sequences corresponding to $\Phi_6(S) = [0.7426,\,0.7428)$ are 
all those that \emph{start} as $\{2,1,0,0,1,3,\ldots\}$. The code value 
$\hat{v} = 0.74267578125$ corresponds to one such infinite sequence, 
and the decoding process can go on forever decoding that particular 
sequence.

There are two practical ways to inform that decoding should stop:
\begin{enumerate}
 \item Provide the number of data symbols ($N$) in the beginning of the
  compressed file.
 \item Use a special symbol as ``end-of-message,'' which is coded only at the
  end of the data sequence, and assign to this symbol the smallest probability
  value allowed by the encoder/decoder.
\end{enumerate}

As we explained above, the decoding procedure will always produce a decoded 
data sequence. However, how do we know that it is the right sequence? This 
can be inferred from the fact that if $S$ and ${S}'$ are sequences with $N$ 
symbols then
\begin{equation}
                                                           \label{eqAritCode13}
 S \ne {S}' \Leftrightarrow \Phi _N (S) \cap \Phi _N ({S}') = \emptyset.
\end{equation}

This guarantees that different sequences cannot produce the same code value. 
In Section~\ref{ssCorrect} we show that, due to approximations, we have
incorrect decoding if~(\ref{eqAritCode13}) is not satisfied.

\section{Optimality of Arithmetic Coding}                     \label{scOptimal}

Information theory~\cite{Shannon48,Gallager68,Jelinek68,McEliece84,Cover91,Sayood99,Salomon00}
shows us that the average number of 
bits needed to code each symbol from a stationary and memoryless source 
$\Omega$ cannot be smaller than its entropy $H(\Omega)$, defined by
\begin{equation}
                                                           \label{eqOptCode1}
 H(\Omega) = -\sum\limits_{m = 0}^{M - 1} {p(m)\log _2 p(m)} \quad \mbox{bits/symbol.}
\end{equation}

We have seen that the arithmetic coding process generates code values that 
are uniformly distributed across the interval [0,~1). This is a necessary 
condition for optimality, but not a sufficient one. In the interval $\Phi_N(S)$
we can choose values that require an arbitrarily large number of bits 
to be represented, or choose code values that can be represented with the 
minimum number of bits, given by equation~(\ref{eqAritCode4}). Now we show that the latter 
choice satisfies the sufficient condition for optimality.

To begin, we have to consider that there is some overhead in a compressed 
file, which may include
\begin{itemize}
 \item Extra bits required for saving $\hat{v}$ with an integer number of bytes.
 \item A fixed or variable number of bits representing the number of symbols coded.
 \item Information about the probabilities (\textbf{p} or \textbf{c}).
\end{itemize}

Assuming that the total overhead is a positive number $\sigma$~bits, we 
conclude from~(\ref{eqAritCode4}) that the number of bits per symbol
used for coding a sequence $S$ should be bounded by
\begin{equation}
                                                           \label{eqOptCode2}
 {\rm B}_S \le \frac{\sigma - \log_2(l_N )}{N} \quad \mbox{bits/symbol}.
\end{equation}

It follows from~(\ref{eqAritCode3}) that
\begin{equation}
                                                           \label{eqOptCode3}
 l_N = \prod\limits_{k=1}^N {p(s_k )} ,
\end{equation}
and thus
\begin{equation}
                                                           \label{eqOptCode4}
 {\rm B}_S \le \frac{\sigma - \sum\limits_{k=1}^N {\log _2 p(s_k )}}{N}
  \quad \mbox{bits/symbol.}
\end{equation}

Defining $E\left\{ \cdot \right\}$ as the expected value operator, the 
expected number of bits per symbol is
\begin{eqnarray}
                                                           \label{eqOptCode5}
 \bar{\rm B} = E\{{\rm B}_S \} & \le & \frac{\sigma - \sum\limits_{k=1}^N
               E\left\{ \log_2 p(s_k) \right\}}{N}
             = \frac{\sigma - \sum\limits_{k=1}^N {\sum\limits_{m=0}^{M - 1}
               {p(m)\log _2 p(m)} } }{N} \\ \nonumber
             & \le & H(\Omega) + \frac{\sigma}{N}
\end{eqnarray}

Since the average number of bits per symbol cannot be smaller than the 
entropy, we have
\begin{equation}
                                                           \label{eqOptCode6}
 H(\Omega ) \le \bar {{\rm B}} \le H(\Omega ) + \frac{\sigma }{N},
\end{equation}
and it follows that
\begin{equation}
                                                           \label{eqOptCode7}
  \mathop {\lim }\limits_{N \to \infty } \left\{ \bar {{\rm B}} \right\} = H(\Omega ),
\end{equation}
which means that arithmetic coding indeed achieves optimal compression 
performance.

At this point we may ask why arithmetic coding creates intervals, instead of 
single code values. The answer lies in the fact that arithmetic coding is 
optimal not only for binary output---but rather for any output alphabet. In 
the final interval we find the different code values that are optimal for 
each output alphabet. Here is an example of use with non-binary outputs.

\begin{Example}
Consider transmitting the data sequence of Example~\ref{exFirstEnc} using
a communications system that conveys information using three levels,
{\{}--V,~0,~+V{\}} (actually used in radio remote controls). Arithmetic
coding with ternary output can simultaneously compress the data and
convert it to the proper transmission format.

The generalization of~(\ref{eqAritCode4}) for a $D$-symbol output alphabet is
\begin{equation}
                                                           \label{eqOptCode8}
 {\rm B}_{\min}(l_N,D) = \left\lceil {-\log_D(l_N )} \right\rceil 
 \quad \mbox{symbols.}
\end{equation}

Thus, using the results in Table~\ref{tbFirsCod}, we conclude that we need
$\left\lceil-\log _3(0.0002) \right\rceil = 8$ ternary symbols. We later show how to 
use standard arithmetic coding to find that the shortest ternary 
representation is $\hat{v}_3=0.20200111_{3} \approx 0.742722146$, 
which means that the sequence $S=\{2,1,0,0,1,3\}$ can be 
transmitted as the sequence of electrical signals 
{\small {\{}+V,~0,~+V,~0,~0,~\mbox{--V},~\mbox{--V},~\mbox{--V}{\}}}.
\end{Example}

\section{Arithmetic Coding Properties}

\subsection{Dynamic Sources}                                   \label{ssDynSrc}

In Section~\ref{ssNotation} we assume that the data source $\Omega$ is 
stationary, so we 
have one set of symbol probabilities for encoding and decoding all symbols 
in the data sequence $S$. Now, with an understanding of the coding process, we 
generalize it for situations where the probabilities change for each symbol 
coded, i.e., the $k$-th symbol in the data sequence $S$ is a random variable with 
probabilities ${\bf p}_k$ and distribution ${\bf c}_k$.

The only required change in the arithmetic coding process is that instead of 
using~(\ref{eqAritCode3}) for interval updating, we should use
\begin{equation}
                                                         \label{eqPropDS1}
  \Phi_k(S) = \Range{b_k}{l_k} = \Range{b_{k-1}+c_k(s_k)\,l_{k-1}}
    {p_k(s_k)\,l_{k-1}},\quad k = 1,2,\,\ldots,N.
\end{equation}

To understand the changes in the decoding process, remember that the process 
of working with updated code values is equivalent to ``erasing'' all 
information about past symbols, and decoding in the [0,~1) interval. Thus, 
the decoder only has to use the right set of probabilities for that symbol 
to decode it correctly. The required changes to (\ref{eqAritCode10})
and~(\ref{eqAritCode11}) yield
\begin{eqnarray}
                                                          \label{eqPropDS2}
  \hat{s}_k(\hat{v}) & = & \left\{\,s:\;c_k(s)\le\frac{\hat{v}-b_{k-1}}{l_{k-1}}<c_k(s+1) \right\},\quad
    k = 1,2,\ldots,N, \\
                                                          \label{eqPropDS3}
  \Phi_k(S) & = & \Range{b_k}{l_k} = \Range{b_{k-1}+c_k(\hat{s}_k(\hat{v}))\,l_{k-1}}
    {p_k(\hat{s}_k(\hat{v}))\,l_{k-1}},\quad k = 1,2,\,\ldots,N.
\end{eqnarray}

Note that the number of symbols used at each instant can change. Instead of 
having a single input alphabet with $M$ symbols, we have a sequence of alphabet 
sizes $\{ M_1,M_2,\ldots,M_N \}$.

\subsection{Encoder and Decoder Synchronized Decisions}

In data compression an encoder can change its behavior (parameters, coding 
algorithm, etc.) while encoding a data sequence, as long as the decoder uses 
the same information and the same rules to change its behavior. In addition, 
these changes must be ``synchronized,'' not in time, but in relation to the 
sequence of data source symbols.

For instance, in Section~\ref{ssDynSrc}, we assume that the encoder and decoder are 
synchronized in their use of varying sets of probabilities. Note that we do 
not have to assume that all the probabilities are available to the decoder 
when it starts decoding. The probability vectors can be updated with any 
rule based on symbol occurrences, as long as ${\bf p}_k$ is computed 
from the data already available to the decoder, i.e.,
$\{\hat{s}_1,\hat{s}_2,\ldots,\hat{s}_{k-1}\}$.
This principle is used for adaptive coding, and it is covered in
Section~\ref{ssAdapCod}.

This concept of synchronization is essential for arithmetic coding because 
it involves a nonlinear dynamic system (Figure~\ref{fgDinSys}), and error accumulation 
leads to incorrect decoding, unless the encoder and decoder use \emph{exactly} the same 
implementation (same precision, number of bits, rounding rules, equations, 
tables, etc.). In other words, we can make arithmetic coding work correctly 
even if the encoder makes coarse approximations, as long as the decoder 
makes exactly the same approximations. We have already seen an example of a 
choice based on numerical stability: equations (\ref{eqAritCode10})
and~(\ref{eqAritCode11}) enable us to 
synchronize the encoder and decoder because they use the same interval 
updating rules used by~(\ref{eqAritCode3}), while (\ref{eqAritCode7})
and~(\ref{eqAritCode8}) use a different recursion.

\subsection{Separation of Coding and Source Modeling}       \label{ssSepCodMod}

There are many advantages for separating the source modeling (probabilities 
estimation) and the coding processes~\cite{Rissanen81,Witten87,Bell90,ITU93,Weinberger96,Moffat98,Witten99}.
For example, it allows us 
to develop complex compression schemes without worrying about the details in 
the coding algorithm, and/or use them with different coding methods and 
implementations.

Figure~\ref{fgSepCodMod} shows how the two processes can be separated in
a complete system 
for arithmetic encoding and decoding. The coding part is responsible only 
for updating the intervals, i.e., the arithmetic encoder implements 
recursion~(\ref{eqPropDS1}), and the arithmetic decoder implements
(\ref{eqPropDS2}) and~(\ref{eqPropDS3}). The 
encoding/decoding processes use the probability distribution vectors as 
input, but do not change them in any manner. The source modeling part is 
responsible for choosing the distribution ${\bf c}_k$ that is used to 
encode/decode symbol $s_k$. Figure~\ref{fgSepCodMod} also shows that a delay of one data 
symbol before the source-modeling block guarantees that encoder and decoder 
use the same information to update ${\bf c}_k$.

Arithmetic coding simplifies considerably the implementation of systems like 
Figure~\ref{fgSepCodMod} because the vector ${\bf c}_k$ is used directly
for coding. With Huffman coding, changes in probabilities require re-computing
the optimal code, or using complex code updating 
techniques~\cite{Gallagher78,Knuth85,Vitter87}.

%
\begin{figure}[tp]
\setlength{\unitlength}{1mm}
\begin{center}
 \begin{picture}(120,70)
%
%
  \thicklines
  \put(0,40){\makebox(20,10){\color{red} \shortstack{\textsf{Data} \\ \textsf{sequence}}}}
  \put(0,20){\makebox(20,10){\color{red} \shortstack{\textsf{Recovered} \\ \textsf{data}}}}
  \multiput(30,2)(0,60){2}{\framebox(20,6){\color{acblue} \textsf{Delay}}}
  \multiput(65,0)(0,60){2}{\framebox(25,10){\color{acblue}\shortstack{\textsf{Source} \\ \textsf{modeling}}}}
  \multiput(92,0)(0,60){2}{\makebox(25,10)[l]{\footnotesize \shortstack[l]{\textsf{Choice of probability} \\ \textsf{distribution}}}}
  \put(65,40){\framebox(25,10){\color{acblue} \shortstack{\textsf{Arithmetic} \\ \textsf{encoding}}}}
  \put(92,40){\makebox(25,10)[l]{\footnotesize \textsf{Interval updating}}}
  \put(65,20){\framebox(25,10){\color{acblue}\shortstack{\textsf{Arithmetic} \\ \textsf{decoding}}}}
  \put(92,20){\makebox(25,10)[l]{\footnotesize \shortstack[l]{\textsf{Interval selection} \\ \textsf{and updating}}}}
%
%
  \multiput(50,5)(0,60){2}{\vector(1,0){15}}
  \put(30,45){\vector(1,0){35}}
  \put(65,25){\vector(-1,0){35}}
  \put(40,45){\vector(0,1){17}}
  \put(40,25){\vector(0,-1){17}}
  \put(77.5,10){\vector(0,1){10}}
  \put(77.5,40){\vector(0,-1){10}}
  \put(77.5,60){\vector(0,-1){10}}
  \multiput(40,25)(0,20){2}{\circle*{1}}
%
%
  \put(25,45){\makebox(0,0){\large $s_k$}}
  \put(25,25){\makebox(0,0){\large $\hat{s}_k$}}
  \multiput(82,15)(0,40){2}{\makebox(0,0){\large ${\bf c}_k$}}
  \put(82,35){\makebox(0,0){\large ${\bf d}$}}
 \end{picture}
\end{center}
 \figurecaption{Separation of coding and source modeling tasks. Arithmetic encoding
  and decoding process intervals, while source modeling chooses
  the probability distribution for each data symbol.}
                                                            \label{fgSepCodMod}
\end{figure}

\subsection{Interval Rescaling}                               \label{ssIntScal}

Figure~\ref{fgCodeIntv} shows graphically one important property of arithmetic
coding: the actual intervals used during coding depend on the initial interval
and the previously coded data, but the proportions within subdivided intervals
do not. For example, if we change the initial interval to
$\Phi_0 = \Range{1}{2} = \Intv{1}{3}$ and apply~(\ref{eqAritCode3}), the coding
process remains the same, except that all intervals are scaled by a factor of
two, and shifted by one.

We can also apply rescaling in the middle of the coding process. Suppose that
at a certain stage $m$ we change the interval according to
\begin{equation}
                                                         \label{eqPropSc1}
  {b}'_m = \gamma \,(b_m-\delta), \quad {l}'_m = \gamma \,l_m,
\end{equation}
and continue the coding process normally (using (\ref{eqAritCode3})
or~(\ref{eqPropDS1})). When we finish 
coding we obtain the interval ${\Phi}'_N (S) = \Range{{b}'_N}{{l}'_N}$ 
and the corresponding code value ${v}'$. We can use the 
following equations to recover the interval and code value that we would 
have obtained without rescaling:
\begin{equation}
                                                         \label{eqPropSc2}
  b_N = \frac{{b}'_N}{\gamma}+\delta, \quad l_N = \frac{{l}'_N}{\gamma},
    \quad \hat{v} = \frac{{v}'}{\gamma}+\delta.
\end{equation}

The decoder needs the original code value $\hat{v}$ to start recovering the 
data symbols. It should also rescale the interval at stage $m$, and thus needs 
to know $m,\;\delta ,\;\gamma$. Furthermore, when it scales the interval 
using~(\ref{eqPropSc1}), it must scale the code value as well, using
\begin{equation}
                                                         \label{eqPropSc3}
  {v}' = \gamma \,(\hat{v}-\delta).
\end{equation}

We can generalize the results above to rescaling at stages
$m \le n \le \ldots \le p$. In general, the scaling process, including
the scaling of the code values is
\begin{equation}
                                                         \label{eqPropSc4}
  \begin{array}{lll}
    {b}'_m  = \gamma_1 \,(b_m-\delta_1), &
    {l}'_m  = \gamma_1 \,l_m, & 
    {v}'    = \gamma_1 \,(\hat{v}-\delta_1), \\
    {b}''_n = \gamma_2 \,({b}'_n-\delta_2), &
    {l}''_n = \gamma_2 \,{l}'_n, &
    {v}''   = \gamma_2 \,({v}'-\delta_2), \\
    \vdots & \vdots & \vdots \\
    b_p^{(T)} = \gamma_T \,(b_p^{(T-1)}-\delta_T), &
    l_p^{(T)} = \gamma_T \,l_p^{(T-1)}, &
    v^{(T)}   = \gamma_T \,(v^{(T-1)}-\delta_T).
  \end{array}
\end{equation}

At the end of the coding process we have interval
$\bar{\Phi}_N(S) = \Range{\bar{b}_N}{\bar{l}_N}$ and code value $\bar{v}$.
We recover original values using
\begin{equation}
                                                         \label{eqPropSc5}
  \Phi_N(S) = \Range{b_N}{l_N} = \Range{ \delta_1+\frac{1}{\gamma_1}
    \left( \delta_2+\frac{1}{\gamma_2} \left( \delta_3+\frac{1}{\gamma_3}
    \left( \cdots \left( \delta_T+\frac{\bar{b}_N}{\gamma_T}
    \right) \right) \right) \right)}
    {\frac{\bar{l}_N}{{\textstyle \prod_{i=1}^T \gamma_i}}} , 
\end{equation}
and
\begin{equation}
                                                         \label{eqPropSc6}
  \hat{v} = \delta_1+\frac{1}{\gamma_1} \left( \delta_2+\frac{1}{\gamma_2}
    \left( \delta_3+\frac{1}{\gamma_3} \left( \cdots
    \left( \delta_T+\frac{\bar {v}}{\gamma_T} \right) \right) \right) 
\right).
\end{equation}

These equations may look awfully complicated, but in some special cases they 
are quite easy to use. For instance, in Section~\ref{scFinPrec} we show
how to use scaling with $\delta_i \in \{0,1/2\}$ and $\gamma_i \equiv 2$, 
and explain the connection between $\delta_i$ and the binary representation 
of $b_N$ and $\hat{v}$. The next example shows another simple application 
of interval rescaling.

\begin{figure}[tp]
\begin{center}
  \includegraphics[scale=1.1]{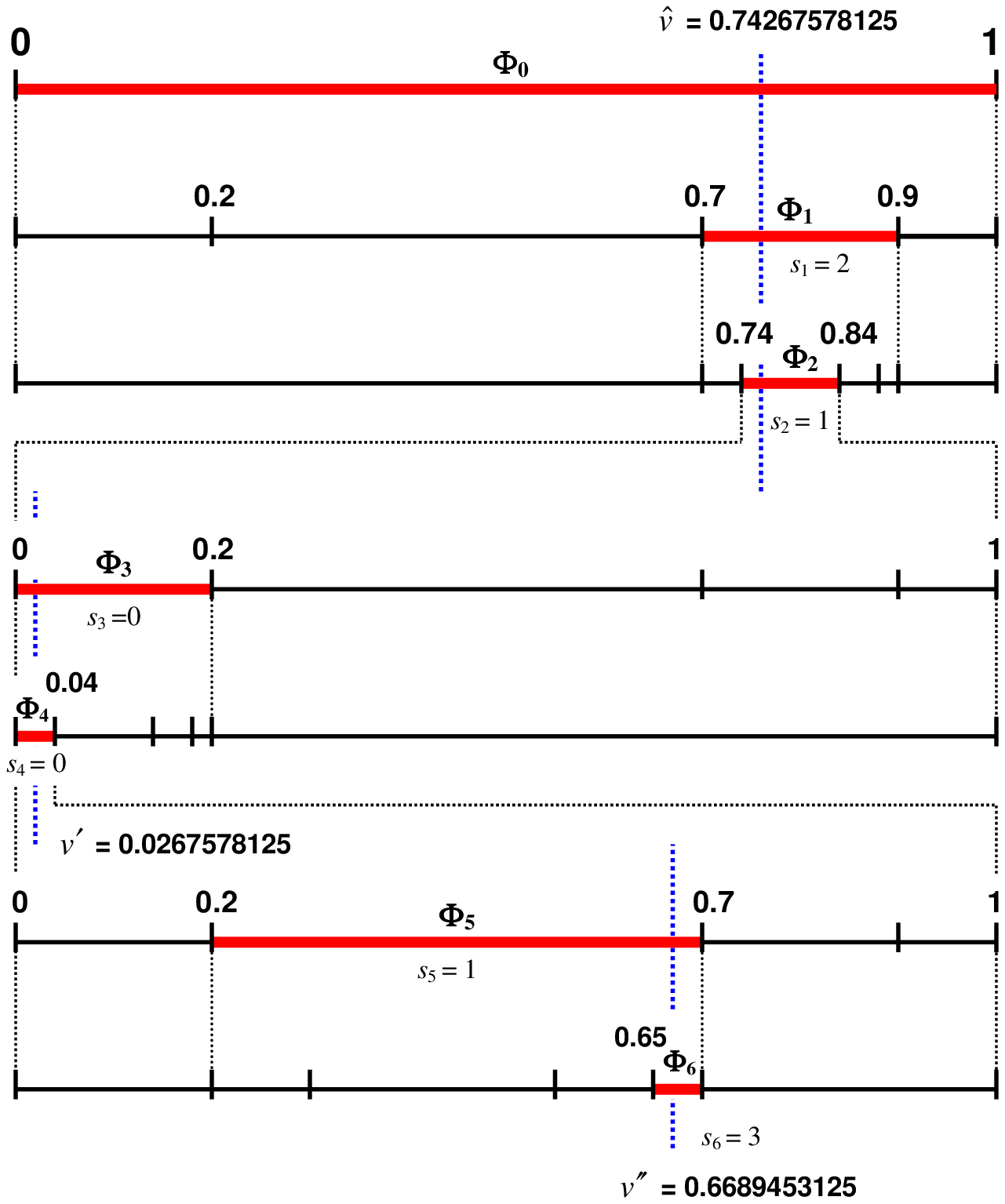}
\end{center}
 \figurecaption{Graphical representation of the arithmetic coding process of
  Example~\ref{exFirstEnc} (Figure~\ref{fgCodeIntv}) using numerical
  rescaling. Note that the code value changes each time the intervals
  are rescaled.}
                                                             \label{fgScalIntv}
\end{figure}

\begin{Example}
Figure~\ref{fgScalIntv} shows rescaling applied to Example~\ref{exFirstEnc}.
It is very similar to Figure~\ref{fgCodeIntv}, but instead of having just
an enlarged view of small intervals, in Figure~\ref{fgScalIntv} the
intervals also change. The rescaling parameters $\delta_1=0.74$ and
$\gamma_1=10$ are used after coding two symbols, and $\delta_2=0$ and
$\gamma_2=25$ after coding two more symbols. 
The final interval is $\bar{\Phi}_6(S)=\Range{0.65}{0.05}$, that
corresponds to
\[
 \Phi_6(S)= \Range{0.74+\frac{1}{10}\left(\frac{0.65}{25}\right)}
  {\frac{0.05}{10\times 25}} = \Range{0.7426}{0.0002},
\]
and which is exactly the interval obtained in Example~\ref{exFirstEnc}.
                                                      \label{exScalEnc}
\end{Example}

%

\subsection{Approximate Arithmetic}                           \label{ssApproxA}

To understand how arithmetic coding can be implemented with fixed-precision 
we should note that the requirements for addition and for multiplication are 
quite different. We show that if we are willing to lose some compression 
efficiency, then we do not need exact multiplications. We use the double 
brackets ($[[\, \cdot \,]])$ around a multiplication to indicate that it is 
an approximation, i.e., $[[\alpha \cdot \beta ]] \approx \alpha \cdot \beta$.
We define truncation as any approximation such that 
$[[\alpha \cdot \beta ]] \le \alpha \cdot \beta $.
The approximation we are considering here can be rounding or truncation 
to any precision. The following example shows an alternative way to 
interpret inexact multiplications.

\begin{Example}
We can see in Figure~\ref{fgDinSys} that the arithmetic coding multiplications
always occur with data from the source model---the probability $p$ and the
cumulative  distribution $c$. Suppose we have $l=0.04$, $c=0.317$, and $p=0.123$,
with
\begin{eqnarray*}
 l \times c & = & 0.04 \times 0.317 = 0.01268, \\
 l \times p & = & 0.04 \times 0.123 = 0.00492.
\end{eqnarray*}

Instead of using exact multiplication we can use an approximation (e.g., 
with table look-up and short registers) such that
\begin{eqnarray*}
 {[[l \times c]]} & = & [[0.04 \times 0.317]] = 0.012, \\
 {[[l \times p]]} & = & [[0.04 \times 0.123]] = 0.0048.
\end{eqnarray*}

Now, suppose that instead of using $p$ and $c$, we had used another model, 
with ${c}'=0.3$ and ${p}'=0.12$. We would have obtained
\begin{eqnarray*}
 l \times {c}' & = & 0.04 \times 0.3 = 0.012, \\
 l \times {p}' & = & 0.04 \times 0.12 = 0.0048,
\end{eqnarray*}
which are exactly the results with approximate multiplications. This shows 
that inexact multiplications are mathematically equivalent to making 
approximations in the source model and then using exact multiplications.
                                                      \label{exApproxMod}
\end{Example}

What we have seen in this example is that whatever the approximation used 
for the multiplications we can always assume that exact multiplications 
occur all the time, but with inexact distributions. We do not have to worry 
about the exact distribution values as long as the decoder is synchronized 
with the encoder, i.e., if the decoder is making exactly the same 
approximations as the encoder, then the encoder and decoder distributions 
must be identical (just like having dynamic sources, as explained in
Section~\ref{ssDynSrc}).

The version of~(\ref{eqAritCode3}) with inexact multiplications is
\begin{small}
\begin{equation}
                                                         \label{eqPropAM1}
  \Phi_k(S) = \Range{b_k}{l_k} = \Range{b_{k-1} + [[ c(s_k) \cdot l_{k-1} ]]}
    {[[p(s_k) \cdot l_{k-1} ]]}, \quad k = 1,2,\,\ldots ,N.
\end{equation}
\end{small}

We must also replace (\ref{eqAritCode10}) and~(\ref{eqAritCode11}) with
\begin{small}
\begin{eqnarray}
                                                         \label{eqPropAM2}
  \hat{s}_k(\hat{v}) & = & \left\{ s: b_{k-1}+[[c(s) \cdot l_{k-1} ]] \le \hat{v} < 
    b_{k-1}+[[c(s+1) \cdot l_{k-1} ]] \right\}, \; k = 1,2,\ldots,N, \\
                                                         \label{eqPropAM3}
  \Phi_k(\hat{v}) & = & \Range{b_k}{l_k} = \Range{b_{k-1} +
    [[c(\hat{s}_k(\hat{v})) \cdot \,l_{k-1} ]]}{[[p(\hat{s}_k(\hat{v})) \cdot l_{k-1} ]]},
    \; k = 1,2,\ldots,N.
\end{eqnarray}
\end{small}

In the next section we explain which conditions must be satisfied by the 
approximate multiplications to have correct decoding.

In equations (\ref{eqPropAM1}) to~(\ref{eqPropAM3}) we have one type of
approximation occurring from 
the multiplication of the interval length by the cumulative distribution, 
and another approximation resulting from the multiplication of the interval 
length by the probability. If we want to use only one type of approximation, 
and avoid multiplications between length and probability, we should update 
interval lengths according to
\begin{equation}
                                                         \label{eqPropAM4}
  l_k = \left( {b_{k-1}+[[c(s_k+1) \cdot l_{k-1} ]]} \right)-\left( 
    {b_{k-1}+[[c(s_k) \cdot l_{k-1} ]]} \right).
\end{equation}

The price to pay for inexact arithmetic is degraded compression performance. 
Arithmetic coding is optimal only as long as the source model probabilities 
are equal to the true data symbol probabilities; any difference reduces the 
compression ratios.

A quick analysis can give us an idea of how much can be lost. If we use a 
model with probability values ${\bf {p}'}$ in a source with 
probabilities ${\bf p}$, the average loss in compression is
\begin{equation}
                                                         \label{eqPropAM5}
  \Delta = \sum_{n = 0}^{M-1} {p(n)} \log_2 \left[ {\frac{p(n)}{{p}'(n)}} \right],
    \quad \mbox{bits/symbol.}
\end{equation}

This formula is similar to the relative entropy~\cite{Cover91}, but in
this case ${\bf {p}'}$ represents the values that would result from the
approximations, and it is possible to have
$\sum\nolimits_{n = 0}^{M-1} {p'(n)} \neq 1$.

Assuming a relative multiplication error within $\varepsilon$, i.e.,
\begin{equation}
                                                         \label{eqPropAM6}
  1-\varepsilon \le \frac{p(n)}{{p}'(n)} \le 1+\varepsilon,
\end{equation}
we have
\begin{equation}
                                                         \label{eqPropAM7}
  \Delta \le \sum\limits_{n = 0}^{M-1} {p(n)} \log_2 (1+\varepsilon) 
    \approx \frac{\varepsilon}{\ln (2)} \approx 1.4\;\varepsilon
    \quad \mbox{bits/symbol.} 
\end{equation}

This is not a very tight bound, but it shows that if we can make 
multiplication accurately to, say 4~digits, the loss in compression 
performance can be reasonably small.

\subsection{Conditions for Correct Decoding}                  \label{ssCorrect}

\begin{figure}[tp]
\begin{center}
  \includegraphics[scale=0.9]{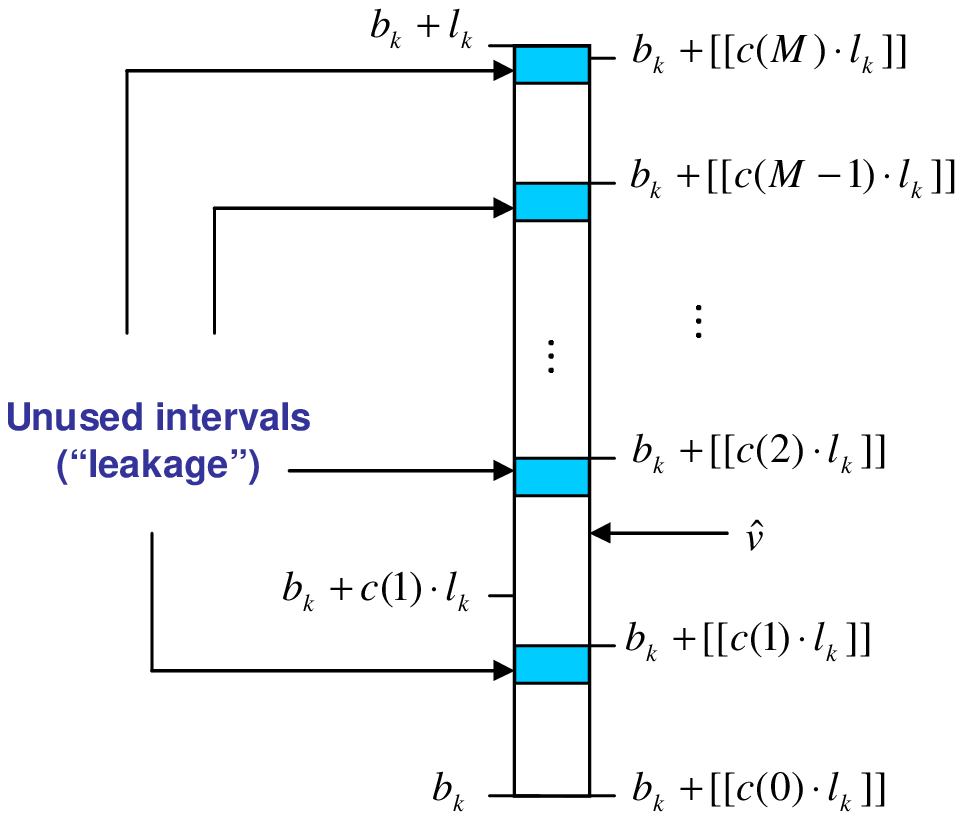}
\end{center}
 \figurecaption{Subdivision of a coding interval with approximate multiplications.
   Due to the fixed-precision arithmetic, we can only guarantee that all
   coding intervals are disjoint if we leave small regions between intervals
   unused for coding.}
                                                            \label{fgLeakDiag}
\end{figure}

Figure~\ref{fgLeakDiag} shows how an interval is subdivided when using inexact 
multiplications. In the figure we show that there can be a substantial
difference between, say, $b_k+ c(1) \cdot l_k$ and $b_k+[[c(1) \cdot l_k ]]$,
but this difference does not lead to decoding errors if the decoder uses
the same approximation.

%

Decoding errors occur when condition (\ref{eqAritCode13}) is not satisfied. 
Below we show the constraints that must be satisfied by approximations, and 
analyze the three main causes of coding error to be avoided.

\paragraph{(a) The interval length must be positive and intervals must be disjoint.} {\ } \par \nopagebreak

\noindent The constraints that guarantee that the intervals do not collapse into a 
single point, and that the interval length does not become larger than the 
allowed interval are
\begin{equation}
                                                         \label{eqPropCD1}
  0 < l_{k+1} = [[p(s) \cdot l_k ]] \le \left( {b_k+[[c(s+1) \cdot l_k ]]} \right)-
  \left( {b_k+[[c(s) \cdot l_k ]]} \right), \; s = 0,1,\ldots,M-1.
\end{equation}

For example, if the approximations can create a situation in which
$[[c(s+1) \cdot l_k ]] < [[c(s) \cdot l_k ]]$, there would
be an non-empty intersection of subintervals assigned for $s+1$ and $s$,
and decoder errors would occur whenever a code value belongs to the
intersection.

If $[[c(s+1) \cdot l_k ]] = [[c(s) \cdot l_k ]]$ then the interval
length collapses to zero, and stays as such, independently of the symbols
coded next.
The interval length may become zero due to arithmetic underflow, when both 
$l_k$ and $p(s) = c(s+1)-c(s)$ are very small. In Section~\ref{scFinPrec} we show 
that interval rescaling is normally used to keep $l_k$ within a certain 
range to avoid this problem, but we also have to be sure that all symbol 
probabilities are larger than a minimum value defined by the arithmetic 
precision (see Sections (\ref{eqIntArith1}) and~(\ref{eqAppendix1})).

Besides the conditions defined by~(\ref{eqPropCD1}), we also need to have
\begin{equation}
                                                         \label{eqPropCD2}
  [[c(0) \cdot l_k ]] \ge 0, \quad \mbox{and} \quad [[c(M) \cdot l_k ]] \le l_k.
\end{equation}
These two condition are easier to satisfy because $c(0) \equiv 0$ and
$c(M) \equiv 1$, and it is easy to make such multiplications exact.

\paragraph{(b) Sub-intervals must be nested.} {\ } \par \nopagebreak

\noindent We have to be sure 
that the accumulation of the approximation errors, as we continue coding 
symbols, does not move the interval base to a point outside all the previous 
intervals. With exact arithmetic, as we code new symbols, the interval base 
increases within the interval assigned to $s_{k+1}$, but it never crosses 
the boundary to the interval assigned to $s_{k+1}+1$, i.e.,
\begin{equation}
                                                         \label{eqPropCD3}
  b_{k+n} = b_k+\sum\limits_{i = k}^{k+n-1} {c(s_{i+1}) \cdot l_i < 
    b_k+c(s_{k+1}+1) \cdot l_k ,\quad \mbox{for all} \quad n \ge 0}.
\end{equation}

The equivalent condition for approximate arithmetic is that for every data 
sequence we must have
\begin{equation}
                                                         \label{eqPropCD4}
b_k+[[c(s_{k+1}+1) \cdot l_k ]] > b_k+[[c(s_{k+1}) \cdot l_k ]]+
\sum\limits_{i = k+1}^\infty {[[c(s_{i+1}) \cdot l_i ]]}.
\end{equation}

To determine when~(\ref{eqPropCD4}) may be violated we have to assume
some limits on the  multiplication approximations. There should be a
non-negative number $\varepsilon$ such that
\begin{equation}
                                                         \label{eqPropCD5}
  [[c(s_{i+1}) \cdot l_i ]](1-\varepsilon) < c(s_{i+1}) \cdot l_i.
\end{equation}

We can combine (\ref{eqPropAM4}), (\ref{eqPropCD4}) and~(\ref{eqPropCD5}) to obtain
\begin{equation}
                                                         \label{eqPropCD6}
  (1-\varepsilon) \cdot l_k > \sum\limits_{i = k+1}^\infty {c(s_{i+1}) \cdot l_i},
\end{equation}
which is equal to
\begin{equation}
                                                         \label{eqPropCD7}
  1-\varepsilon > c(s_{k+2})+p(s_{k+3})\left( {c(s_{k+3})+
    p(s_{k+3})\left( {c(s_{k+4})+p(s_{k+4})\left( \cdots \right)} 
    \right)} \right).
\end{equation}

To find the maximum for the right-hand side of~(\ref{eqPropCD7})
we only have to consider 
the case $s_{k+2} = s_{k+3} = \ldots = M-1$ to find
\begin{equation}
                                                         \label{eqPropCD8}
  1-\varepsilon > c(M-1)+p(M-1)\left( {c(M-1)+p(M-1)\left( {c(M-1)+p(M-1)
    \left( \cdots \right)} \right)} \right),
\end{equation}
which is equivalent to
\begin{equation}
                                                         \label{eqPropCD9}
  1-\varepsilon > c(M-1)+p(M-1).
\end{equation}

But we know from~(\ref{eqNota3}) that by definition $c(M-1)+p(M-1) \equiv 1$! The 
answer to this contradiction lies in the fact that with exact arithmetic we 
would have equality in~(\ref{eqPropCD3}) only after an infinite number of symbols.
With inexact arithmetic it is impossible to
have semi-open intervals that are fully used and match perfectly, so we need to take some 
extra precautions to be sure that~(\ref{eqPropCD4}) is always satisfied. What 
equation~(\ref{eqPropCD9}) tells us is that we solve the problem if we artificially 
decrease the interval range assigned for $p(M-1)$. This is equivalent to setting
aside small regions, indicated as gray areas in Figure~\ref{fgLeakDiag}, that are not
used for coding, and serve as a ``safety net.''

This extra space can be intentionally added, for example, by replacing~(\ref{eqPropAM4})
with
\begin{equation}
                                                         \label{eqPropCD11}
  l_k = \left( {b_{k - 1}+[[c(s_k+1) \cdot l_{k-1} ]]} \right)-\left( {b_{k-1}+
[[c(s_k) \cdot l_{k-1} ]]} \right)-\zeta 
\end{equation}
where $0 < \zeta \ll 1$ is chosen to guarantee correct coding and small compression
loss.

The loss in compression caused by these unused subintervals is called 
``leakage'' because a certain fraction of bits is ``wasted'' whenever a 
symbol is coded. This fraction is on average
\begin{equation}
                                                         \label{eqPropCD10}
  \Delta_s = p(s)\log_2 \left( {\frac{p(s)}{{p}'(s)}} \right) \quad \mbox{bits,}
\end{equation}
where $p(s) / {p}'(s) > 1$ is the ratio between the symbol probability and 
the size of interval minus the unused region. With reasonable precision, leakage
can be made extremely small. For instance, if $p(s) / {p}'(s) = 1.001$ (low precision)
then leakage is less than 0.0015~bits/symbol.

\paragraph{(c) Inverse arithmetic operations must not produce error accumulation.} {\ } \par \nopagebreak

\noindent Note that in~(\ref{eqPropAM2}) we 
define decoding assuming only the additions and multiplications used by the 
encoder. We could have used
\begin{equation}
                                                         \label{eqPropCD12}
  \hat{s}_k(\hat{v}) = \left\{ {s:c(s) \le \left[ {\left[ {\frac{\hat{v}-b_{k-1} }{l_{k-1}}}
  \right]} \right] < c(s+1)} \right\},\quad k = 1,2,\ldots,N.
\end{equation}

However, this introduces approximate subtraction and division, which have to 
be consistent with the encoder's approximations. Here we cannot possibly 
cover all problems related to inverse operations, but we should say that the 
main point is to observe error accumulation.

For example, we can exploit the fact that in~(\ref{eqAritCode10}) decoding
only uses the difference $\varpi_k \equiv \hat{v}-b_k $, and use the following
recursions. 
\begin{eqnarray}
                                                         \label{eqPropCD13}
  \Range{\varpi_0}{l_0} & = & \Range{\hat{v}}{1}, \\
                                                         \label{eqPropCD14}
  \hat{s}_k & = & \left\{ {s:[[c(s) \cdot l_{k-1} ]] \le \varpi_k <
    [[c(s+1) \cdot l_{k-1} ]]} \right\}, \quad k = 1,2,\ldots,N. \\
                                                         \label{eqPropCD15}
  \Range{\varpi_k}{l_k} & = & \Range{\varpi_{k-1}-[[c(\hat{s}_k) \cdot \,l_{k-1} ]]}
    {[[p(\hat{s}_k) \cdot l_{k-1} ]]}, \quad k = 1,2,\ldots,N.
\end{eqnarray}

However, because we are using a sequence of subtractions in~(\ref{eqPropCD15}),
this technique works with integer arithmetic implementations (see Appendix~A), 
but it may not work with floating-point implementations because of error 
accumulation.

%
%
\chapter{Arithmetic Coding Implementation}

In this second part, we present the practical implementations of arithmetic 
coding. We show how to exploit all the arithmetic coding properties 
presented in the previous sections and develop a version that works with 
fixed-precision arithmetic. First, we explain how to implement binary 
extended-precision additions that exploit the arithmetic coding properties, 
including the carry propagation process. Next, we present complete encoding 
and decoding algorithms based on an efficient and simple form of interval 
rescaling. We provide the description for both floating-point and integer 
arithmetic, and present some alternative ways of implementing the coding,
including different scaling and carry propagation strategies. After covering
the details of the coding process, we study the symbol probability estimation
problem, and explain how to implement adaptive coding by integrating coding
and source modeling. At the end, we analyze the computational complexity
of arithmetic coding.

\section{Coding with Fixed-Precision Arithmetic}              \label{scFinPrec}

Our first practical problem is that the number of digits (or bits) 
required to represent the interval length exactly grows when a symbol is 
coded. For example, if we had $p(0) = 0.99$ and we repeatedly code symbol 0, 
we would have
\[
  l_0 = 1, \quad l_1 = 0.99, \quad l_2 = 0.9801, \quad l_3 = 0.970299,
    \quad l_4 = 0.96059601, \quad \ldots 
\]

We solve this problem using the fact we do not need exact 
multiplications by the interval length (Section~\ref{ssApproxA}). Practical 
implementations use $P$-bit registers to store approximations of the mantissa 
of the interval length and the results of the multiplications. All bits with
significance smaller than those in the register are assumed to be zero.

With the multiplication precision problem solved, we still have the problem 
of implementing the additions in~(\ref{eqPropAM1}) when there is a large difference 
between the magnitudes of the interval base and interval length. We show 
that rescaling solves the problem, simultaneously enabling exact addition, and
reducing loss of multiplication accuracy. For a binary output, we can use 
rescaling in the form of~(\ref{eqPropSc1}), with $\delta \in \{0,1/2\}$ and $\gamma=2$ 
whenever the length of the interval is below 1/2. Since the decoder needs to 
know the rescaling parameters, they are saved in the data buffer ${\bf d}$,
using bits ``0'' or ``1'' to indicate whether $\delta=0$ or $\delta=1/2$. 
Special case $\delta=1$ and $\gamma=1$, corresponding to a carry in the
binary representation, is explained later.

To simplify the notation we represent the rescaled intervals simply as 
$\Range{b}{l}$ (no subscripts), and the rescaled code value as $v$.
\begin{eqnarray}
 l & = & 2^{t(l_k)} l_k, \nonumber \\
                                                         \label{eqFixPrec1}
 b & = & \mbox{frac}(2^{t(l_k)}b_k) = 2^{t(l_k)} b_k - \left\lfloor 2^{t(l_k)} b_k \right\rfloor, \\
 v & = & \mbox{frac}(2^{t(l_k)}\hat{v}), \nonumber
\end{eqnarray}
where $\mbox{frac}(\cdot)$ is the fractional part of a number and
\begin{equation}
                                                         \label{eqFixPrec2}
  t(x) = \{n:2^{-n-1} < x \le 2^{-n}\} = \left\lfloor {-\log_2 (x)}  \right\rfloor .
\end{equation}

Note that under these conditions we have $b \in [0,\,1)$ and $l \in (0.5,\;1]$,
and for that reason the rescaling process is called \emph{renormalization}.
Of course, $l$, $b$, and $v$ change with $k$, but this new notation is more convenient
to represent variables in algorithm descriptions or computer programs.

The binary representations of the interval base and length have the 
following structure:
\begin{equation}
                                                         \label{eqFixPrec3}
  \begin{array}{rlllc}
   l_k = & 0.0000\,\ldots 00 & 0000\ldots 00 & 
     \overbrace{1aaa\ldots aa}^{(L=2^Pl)} & 000000\ldots _2  \\
   b_k = & \underbrace{0.aaaa\,\ldots aa}_{\mbox{\small settled}} &
     \underbrace{\mbox{0111}\ldots \mbox{11}}_{\mbox{\small outstanding}} &
     \underbrace{aaaa\ldots aa}_{\stackrel{\scriptstyle (B=2^Pb)}{\mbox{\small active}}} &
     \underbrace{000000\ldots _2 }_{\mbox{\small trailing zeros}}
  \end{array}
\end{equation}
where symbol $a$ represents an arbitrary bit value.

We can see in~(\ref{eqFixPrec3}) that there is ``window'' of $P$ \emph{active} bits, forming
integers $L$ and $B$, corresponding to the nonzero bits of $l_k$, and the renormalized length $l$. 
Because the value of $l$ is truncated to $P$-bit precision, there is a set of 
trailing zeros that does not affect the additions. The bits to the left of 
the active bits are those that had been saved in the data buffer ${\bf d}$ 
during renormalization, and they are divided in two sets.

The first set to the left is the set of \emph{outstanding} bits: those that can be changed
due to a carry from the active bits when new symbols are encoded. The second is the 
set of bits that have been \emph{settled}, i.e., they stay constant until the end of the 
encoding process. This happens because intervals are nested, i.e., 
the code value cannot exceed
\begin{equation}
                                                         \label{eqFixPrec4}
  b_{k+n} < b_k + l_k \le 
      \underbrace{0.aaaa\ldots aa}_{\mbox{\small settled}} 
      \underbrace{1000\ldots 00}_{\mbox{\small changed by carry}} 
      \underbrace{aaaa\ldots aa}_{\mbox{\small active}}
      \quad 00000\ldots _2 \hfill \\
\end{equation}

This equation shows that only the outstanding bits may change due to a carry 
from the active bits. Furthermore, inequality~(\ref{eqFixPrec4}) also shows
that there can be only 
one carry that would change these bits. If there is a carry, or when it is 
found that there can be no carry, these bits become settled. For that 
reason, the set of outstanding bits always start with~0, and is possibly 
followed only by 1s. As new symbols are encoded, all sets move to the right.

\subsection{Implementation with Buffer Carries}              \label{ssFPImplem}

Combining all we have seen, we can present an encoding algorithm that works 
with fixed-precision arithmetic. Algorithm~\ref{alFPEnc} shows a function
\mbox{\textsf{Arithmetic{\_}Encoder}} to encode a 
sequence $S$ of $N$ data symbols, following the notation of Section~\ref{ssNotation}. 
This algorithm is very similar to the encoding process that we used in 
Section~\ref{ssAritCode}, but with a renormalization stage after each time a symbol is 
coded, and the settled and outstanding bits being saved in the buffer 
${\bf d}$. The function returns the number of bits used to compress $S$. 

\begin{Algorithm}{{Function Arithmetic{\_}Encoder} $(N,S,M,{\bf c},{\bf d})$}
1. set \{ \= $b \leftarrow 0; \quad l \leftarrow 1$; \COM{Initialize interval}
          \> $t \leftarrow 0$; \} \COM{and bit counter}
\pushtabs
2. \= for $k$ = 1 to $N$ do \COM{Encode N data symbols}
   \> 2.1. \= \textsf{Interval{\_}Update} $(s_k,b,l,M,{\bf c})$;  \CON{Update interval according to symbol}
   \> 2.2. if $b \ge 1$ then \CON{Check for carry}
   \> \> 2.2.1. set \{ \= $b \leftarrow b-1$; \COO{Shift interval base}
   \> \> \> \textsf{Propagate{\_}Carry} $(t,\bf{d})$; \} \COO{Propagate carry on buffer}
   \> 2.3. if $l \le 0.5$ then \CON{If interval is small enough}
   \> \> 2.3.1. \textsf{Encoder{\_}Renormalization} $(b,l,t,{\bf d})$; \COO{then renormalize interval}
3. \textsf{Code{\_}Value{\_}Selection} $(b,t,{\bf d})$;  \COM{Choose final code value}
\poptabs
4. return $t$. \COM{Return number of code bits}
                                                        \label{alFPEnc}
\end{Algorithm}

In Algorithm~\ref{alFPEnc}, Step~1 sets the initial interval equal to [0,~1),
and initializes the bit 
counter~$t$ to zero. Note that we use curly braces (\{~\}) to enclose a set 
of assignments, and use symbol ``$\star$'' before comments. In Step~2, we have the 
sequential repetition of interval resizing and renormalizations.
Immediately after updating the interval we find out if there is a carry, i.e.,
if $b \ge 1$, and next we check if further renormalization is necessary.
The encoding process finishes in Step~3, when the final code value $v$ that 
minimizes the number of code bits is chosen. In all our algorithms, we 
assume that functions receive references, i.e., variables can be changed 
inside the function called. Below we describe each of the functions used by 
Algorithm~\ref{alFPEnc}.

There are many mathematically equivalent ways of updating the interval 
$\Range{b}{l}$. We do not need to have both vectors ${\bf p}$ and ${\bf c}$
stored to use~(\ref{eqAritCode3}). In Algorithm~\ref{alFPIntUpdt} we use~(\ref{eqPropAM4})
to update length as a difference, and we avoid multiplication for the
last symbol ($s=M-1$), since it is more 
efficient to do the same at the decoder. 
To simplify notation, we do not use double brackets to indicate 
inexact multiplications, but it should be clear that here all numbers 
represent the content of CPU registers.

\begin{Algorithm}{{Procedure Interval{\_}Update} $(s,b,l,M,{\bf c})$}
1. \= if $s = M - 1$ \COM{Special case for last symbol}
   \> then \= set $y \leftarrow b+l$; \CON{end of interval}
   \> else \> set $y \leftarrow b+l \cdot c(s+1)$; \CON{base of next subinterval}
2. \> set \{ \> $b \leftarrow b+l \cdot c(s)$; \COM{Update interval base}
   \> \> $l \leftarrow y-b$; \} \COM{Update interval length as difference}
3. return.
                                                    \label{alFPIntUpdt}
\end{Algorithm}

In Step~2.2.1 of Algorithm~\ref{alFPEnc}, the function to propagate the carry in
the buffer ${\bf d}$ is called, changing bits that have been added to ${\bf d}$
previously, and we shift the interval to have $b < 1$.
Figure~\ref{fgCarryProp} shows the carry propagation process. Active bits
are shown in bold and outstanding bits are underlined. Whenever there is a
carry, starting from the most recent bits added to buffer ${\bf d}$,
we complement all bits 
until the first~0-bit is complemented, as in Algorithm~\ref{alFPPrCarry}.

%
\setlength{\unitlength}{1.2mm}
\begin{figure}[tp]
\begin{center}
 \begin{picture}(100,30)
%
%
  \thinlines
  \multiput(0,5)(0,10){2}{\framebox(5,5){\color{acblue} 0}}
  \multiput(5,5)(0,10){2}{\framebox(5,5){\color{acblue} 1}}
  \multiput(10,5)(0,10){2}{\framebox(5,5){\color{acblue} 1}}
  \multiput(15,5)(0,10){2}{\framebox(5,5){\color{acblue} 0}}
  \put(20,5){\framebox(5,5){\color{acblue} 1}}
  \put(20,15){\framebox(5,5){\underline{0}}}
  \multiput(25,15)(5,0){5}{\framebox(5,5){\underline{1}}}
  \multiput(25,5)(5,0){4}{\framebox(5,5){\color{acblue} 0}}
  \put(45,5){\framebox(5,5){\underline{0}}}
  \multiput(55,15)(5,0){2}{\framebox(5,5){\textbf{1}}}
  \multiput(55,5)(5,0){2}{\framebox(5,5){\textbf{0}}}
  \multiput(65,15)(5,0){2}{\framebox(5,5){\textbf{0}}}
  \multiput(65,5)(5,0){2}{\framebox(5,5){\textbf{1}}}
  \multiput(50,5)(0,5){3}{\framebox(5,5){\textbf{1}}}
  \put(55,10){\framebox(5,5){\textbf{0}}}
  \multiput(60,10)(5,0){3}{\framebox(5,5){\textbf{1}}}
%
%
  \thicklines
  \put(77,17){\makebox(0,0)[l]{$b_{k-1}$}}
  \put(77,12){\makebox(0,0)[l]{$[[l_{k-1} \cdot c(s_k)]]$}}
  \put(77,7){\makebox(0,0)[l]{$b_k = b_{k-1} + [[l_{k-1} \cdot c(s_k)]]$}}
  \put(38,0){\makebox(0,0){\color{acgreen} \textsf{carry propagation}}}
  \put(53,3){\vector(-1,0){30}}

 \end{picture}
\end{center}
 \figurecaption{Carry propagation process. Bold letters indicate the active bits,
   outstanding bits are underlined, and leftmost bits are settled.}
                                                     \label{fgCarryProp}
\end{figure}

\begin{Algorithm}{{Procedure Propagate{\_}Carry} $(t,{\bf d})$}
1. set $n \leftarrow t$; \COM{Initialize pointer to last outstanding bit}
2. \= while $d(n)=1$ do \COM{While carry propagation}
   \> 2.1. set \{ \= $d(n) \leftarrow 0$; \CON{complement outstanding 1-bit and}
   \> \> $n \leftarrow n-1$; \} \CON{move to previous bit}
3. set $d(n) \leftarrow 1$; \COM{Complement outstanding 0-bit}
4. return.
                                                    \label{alFPPrCarry}
\end{Algorithm}

Algorithm~\ref{alFPENormal} implements interval renormalization, where
we test if active 
bits became outstanding or settled. While the interval length is smaller 
than 0.5, the interval is rescaled by a factor of two, and a bit is added to 
the bit buffer~${\bf d}$.

\begin{Algorithm}{{Procedure Encoder{\_}Renormalization} $(b,l,t,{\bf d})$}
1. \= while $l \le 0.5$ do \COM{Renormalization loop}
   \> 1.1. \= set \{ \= $t \leftarrow t+1$;        \CON{Increment bit counter and}
   \> \> \> $l \leftarrow 2l$; \}            \CON{scale interval length}
   \> 1.2. if $b \ge 0.5$ \CON{Test most significant bit of interval base}
   \> \> then \= set \{ \= $d(t) \leftarrow 1$; \COO{Output bit 1}
   \> \> \> \> $b \leftarrow 2(b-0.5)$; \}  \COO{shift and scale interval base}
   \> \> else \> set \{ $d(t) \leftarrow 0$; \COO{Output bit 0}
   \> \> \> \> $b \leftarrow 2b$; \}            \COO{scale interval base}
2. return.
                                                    \label{alFPENormal}
\end{Algorithm}

The final encoding stage is the selection of the code value. We use 
basically the same process explained at the end of Section~\ref{ssEncProc}, but here we 
choose a code value belonging to the rescaled interval. Our choice is made 
easy because we know that, after renormalization, we always have $0.5 < l \leq 1$
(see~(\ref{eqFixPrec1})),
meaning that we only need an extra bit to define the final code value. In other 
words: all bits that define the code value are already in buffer ${\bf d}$, 
and we only need to choose one more bit. The only two choices to consider in 
the rescaled interval are $v=0.5$ or $v=1$.

\begin{Algorithm}{{Procedure Code{\_}Value{\_}Selection} $(b,t,{\bf d})$}
1. \= set $t \leftarrow t+1$; \COM{Increment bit counter}
2. \= if $b \le 0.5$ \COM{Eenormalized code value selection}
   \> then \= set \hspace{1ex} \= $d(t) \leftarrow 1$; \CON{Choose $v=0.5$: output bit 1}
   \> else \> set \{ \> $d(t) \leftarrow 0$; \CON{Choose $v=1.0$: output bit 0 and}
   \> \> \> Propagate{\_}Carry $(t-1,{\bf d})$; \} \CON{propagate carry}
3. return.
                                                    \label{alFPValSelc}
\end{Algorithm}

The decoding procedure, shown in Algorithm~\ref{alFPDec}, gets as input
the number of compressed symbols, $N$, the number of data symbols, $M$,
and their cumulative distribution ${\bf c}$, and the array with the
compressed data bits, ${\bf d}$. Its output is the recovered data sequence
$\hat{S}$.
The decoder must keep the $P$-bit register with the code value updated, so it 
will read $P$ extra bits at the end of ${\bf d}$. We assume that this can be 
done without problems, and that those bits had been set to zero.

\begin{Algorithm}{{Procedure Arithmetic{\_}Decoder} $(N,M,{\bf c},{\bf d},\hat{S})$}
1. \= set \{ \= $b \leftarrow 0;\; l \leftarrow 1$; \COM{Initialize interval}
   \> \> $v = \sum\nolimits_{n=1}^P {2^{-n}d(n)}$; \COM{Read P bits of code value}
   \> \> $t \leftarrow P$; \} \COM{Initialize bit counter}
2. for $k=1$ to $N$ do \COM{Decode $N$ data symbols}
   \> 2.1. \= $\hat{s}_k = $ \textsf{Interval{\_}Selection} $(v,b,l,M,{\bf c})$; \CON{Decode symbol and update interval}
   \> 2.2. if $b \ge 1$ then \CON{Check for ``overflow''}
   \> \> 2.2.1. set \{ \= $b \leftarrow b-1$; \COO{shift interval base}
   \> \> \> $v \leftarrow v-1$; \} \COO{shift code value}
   \> 2.3. \= if $l \le 0.5$ then \CON{If interval is small enough}
   \> \> 2.3.1. \textsf{Decoder{\_}Renormalization} $(v,b,l,t,{\bf d})$; \COO{then renormalize interval}
3. return.
                                                        \label{alFPDec}
\end{Algorithm}

The interval selection is basically an implementation of~(\ref{eqPropAM2}):
we want to find the subinterval that contains the code value $v$. The
implementation that we show in Algorithm~\ref{alFPIntSel} has one small
shortcut: it combines the symbol decoding with interval
updating~(\ref{eqPropAM3}) in a single function. We do a sequential
search, starting from the last symbol ($s=M-1$), because we assume
that symbols are sorted by increasing probability. The advantages of
sorting symbols and more efficient searches are explained in
Section~\ref{ssAdapCod}.

\begin{Algorithm}{{Function Interval{\_}Selection} $(v,b,l,M,{\bf c})$}
1. \= set \{ \= $s \leftarrow M-1$; \COM{Start search from last symbol}
   \> \> $x \leftarrow b + l \cdot c(M-1)$; \COM{Base of search interval}
   \> \> $y \leftarrow b+l$; \} \COM{End of search interval}
\pushtabs
2. \= while $x > v$ do \COM{Sequential search for correct interval}
   \> 2.1. set \{ \= $s \leftarrow s-1$; \CON{Decrement symbol by one}
   \> \> $y \leftarrow x$;  \CON{move interval end}
   \> \> $x \leftarrow b+l \cdot c(s)$; \} \CON{compute new interval base}
\poptabs
3. set \{ $b \leftarrow x$; \COM{Update interval base}
   \> \> $l \leftarrow y-b$; \} \COM{Update interval length as difference}
4. return $s$.
                                                     \label{alFPIntSel}
\end{Algorithm}

In Algorithm~\ref{alFPIntSel} we use only arithmetic operations that are
exactly equal to those used by the encoder. This way we can easily guarantee
that the encoder and decoder approximations are exactly the same. Several
simplifications can be used to reduce the number of arithmetic operations
(see Appendix~A).

The renormalization in Algorithm~\ref{alFPDNormal} is
similar to Algorithm~\ref{alFPENormal}, and in fact all its decisions
(comparisons) are meant to be based on exactly the same values used by
the encoder. However, it also needs to rescale the code value in the 
same manner as the interval base (compare (\ref{eqPropSc5})
and~(\ref{eqPropSc6})), and it reads its least significant bit
(with value $2^{-P})$.

\begin{Algorithm}{{Procedure Decoder{\_}Renormalization} $(v,b,l,t,{\bf d})$}
1. \= while $l \le 0.5$ do \COM{Renormalization loop}
   \> 1.1. \= if $b \ge 0.5$ \CON{Remove most significant bit}
   \> \> then \= set \{ \= $b \leftarrow 2(b-0.5)$; \COO{shift and scale interval base}
   \> \> \> \> $v \leftarrow 2(v-0.5)$; \} \COO{shift and scale code value}
   \> \> else \> set \{ \> $b \leftarrow 2b$; \COO{scale interval base}
   \> \> \> \> $v \leftarrow 2v$; \} \COO{scale code value}
   \> 1.2. set \{ $t \leftarrow t+1$; \CON{Increment bit counter}
   \> \> \> $v \leftarrow v+2^{-P}d(t)$; \CON{Set least significant bit of code value}
   \> \> \> $l \leftarrow 2l$; \} \CON{Scale interval length}
2. return.
                                                     \label{alFPDNormal}
\end{Algorithm}

\begin{Example}
We applied Algorithms \ref{alFPEnc} to \ref{alFPDNormal} to the source
and data sequence of Example~\ref{exFirstEnc}, and the results are shown
in Table~\ref{tbRenormEx}. The first column shows what event caused the
change in the interval. If a new symbol $\sigma$ is coded, then we show it
as $s=\sigma$, and show the results of Algorithm~\ref{alFPIntUpdt}. We
indicate the interval changes during renormalization
(Algorithm~\ref{alFPENormal}) by showing the value of $\delta$ used for
rescaling, according to~(\ref{eqPropSc1}).

Comparing these results with those in Table~\ref{tbFirsCod}, we see how
renormalization keeps all numerical values within a range that maximizes
numerical accuracy. In fact, the results in Table~\ref{tbRenormEx} are
exact, and can be shown with a few significant digits.
Table~\ref{tbRenormEx} also shows the decoder's updated code value. 
Again, these results are exact and agree with the results shown in
Table~\ref{tbFirsCod}. The third column in Table~\ref{tbRenormEx} shows
the contents of the bit buffer~${\bf d}$, and the bits that are added
to this buffer every time the interval is rescaled. Note that carry
propagation occurs twice: when $s=3$, and when the final code value
$v=1$ is chosen by Algorithm~\ref{alFPValSelc}.
                                                      \label{exNormalEnc}
\end{Example}

%

\begin{table}[t]
\begin{center}
  \begin{tabular}{|l|l|l|l|l|l|} \hline
  \multicolumn{1}{|c|}{\bf Event} & \multicolumn{1}{|c|}{\bf Scaled} &
    \multicolumn{1}{|c|}{\bf Scaled} & \multicolumn{1}{|c|}{\bf Bit buffer} &
    \multicolumn{1}{|c|}{\bf Scaled} & \multicolumn{1}{|c|}{\bf Normalized} \\
  & \multicolumn{1}{|c|}{\bf base} & \multicolumn{1}{|c|}{\bf length} & &
    \multicolumn{1}{|c|}{\bf code value} & \multicolumn{1}{|c|}{\bf code value} \\
  & \multicolumn{1}{|c|}{$b$} & \multicolumn{1}{|c|}{$l$} & \multicolumn{1}{|c|}{\bf d} &
    \multicolumn{1}{|c|}{$v$} & \multicolumn{1}{|c|}{$(v-b)/l$} \\ \hline \hline
 ---         & 0      & 1      &               & 0.74267578125 & 0.74267578125 \\ \hline
$s=2$        & 0.7    & 0.2    &               & 0.74267578125 &               \\ \hline
$\delta=0.5$ & 0.4    & 0.4    & 1             & 0.4853515625  &               \\ \hline
$\delta=0$   & 0.8    & 0.8    & 10            & 0.970703125   & 0.21337890625 \\ \hline
$s=1$        & 0.96   & 0.4    & 10            & 0.970703125   &               \\ \hline
$\delta=0.5$ & 0.92   & 0.8    & 101           & 0.94140625    & 0.0267578125  \\ \hline
$s=0$        & 0.92   & 0.16   & 101           & 0.94140625    &               \\ \hline
$\delta=0.5$ & 0.84   & 0.32   & 1011          & 0.8828125     &               \\ \hline
$\delta=0.5$ & 0.68   & 0.64   & 10111         & 0.765625      & 0.1337890625  \\ \hline
$s=0$        & 0.68   & 0.128  & 10111         & 0.765625      &               \\ \hline
$\delta=0.5$ & 0.36   & 0.256  & 101111        & 0.53125       &               \\ \hline
$\delta=0$   & 0.72   & 0.512  & 1011110       & 1.0625        & 0.6689453125  \\ \hline
$s=1$        & 0.8224 & 0.256  & 1011110       & 1.0625        &               \\ \hline
$\delta=0.5$ & 0.6448 & 0.512  & 10111101      & 1.125         & 0.937890625   \\ \hline
$s=3$        & 1.1056 & 0.0512 & 10111101      & 1.125         &               \\ \hline
$\delta=1$   & 0.1056 & 0.0512 & 10111110      & 0.125         &               \\ \hline
$\delta=0$   & 0.2112 & 0.1024 & 101111100     & 0.25          &               \\ \hline
$\delta=0$   & 0.4224 & 0.2048 & 1011111000    & 0.5           &               \\ \hline
$\delta=0$   & 0.8448 & 0.4096 & 10111110000   & 1             &               \\ \hline
$\delta=0.5$ & 0.6896 & 0.8192 & 101111100001  & 1             &               \\ \hline
$v=1$        & 1      & ---    & 101111100001  & ---           &               \\ \hline
$\delta=1$   & 0      & ---    & 101111100010  & ---           &               \\ \hline
$\delta=0$   & 0      & ---    & 1011111000100 & ---           &               \\ \hline
\end{tabular}
\end{center}
 \tablecaption{Results of arithmetic encoding and decoding, with renormalization,
   applied to source and data sequence of Example~\ref{exFirstEnc}. Final code value is 
   $\hat{v}=0.1011111000100_2$ = 0.74267578125.}
                                                           \label{tbRenormEx}
\end{table}

\subsection{Implementation with Integer Arithmetic}           \label{ssIntArit}

Even though we use real numbers to describe the principles of arithmetic 
coding, most practical implementations use only integer arithmetic. The 
adaptation is quite simple, as we just have to assume that the $P$-bit integers 
contain the fractional part of the real numbers, with the following 
adaptations. (Appendix~A has the details.)

\begin{itemize}
 \item Define $B=2^Pb$, $L=2^Pl$, $V=2^Pv$, and $C(s)=2^Pc(s)$. Products 
  can be computed with $2P$ bits, and the $P$ least-significant bits are discarded. 
  For example, when updating the interval length we compute $L \leftarrow 
  \left\lfloor {L \cdot [C(s + 1) - C(s)] \cdot 2^{-P}} \right\rfloor $. The 
  length value $l=1$ cannot be represented in this form, but this is not a 
  real problem. We only need to initialize the scaled length with $L 
  \leftarrow 2^P-1$, and apply renormalization only when $l<0.5$ (strict 
  inequality).
 \item The carry condition $b \ge 1$ is equivalent to $B \ge 2^P$, which can
  mean integer overflow. It can be 
  detected accessing the CPU's carry flag, or equivalently, checking when the 
  value of $B$ decreases.
 \item Since $l>0$ we can work with a scaled length equal to ${L}'=2^Pl-1$. 
  This way we can represent the value $l=1$ and have some extra precision if 
  $P$ is small. On the other hand, updating the length using
  ${L}' \leftarrow \left\lfloor {({L}'+1) \cdot [C(s+1)-C(s)] \cdot 2^{-P}-1} \right\rfloor$
  requires two more additions.
\end{itemize}

When multiplication is computed with $2P$ bits, we can determine what is the
smallest allowable probability to avoid length underflow. Since renormalization
guarantees that $L \geq 2^{P-1}$, we have~(\ref{eqPropCD1}) satisfied if
\begin{equation}
                                                         \label{eqIntArith1}
  \left\lfloor [C(s+1) - C(s)]\; 2^{-P}\,L \right\rfloor \geq
  \left\lfloor [C(s+1) - C(s)]\; 2^{-1} \right\rfloor \geq 1
  \Rightarrow C(s+1) - C(s) \geq 2,
\end{equation}
Meaning that the minimum probability supported by such implementations with
$P$-bit integers is $p(s) \geq 2^{1-P}$.

\begin{Example}
Table~\ref{tbByteReg} shows the results of an 8-bit register
implementation ($P=8$) applied to the data sequence and source used in
Example~\ref{exFirstEnc}. The table shows the interval $\Range{b_k}{l_k}$
in base-2 notation to make clear that even though we use 8-bit arithmetic,
we are actually implementing \emph{exact} additions. Following the
conventions in~(\ref{eqFixPrec3}) and Figure~\ref{fgCarryProp},
we used bold letters to indicate active bits, and underlined for outstanding
bits. We also show the approximate results of the multiplications
$l_{k-1} \cdot c(s_k)$. The last column shows the binary contents of the
registers with active bits. We used ${L}'=2^Pl-1$ to represent the length.
The results are also shown in decimal notation so that they can be
compared with the exact results in Table~\ref{tbFirsCod}. 
Note that the approximations change the code value after only 7~bits, but 
the number of bits required to represent the final code value is still
13~bits.
                                                      \label{exByteReg}
\end{Example}

%

\begin{table}[tbp]
\begin{center}
  \begin{tabular}{|c|c|l|l|c|} \hline
    \textbf{Input} & \textbf{Encoder} & \multicolumn{1}{|c|}{\textbf{Binary}} & \multicolumn{1}{c|}{\textbf{Decimal}} & \textbf{8-bit} \\
    $s_k$ & \textbf{Data} & \multicolumn{1}{|c|}{\textbf{Representation}} & \multicolumn{1}{c|}{\textbf{Representation}} &  \textbf{Registers} \\ \hline \hline
--- & $b_o$          & $0.{\bf 00000000}_2$                         & 0                   & 00000000 \\ \cline{2-5} 
  & $l_0$            & ${\bf 1.00000000}_2$                         & 1                   & 11111111 \\ \hline
  & $l_0 \cdot c(2)$ & $0.{\bf 10110011}_2$                         & 0.69921875          & 10110011 \\ \cline{2-5} 
2 & $b_1$            & $0.1\underline{0}{\bf 11001100}_2$           & 0.69921875          & 11001100 \\\cline{2-5} 
  & $l_1$            & $0.00{\bf 11001100}_2$                       & 0.19921875          & 11001011 \\ \hline
  & $l_1 \cdot c(1)$ & $0.00{\bf 00101000}_2$                       & 0.0390625           & 00101000 \\ \cline{2-5} 
1 & $b_2$            & $0.1\underline{01}{\bf 11101000}_2$          & 0.73828125          & 11101000 \\ \cline{2-5} 
  & $l_2$            & $0.000{\bf 11001100}_2$                      & 0.099609375         & 11001011 \\ \hline
  & $l_2 \cdot c(0)$ & $0.000{\bf 00000000}_2$                      & 0                   & 00000000 \\ \cline{2-5} 
0 & $b_3 $           & $0.1\underline{0111}{\bf 10100000}_2$        & 0.73828125          & 10100000 \\ \cline{2-5}
  & $l_3 $           & $0.00000{\bf 10100000}_2$                    & 0.01953125          & 10011111 \\ \hline
  & $l_3 \cdot c(0)$ & $0.00000{\bf 00000000}_2$                    & 0                   & 00000000 \\ \cline{2-5} 
0 & $b_4 $           & $0.101111\underline{01}{\bf 00000000}_2$     & 0.73828125          & 00000000 \\ \cline{2-5} 
  & $l_4 $           & $0.00000000{\bf 11111000}_2$                 & 0.0037841796875     & 11110111 \\ \hline
  & $l_4 \cdot c(1)$ & $0.00000000{\bf 00110001}_2$                 & 0.0007476806640625  & 00110001 \\ \cline{2-5} 
1 & $b_5 $           & $0.10111101\underline{0}{\bf 01100010}_2$    & 0.7390289306640625  & 01100010 \\ \cline{2-5} 
  & $l_5 $           & $0.000000000{\bf 11111000}_2$                & 0.00189208984375    & 11110111 \\ \hline
  & $l_5 \cdot c(3)$ & $0.000000000{\bf 11011110}_2$                & 0.0016937255859375  & 11011110 \\ \cline{2-5} 
3 & $b_6 $           & $0.10111101101\underline{0}{\bf 00000000}_2$ & 0.74072265625       & 00000000 \\ \cline{2-5} 
  & $l_6 $           & $0.000000000000{\bf 11001000}_2$             & 0.00019073486328125 & 11000111 \\ \hline
--- & $\hat{v}$      & $0.1011110110101_2$                          & 0.7408447265625     & ---      \\ \hline
\end{tabular}
\end{center}
 \tablecaption{Results of arithmetic encoding and decoding for
  Example~\ref{exFirstEnc} using 8-bit precision for the arithmetic operations.}
                                                           \label{tbByteReg}
\end{table}

\subsection{Efficient Output}                                \label{ssEfficOut}

Implementations with short registers (as in Example~\ref{exByteReg})
require renormalizing the intervals as soon as possible to avoid losing 
accuracy and compression efficacy. We can see in Table~\ref{tbRenormEx}
that, as a consequence, intervals
may be rescaled many times whenever a symbol is coded. Even though
rescaling can be done with bit shifts instead of multiplications, this
process still requires many CPU cycles (see Section~\ref{ssComplexA}).

If we use longer registers (e.g., 16~bits or more), we can increase 
efficiency significantly by moving more than one bit to the output
buffer ${\bf d}$ whenever rescaling. This process is equivalent to
have an encoder output alphabet with $D$ symbols, where $D$ is a power
of two. For example, moving groups of 1, 2, 4, or 8~bits at a time,
corresponds to output alphabets with $D=2$, 4, 16, and 256~symbols,
respectively. Carry propagation and the use of the output buffer ${\bf d}$
also become more efficient with larger $D$.

It is not difficult to modify the algorithms in Section~\ref{ssFPImplem}
for a \mbox{$D$-symbol} output. Renormalization is the most important change.
The rescaling parameters defined by~(\ref{eqPropSc1}) are such that
$\delta \in \{0,1/D,2/D,\ldots,1\}$ and $\gamma \equiv D$. Again, 
$\delta=1$ corresponds to a carry.
Algorithm~\ref{alEONormal} has the required changes to Algorithm~\ref{alFPENormal},
and the corresponding changes in the decoder renormalization are shown in
Algorithm~\ref{alDONormal}.
In Appendix~A we have an integer arithmetic implementation with \mbox{$D$-symbol}
output, with the carry propagation in Algorithm~\ref{alINTPrCarry}.

We also need to change the equations that define the normalized intervals,
(\ref{eqFixPrec1}) and~(\ref{eqFixPrec2}), to
\begin{eqnarray}
 l & = & D^{t(l_k)} l_k, \nonumber \\
                                                         \label{eqFixPrec5}
 b & = & \mbox{frac}(D^{t(l_k)}b_k) = D^{t(l_k)} b_k - \left\lfloor D^{t(l_k)} b_k \right\rfloor, \\
 v & = & \mbox{frac}(D^{t(l_k)}\hat{v}), \nonumber
\end{eqnarray}
and
\begin{equation}
                                                         \label{eqFixPrec6}
  t(x) = \{n:\;D^{-n-1} < x \le D^{-n}\} = \left\lfloor {-\log_D (x)}  \right\rfloor .
\end{equation}

\begin{Algorithm}{{Procedure Encoder{\_}Renormalization} $(b,l,t,{\bf d})$}
1. \= while $l \le 1 / D$ do \COM{Renormalization loop}
   \> 1.1. set \{ \= $t \leftarrow t + 1$; \} \CON{Increment symbol counter}
   \> \> $d(t) \leftarrow \left\lfloor {D \cdot b} \right\rfloor $; \CON{Output symbol from most significant bits}
   \> \> $b \leftarrow D \cdot b - d(t)$; \CON{Update interval base}
   \> \> $l \leftarrow D \cdot l$; \} \CON{Scale interval length}
2. return.
                                                     \label{alEONormal}
\end{Algorithm}

\begin{Algorithm}{{Procedure Decoder{\_}Renormalization} $(v,b,l,t,{\bf d})$}
1. \= while $l \le 1 / D$ do \COM{Renormalization loop}
   \> 1.1. set \{ \= $t \leftarrow t + 1$; \} \CON{Increment symbol counter}
   \> \> $a \leftarrow \left\lfloor {D \cdot b} \right\rfloor$; \CON{Most significant digit}
   \> \> $b \leftarrow D \cdot b - a$; \CON{Update interval base}
   \> \> $v \leftarrow D \cdot v - a + D^{-P}d(t)$; \CON{Update code value}
   \> \> $l \leftarrow D \cdot l$; \} \CON{Scale interval length}
3. return.
                                                     \label{alDONormal}
\end{Algorithm}

\begin{Example}
Table~\ref{tbByteOut} shows an example of how the arithmetic coding
renormalization changes when $D=16$. Again, the source and data sequence
are those of Example~\ref{exFirstEnc}. Comparing these results with those
of Example~\ref{exNormalEnc} (Table~\ref{tbRenormEx}), we 
can see the substantial reduction in the number of times the intervals are 
rescaled.
                                                      \label{exByteOut}
\end{Example}

%

\begin{table}[tbp]
\begin{center}
  \begin{tabular}{|l|l|l|l|l|} \hline
  \multicolumn{1}{|c|}{\bf Event} & \multicolumn{1}{|c|}{\bf Scaled} &
    \multicolumn{1}{|c|}{\bf Scaled} & \multicolumn{1}{|c|}{\bf Bit buffer} &
    \multicolumn{1}{|c|}{\bf Scaled code} \\
  & \multicolumn{1}{|c|}{\bf base} & \multicolumn{1}{|c|}{\bf length} & &
    \multicolumn{1}{|c|}{\bf value} \\
  & \multicolumn{1}{|c|}{$b$} & \multicolumn{1}{|c|}{$l$} & \multicolumn{1}{|c|}{\bf d} &
    \multicolumn{1}{|c|}{$v$} \\ \hline \hline
     ---       & 0      & 1      &      & 0.74267578125 \\ \hline
    $s=2$      & 0.7    & 0.2    &      & 0.74267578125 \\\hline
    $s=1$      & 0.74   & 0.1    &      & 0.74267578125 \\ \hline
    $s=0$      & 0.74   & 0.02   &      & 0.74267578125 \\ \hline
$\delta=11/16$ & 0.84   & 0.32   & B    & 0.8828125     \\ \hline
    $s=0$      & 0.84   & 0.064  & B    & 0.8828125     \\ \hline
    $s=1$      & 0.8528 & 0.032  & B    & 0.8828125     \\ \hline
$\delta=13/16$ & 0.6448 & 0.512  & BD   & 1.125         \\ \hline
    $s=3$      & 1.1056 & 0.0512 & BD   & 1.125         \\ \hline
$\delta=1$     & 0.1056 & 0.0512 & BE   & 0.125         \\ \hline
$\delta=1/16$  & 0.6896 & 0.8192 & BE1  & 1             \\ \hline
    $v=1$      & 1      & ---    & BE1  & ---           \\ \hline
$\delta=1$     & 0      & ---    & BE2  & ---           \\ \hline
$\delta=0$     & ---    & ---    & BE20 & ---           \\ \hline
\end{tabular}
\end{center}
 \tablecaption{Results of arithmetic encoding and decoding for Example~\ref{exFirstEnc},
   with renormalization and a \mbox{16-symbol} (hexadecimal) output alphabet. 
   Final code value is $\hat{v}= \mbox{0.BE20}_{16} = 0.1011\,1110\,0010\,0000_2 = 0.74267578125$.}
                                                       \label{tbByteOut}
\end{table}

Algorithms \ref{alEONormal} and \ref{alDONormal} can also be used for values
of $D$ that are not powers of two, but with inefficient radix-$D$
arithmetic. The problem is that while the multiplications by the interval
length can be approximated, rescaling has to be exact. For example, if
$D=2$ then multiplication by $2^{-P}$ is computed exactly with bit-shifts,
but exact multiplication by $3^{-P}$ requires special functions~\cite{Press93}.

A better alternative is to use a small trick, shown in Figure~\ref{fgOutConv}.
We can  compress the data sequence using a standard encoder/decoder with binary 
arithmetic. Next, we ``decode'' the binary compressed data using a 
\mbox{$D$-symbol} alphabet with uniform probability distribution
${\bf p} = [\, 1/D \quad 1/D \quad \cdots \quad 1/D \,]$. The uniform
distribution does not change the distribution of the code values (and thus
will not alter compression), but 
converts the data to the desired alphabet. Since both processes implemented 
by the decoder are perfectly reversible, the decoder only has to implement 
the inverse processes. This takes twice the time for encoding and decoding, 
but is significantly faster than using radix-$D$ arithmetic.

%
\setlength{\unitlength}{1.2mm}
\begin{figure}[tp]
\begin{center}
 \begin{picture}(115,45)
%
%
  \thicklines
  \put(0,30){\framebox(20,10){\color{red} \shortstack{\textsf{Data} \\ \textsf{source}}}}
  \put(0,5){\framebox(20,10){\color{red} \shortstack{\textsf{Decoded} \\ \textsf{data}}}}
  \multiput(40,30)(45,-25){2}{\framebox(25,10){\color{acblue} \shortstack{\textsf{Arithmetic} \\ \textsf{encoder}}}}
  \multiput(40,5)(45,25){2}{\framebox(25,10){\color{acblue} \shortstack{\textsf{Arithmetic} \\ \textsf{decoder}}}}
%
%
  \multiput(20,35)(45,0){2}{\vector(1,0){20}}
  \multiput(85,10)(-45,0){2}{\vector(-1,0){20}}
  \put(97.5,30){\vector(0,-1){15}}
%
%
  \thinlines
  \put(10,44){\large \makebox(0,0){$\Omega$}}
  \multiput(50,1)(0,43){2}{\makebox(0,0){\footnotesize ${\bf p}(\Omega) = [p(0)\;p(1)\;\cdots\;p(M-1)]$}}
  \multiput(100,1)(0,43){2}{\makebox(0,0){\footnotesize ${\bf p}(D) = [1/D\;1/D\;\cdots\;1/D]$}}
  \multiput(30,15)(0,15){2}{\makebox(0,0){\color{acgreen} \footnotesize $M$ symbols}}
  \multiput(75,15)(0,15){2}{\makebox(0,0){\color{acgreen} \footnotesize binary}}
  \put(99,22.5){\makebox(0,0)[l]{\color{acgreen} \footnotesize $D$ symbols}}
 \end{picture}
\end{center}
 \figurecaption{Configuration for using a standard (binary output) arithmetic
  encoder to replace an encoder with \mbox{$D$-symbol} (ternary, decimal, etc.)
  output.}
                                                         \label{fgOutConv}
\end{figure}

\subsection{Care with Carries}

We have seen in Section~\ref{scFinPrec} that carry propagation is applied only to the 
set of outstanding bits, which always start with a 0-bit, and is followed 
possibly by 1-bits. Examples of set of outstanding bits are
\[
\emptyset,\{0\},\{0,1\},\{0,1,1\},\{0,1,1,1\},\cdots,\{0,1,1,1,\ldots,1\}.
\]

Clearly, with such simple structure we do not have to save the outstanding bits to 
know what they are. We can just keep a counter with the number of 
outstanding bits, which is incremented as new bits are defined during 
renormalization. We can output (or ``flush'') these bits whenever a carry 
occurs or when a new outstanding 0-bit comes from renormalization. 

Note that not all bits put out by rescaling have to be outstanding before 
becoming settled. For example, if we have $b < b+l \le 0.5,$ we not only 
know that the next bit is zero, but we know that it cannot possibly be 
changed by a carry, and is thus settled. We can disregard these details when 
implementing carry propagation in buffers, but not when using counters.

With \mbox{$D$-symbol} output the set of outstanding symbols starts
with a symbol $\sigma \neq D-1$, and is possibly followed by 
several occurrences of the symbol $D-1$, as shown below.
\[
  \emptyset,\{\sigma\},\{\sigma,D-1\},\{\sigma,D-1,D-1\},\cdots,
    \{\sigma,D-1,D-1,\ldots,D-1\}.
\]
In this case we can only keep the first symbol $\sigma$ and a counter
with the number of outstanding symbols.

It is important to know that the number of outstanding symbols can grow
indefinitely, i.e., we can create distributions and infinitely long
data sequences such that bits never become settled.
The final code value selection (Algorithm~\ref{alFPValSelc}) settles all
outstanding symbols, so that we can periodically restart the encoder to limit
the number of outstanding symbols.

There are many choices for dealing with carry propagation. The most common
are:

\begin{enumerate}
 \item Save the outstanding symbols temporarily to a buffer, and then implement 
  carry propagation in the buffer. This simple technique is efficient
  when working with bytes ($D=256$), but can be inefficient when $D$ is small.
  It can only be used if we know that the buffer is large enough to fit all
  the compressed symbols, since all of them can be outstanding.
 \item Use a counter to keep track of the outstanding symbols, saving all
  symbols to a buffer or file as soon as they become settled. There is chance
  of overflow depending 
  on the number of bits used by the counter (mantissas with 32, 53, and 64 
  bits allow, respectively, $4 \cdot 10^9$, $9 \cdot 10^{15}$, and $2 \cdot 
  10^{19}$ outstanding symbols). In practical applications a counter overflow is
  extremely improbable, specially with adaptive coders.
 \item Use carry propagation in a buffer, plus ``bit-stuffing''
  \cite{Langdon81,Pennebaker88a} in the 
  following form. Outstanding bits are moved out from the buffer as soon as 
  they become settled. Whenever the number of 1-bits exceeds a threshold 
  (e.g.,~16), an artificial zero is added to the compressed bit sequence, 
  forcing the outstanding bits to become settled. The decoder can identify 
  when this happens by comparing the number of consecutive 1-bits read. When 
  it exceeds the threshold, the decoder interprets the next bit not as data, 
  but as carry information. If it is 1, then a carry is propagated in the 
  decoder buffer, and the decoding process continues normally. This technique 
  is used by the Q-coder~\cite{Pennebaker88a}.
\end{enumerate}

\subsection{Alternative Renormalizations}

We show in Section~\ref{ssIntScal} that there is a great deal of flexibility in 
rescaling the intervals, and we present in Section~\ref{ssFPImplem} an implementation 
based on a particular form of renormalization. Other choices lead to 
renormalized intervals with distinct properties, which had been exploited in 
several manners by different arithmetic coding implementations. Below we 
show a few examples.

\begin{itemize}
 \item We have chosen renormalization~(\ref{eqFixPrec5}), which produces intervals
  such that $b \in [0,\,1)$, $l \in (1/D,\;1]$, and $b+l \in (1/D,\;2)$. Its main
  advantage is that it simplifies carry detection and renormalization, especially
  when $D>2$. Note that it has a criterion for when to renormalize that is based
  only on the interval length.
 \item The decision of when to renormalize can be based on settled symbols. For
  example, the method by Rubin~\cite{Rubin79} keeps intervals such that
  $b \in [0,\,1)$, and $b + l \in (0,\,1]$, which 
  are renormalized when the most significant symbol becomes settled, i.e., 
  $\left\lfloor {Db} \right\rfloor = \left\lfloor {D(b+l)} \right\rfloor$, 
  meaning that the outstanding symbols are kept in the registers. To avoid 
  the interval length eventually collapsing to zero, the encoder and decoder
  rescale and shift the interval when its length gets too small, forcing
  outstanding symbols to become settled.
 \item Witten, Neal, and Cleary~\cite{Witten87,Moffat98} proposed an arithmetic
  coding implementation that became quite popular. Instead of the base length, it
  uses the extremes points to represent the interval, and keeps it renormalized in 
  such a manner that $b + l \in (0,\,1]$. Renormalization 
  occurs whenever bits are settled, and also when $0.25 \le b < b+l \le 0.75$.
  Thus, it avoids the precision loss that occurs in~\cite{Rubin79}, and uses 
  counters to keep track of the number of outstanding bits. This technique
  can be adapted to \mbox{$D$-symbol} output, but it is not as simple as
  what we have in Appendix~A.
 \item The binary Q-coder developed by IBM~\cite{Pennebaker88a,Pennebaker88b},
  keeps intervals with $l \in (0.75,\,1.5]$. This way we normally have $l \approx 1$,
  and we can approximate multiplications with $p \cdot l \approx p$, and
  $(1-p) \cdot l \approx l-p$. Variations include the QM~\cite{Pennebaker92} and
  MQ-coder~\cite{Taubman02}.
\end{itemize}

\section{Adaptive Coding}                                     \label{ssAdapCod}

Since typical information sources tend to be quite complex, we must have a 
good model of the data source to achieve optimal compression. There are many 
techniques for modeling complex sources
\cite{Rissanen81,Witten87,Bell90,Said95,Weinberger96,Wu96,LoPresto97}
that decompose the source data in different categories, under the assumption that in each 
category the source symbols are approximately independent and identically 
distributed, and thus well suited to be compressed with arithmetic coding. 
In general, we do not know the probabilities of the symbols in each 
category. Adaptive coding is the estimation of the probabilities of the 
source symbols during the coding process. In this section we study 
techniques to efficiently combine arithmetic coding with dynamic probability 
estimation.

\subsection{Strategies for Computing Symbol Distributions}

The most efficient technique for computing distributions depends on the data 
type. When we are dealing with completely unknown data we may want 
adaptation to work in a completely automatic manner. In other cases, we can 
use some knowledge of the data properties to reduce or eliminate the 
adaptation effort. Below we explain the features of some of the most common 
strategies for estimating distributions.

\begin{itemize}
\item Use a constant distribution that is available before encoding and decoding, 
normally estimated by gathering statistics in a large number of typical 
samples. This approach can be used for sources such as English text, or 
weather data, but it rarely yields the best results because few information 
sources are so simple as to be modeled by a single distribution. 
Furthermore, there is very little flexibility (e.g., statistics for English 
text do not fit well Spanish text). On the other hand, it may work well if 
the source model is very detailed, and in fact it is the only alternative in 
some very complex models in which meaningful statistics can only be gathered 
from a very large amount of data.

\item Use pre-defined distributions with adaptive parameter estimation. For 
instance, we can assume that the data has Gaussian distribution, and 
estimate only the mean and variance of each symbol. If we allow only a few 
values for the distribution parameters, then the encoder and decoder can 
create several vectors with all the distribution values, and use them 
according to their common parameter estimation. See ref.~\cite{LoPresto97} for an 
example.

\item Use two-pass encoding. A first pass gathers the statistics of the source, 
and the second pass codes the data with the collected statistics. For 
decoding, a scaled version of vectors ${\bf p}$ or ${\bf c}$ must be 
included at the beginning of the compressed data. For example, a book can be 
archived (compressed) together with its particular symbol statistics. It is 
possible to reduce the computational overhead by sharing processes between 
passes. For example, the first pass can simultaneously gather statistics and 
convert the data to run-lengths.

\item Use a distribution based on the occurrence of symbols previously coded, 
updating ${\bf c}$ with each symbol encoded. We can start with a very 
approximate distribution (e.g., uniform), and if the probabilities change 
frequently, we can reset the estimates periodically. This technique,
explained in the next section, is quite effective and the most convenient
and versatile. However, the constant 
update of the cumulative distribution can increase the computational 
complexity considerably. An alternative is to update only the probability 
vector ${\bf p}$ after each encoded symbol, and update the cumulative 
distribution ${\bf c}$ less frequently (Section~\ref{ssPerUpdt}).
\end{itemize}

\subsection{Direct Update of Cumulative Distributions}        \label{ssDirUpdt}

After encoding/decoding $k$ symbols the encoder/decoder can estimate the 
probability of a symbol as
\begin{equation}
                                                       \label{eqAdapCod1}
  p(m) = \frac{\tilde{P}(m)}{k+M},\quad m = 0,1,2,\ldots,M-1,
\end{equation}
where $\tilde{P}(m)>0$ is the number of times symbol $m$ was encoded/decoded 
plus one (added to avoid zero probabilities). The symbol occurrence counters
are initialized with $\tilde{P}(m)=1$, and incremented after a symbol is
encoded. We define the cumulative sum of occurrences as
\begin{equation}
                                                       \label{eqAdapCod2}
  \tilde{C}(m) = \sum\limits_{i=0}^{m-1} {\tilde{P}(i)}, \quad
    m = 0,1,2,\ldots,M,
\end{equation}
and the cumulative distribution as
\begin{equation}
                                                       \label{eqAdapCod3}
  c(m) = \frac{\tilde{C}(m)}{k+M} = \frac{\tilde{C}(m)}{\tilde{C}(M)}, \quad
    m = 0,1,2,\ldots,M
\end{equation}

Only a few changes to the algorithms of Section~\ref{ssFPImplem} are sufficient
to  include the ability to dynamically update the cumulative distribution. 
First, we may use~(\ref{eqAdapCod3}) to change all multiplications by $c(s)$, as
follows. 
\begin{equation}
                                                       \label{eqAdapCod4}
  b \leftarrow b + \frac{l \cdot \tilde{C}(s)}{\tilde{C}(M)}
\end{equation}
Similarly, the changes to the integer arithmetic version in Appendix~A may be in the 
form.
\begin{equation}
                                                       \label{eqAdapCod5}
  B \leftarrow B + \left\lfloor {\frac{L \cdot \tilde{C}(s)}{\tilde{C}(M)}} \right\rfloor 
\end{equation}

However, since divisions are much more expensive than additions and
multiplications, it is better to compute $\gamma = l / \tilde{C}(M)$ once, and
implement interval updating as shown in Algorithm~\ref{alADCIntUpdt} \cite{Moffat97,Moffat98}.
The corresponding changes to Algorithm~\ref{alFPIntSel} is shown in
Algorithm~\ref{alADCIntSel}.

\begin{Algorithm}{{Procedure Interval{\_}Update} $(s,b,l,M,{\bf \tilde{C}})$}
1. \= set $\gamma = l / \tilde{C}(M)$ \COM{Compute division result}
2. \= if $s = M - 1$ \COM{Special case for last symbol}
   \> then \= set $y \leftarrow b+l$; \CON{end of interval}
   \> else \> set $y \leftarrow b+\gamma \cdot \tilde{C}(s+1)$; \CON{base of next subinterval}
3. \> set \{ \> $b \leftarrow b+\gamma \cdot \tilde{C}(s)$; \COM{Update interval base}
   \> \> $l \leftarrow y-b$; \} \COM{Update interval length as difference}
4. return.
                                                     \label{alADCIntUpdt}
\end{Algorithm}

\begin{Algorithm}{{Function Interval{\_}Selection} $(v,b,l,M,{\bf \tilde{C}})$}
1. \= set \{ \= $s \leftarrow M-1$; \COM{Start search from last symbol}
   \> \> $\gamma = l / \tilde{C}(M)$ \COM{Compute division result}
   \> \> $x \leftarrow b + \gamma \cdot \tilde{C}(M-1)$; \COM{Base of search interval}
   \> \> $y \leftarrow b+l$; \} \COM{End of search interval}
\pushtabs
2. \= while $x > v$ do \COM{Sequential search for correct interval}
   \> 2.1. set \{ \= $s \leftarrow s-1$; \CON{Decrement symbol by one}
   \> \> $y \leftarrow x$;  \CON{move interval end}
   \> \> $x \leftarrow b+\gamma \cdot \tilde{C}(s)$; \} \CON{compute new interval base}
\poptabs
3. set \{ $b \leftarrow x$; \COM{Update interval base}
   \> \> $l \leftarrow y-b$; \} \COM{Update interval length as difference}
4. return $s$.
                                                     \label{alADCIntSel}
\end{Algorithm}

In Algorithms \ref{alFPIntSel} and~\ref{alADCIntSel} we may need to
compute several multiplications, under the assumption they are faster
than a single division (see Algorithm~\ref{alSSIntSel} for a
more efficient search method). If this is not true, Algorithm~\ref{alASCIntSel}
shows how to replace those multiplications by one extra division.

After the modifications above, Algorithms \ref{alFPEnc} and~\ref{alFPDec}
are made adaptive by adding the following line:

2.4. \textsf{Update{\_}Distribution} $(s_k,M,{\bf \tilde{C}})$;

\begin{Algorithm}{{Function Interval{\_}Selection} $(v,b,l,M,{\bf \tilde{C}})$}
1. \= set \{ \= $s \leftarrow M-1$; \COM{Initialize search from last symbol}
   \> \> $\gamma = l / \tilde{C}(M)$ \COM{Compute division result}
   \> \> $W \leftarrow (v - b) / \gamma$; \} \COM{Code value scaled by $\tilde{C}(M)$}
2. while $\tilde{C}(s) > W$ do \COM{Look for correct interval}
   \> 2.1. set $s \leftarrow s - 1$; \CON{decrement symbol by one}
3. set \{ $b \leftarrow b + \gamma \cdot \tilde{C}(s)$; \COM{Update interval base}
   \> \> $l \leftarrow \gamma \cdot [\tilde{C}(s+1) - \tilde{C}(s)]$; \} \COM{Update interval length}
4. return $s$.
                                                     \label{alASCIntSel}
\end{Algorithm}

The procedure to update the cumulative distribution is shown in
Algorithm~\ref{alLinUpd}. Note that in this algorithm it
is necessary to compute up to $M$ additions. Similarly, in Step~2 of
Algorithm~\ref{alASCIntSel} we have to perform up to $M$ comparisons 
and subtractions. Since the number of operations in both cases decreases 
with the symbol number, it is good to sort the symbol by increasing 
probability. Reference~\cite{Witten87} presents an implementation that
simultaneously updates the distribution while keeping it sorted.

\begin{Algorithm}{{Procedure Update{\_}Distribution} $(s,M,{\bf \tilde{C}})$}
1. \= for $m=s+1$ to $M$ do \COM{For all symbols larger than s}
   \> 1.1 set $\tilde{C}(s) \leftarrow \tilde{C}(s) + 1$; \CON{increment cumulative distribution}
2. return.
                                                     \label{alLinUpd}
\end{Algorithm}

\subsection{Binary Arithmetic Coding}                          \label{ssBinCod}

Binary arithmetic coders work only with a binary source alphabet ($M=2$). 
This is an important type of encoder because it helps to solve many of the 
complexity issues created with the dynamic update of the cumulative 
distributions (Algorithm~\ref{alLinUpd}). When $M=2$ the cumulative
distribution vector is simply ${\bf c} = [\, 0 \; p(0) \; 1 \, ]$,
which makes coding and updating the cumulative distribution much simpler tasks.

However, it is important to observe that there is a performance trade-off
here. While binary arithmetic coding greatly simplifies coding each binary data
symbol, its final throughput of information (actual information bits) cannot be
larger than one bit per coded symbol, which normally means one bit per few CPU
clock cycles~\cite{Said04}. Consequently, they are not as attractive for fast
coding as they used to be, but there is no reason not to use their special
properties when coding binary sources.

Algorithms \ref{alBNIntUpdt}, \ref{alBNIntSel}, and \ref{alBinUpdt}, 
have the procedures for, respectively, binary encoding (interval update), decoding
(interval selection and update), and distribution update. Note that instead 
of using the full vector ${\bf \tilde{C}}$ we use only
$\tilde{C}(1) = \tilde{P}(0)$ and $\tilde{C}(2) = \tilde{P}(0)+\tilde{P}(1)$. 
The renormalization procedures do not have to be changed for binary
arithmetic coding.

\begin{Algorithm}{{Procedure Binary{\_}Interval{\_}Update} $(s,b,l,\tilde{C}(1),\tilde{C}(2))$}
1. set $x \leftarrow l \cdot \tilde{C}(1) / \tilde{C}(2)$; \COM{Point for interval division}
2. if \= $s = 0$ \COM{If symbol is zero}
   \> then \= set \hspace{1ex} \=$l \leftarrow x$; \CON{Update interval length,}
   \> else \> set \{ \> $b \leftarrow b + x$; \CON{move interval base and}
   \> \> \> $l \leftarrow l - x$; \} \CON{update interval length}
3. return.
                                                     \label{alBNIntUpdt}
\end{Algorithm}

\begin{Algorithm}{{Function Binary{\_}Interval{\_}Selection} $(v,b,l,\tilde{C}(1),\tilde{C}(2))$}
1. set $x \leftarrow l \cdot \tilde{C}(1) / \tilde{C}(2)$; \COM{Point for interval division}
2. if \= $b + x > v$ \COM{Look for correct interval}
   \> then \= set \{ \= $s \leftarrow 0$; \CON{Symbol is 0: no change to interval base}
   \> \> \> $l \leftarrow x$; \} \CON{update interval length}
   \> else \> set \{ \> $s \leftarrow 1$; \CON{Symbol is 1:}
   \> \> \> $b \leftarrow b + x$; \CON{move interval base and}
   \> \> \> $l \leftarrow l - x$; \} \CON{update interval length}
3. return $s$.
                                                     \label{alBNIntSel}
\end{Algorithm}

\begin{Algorithm}{{Procedure Update{\_}Binary{\_}Distribution} $(s,{\bf\tilde{C}})$}
1. if $s = 0$ then set $\tilde{C}(1) \leftarrow \tilde{C}(1) + 1$; \COM{If $s=0$ then increment 0-symbol counter}
2. set $\tilde{C}(2) \leftarrow \tilde{C}(2) + 1$; \COM{Increment symbol counter}
3. return. 
                                                     \label{alBinUpdt}
\end{Algorithm}

\begin{Example}
Binary arithmetic coding has universal application because, just as any
number can be represented using bits, data symbols from any alphabet can
be coded as a sequence of binary symbols.
Figure~\ref{fgFDecTree} shows how the process of coding data from a
\mbox{6-symbol} source can be decomposed in a series of binary decisions,
which can be represented as a \emph{binary search tree}~\cite{Cormen90}. The
leaf nodes correspond to the date source symbols, and 
intermediate nodes correspond to the decisions shown below them.

Underlined numbers are used for the intermediate nodes, and their value 
corresponds to the number used for the comparison (they are the 
``\emph{keys}'' for the binary search tree~\cite{Cormen90}). For instance,
node~$\underline{m}$ 
corresponds to test ``$s < m$?'' Symbols are coded starting from the root 
of the tree, and continue until a leaf node is encountered. For example, if 
we want to code the symbol $s=2$, we start coding the information ``$s < 3$?''
indicated by node~\underline{3}; next we go to node~\underline{1}, 
and code the information ``$s < 1$?'', and finally move to 
node~\underline{2} to code ``$s < 2$?''. At each node the information is 
coded with a different set of probabilities, which in Figure~\ref{fgFDecTree} are shown 
below the tree nodes. These probabilities, based on the number of symbol
occurrences, are updated with Algorithm~\ref{alBinUpdt}. An alphabet with $M$ symbols 
needs $M-1$ probability estimates for the intermediate nodes. The decoder 
follows the same order, using the same set of probabilities. (Note that
this is equivalent to using the scheme explained in Section~\ref{ssSepCodMod}, and shown in 
Figure~\ref{fgSepCodMod}.)

There is no loss in compression in such scheme. For example, when coding 
symbol $s=2$ we can compute the symbol probability as a product of conditional
probabilities~\cite{Papoulis84}.
\begin{eqnarray}
 \mbox{Prob}(s=2) & = & \mbox{Prob}(s<3) \cdot \mbox{Prob}(s=2|s<3) \nonumber \\ 
   & = & \mbox{Prob}(s<3) \cdot \mbox{Prob}(s \ge 1|s<3) \cdot \mbox{Prob}(s=2|1\le s<3) \\
                                                       \label{eqAdapCod6}
   & = & \mbox{Prob}(s<3) \cdot \mbox{Prob}(s \ge 1|s<3) \cdot \mbox{Prob}(s\ge 2|s\ge 1,s<3) \nonumber
\end{eqnarray}

This means that
\begin{eqnarray}
                                                       \label{eqAdapCod7}
  \log_2 \left[ \mbox{Prob}(s=2) \right] & = & 
    \log_2 \left[ \mbox{Prob}(s<3) \right] + \log_2 \left[ \mbox{Prob}(s \ge 1|s<3) \right] + \\
  & & + \log_2 \left[ \mbox{Prob}(s\ge 2|1\le s<3) \right] \nonumber
\end{eqnarray}

The left-hand-side of~(\ref{eqAdapCod7}) is the number of bits required to 
code symbol $s=2$ directly with a \mbox{6-symbol} model, which is equal to 
the sum of the bits used to code the same symbol by successively coding 
the binary symbols in the binary-search tree (see Figure~\ref{fgFDecTree}).
We do not have to worry about computing explicit 
values for the conditional probabilities because, when we use a different 
adaptive binary model for each node, we automatically get the estimate of 
the proper conditional probabilities. This property is valid for
binary-tree decompositions of any data alphabet.
                                                      \label{exTreeDec}
\end{Example}

%
 \begin{figure}[tp]
 \setlength{\unitlength}{1.5mm}
 \begin{center}
 \begin{picture}(80,55)
%
%
  \thicklines
  \put(5,30){\makebox(0,0){\textbf{\color{red} 0}}}
  \put(5,30){\circle{6}}
  \put(5,35){\makebox(0,0){\small \textsf{\color{acblue} 4}}}
  \put(15,40){\makebox(0,0){\textbf{\color{red} 1}}}
  \put(15,40){\circle{6}}
  \put(15,45){\makebox(0,0){\small \textsf{\color{acblue} 8}}}
  \put(35,40){\makebox(0,0){\textbf{\color{red} 2}}}
  \put(35,40){\circle{6}}
  \put(35,45){\makebox(0,0){\small \textsf{\color{acblue} 10}}}
  \put(45,30){\makebox(0,0){\textbf{\color{red} 3}}}
  \put(45,30){\circle{6}}
  \put(45,35){\makebox(0,0){\small \textsf{\color{acblue} 15}}}
  \put(55,40){\makebox(0,0){\textbf{\color{red} 4}}}
  \put(55,40){\circle{6}}
  \put(55,45){\makebox(0,0){\small \textsf{\color{acblue} 19}}}
  \put(75,40){\makebox(0,0){\textbf{\color{red} 5}}}
  \put(75,40){\circle{6}}
  \put(75,45){\makebox(0,0){\small \textsf{\color{acblue} 65}}}
  \put(15,20){\makebox(0,0){\underline{\textbf{1}}}}
  \put(15,20){\circle{6}}
  \put(15,16){\makebox(0,0)[t]{\scriptsize \shortstack{\textbf{$s < 1$?} \\ (4/22)}}}
  \put(11,21.4){\makebox(0,0){\tiny \textsf{Y}}}
  \put(19,21.4){\makebox(0,0){\tiny \textsf{N}}}
  \put(13,22){\vector(-1,1){6}}
  \put(17,22){\vector(1,1){6}}
  \put(25,30){\makebox(0,0){\textbf{\underline{2}}}}
  \put(25,30){\circle{6}}
  \put(27,26){\makebox(0,0)[t]{\scriptsize \shortstack{\textbf{$s < 2$?} \\ (8/18)}}}
  \put(21,31.4){\makebox(0,0){\tiny \textsf{Y}}}
  \put(29,31.4){\makebox(0,0){\tiny \textsf{N}}}
  \put(23,32){\vector(-1,1){6}}
  \put(27,32){\vector(1,1){6}}
  \put(55,20){\makebox(0,0){\underline{\textbf{4}}}}
  \put(55,20){\circle{6}}
  \put(55,16){\makebox(0,0)[t]{\scriptsize \shortstack{\textbf{$s < 4$?} \\ (15/99)}}}
  \put(51,21.4){\makebox(0,0){\tiny \textsf{Y}}}
  \put(59,21.4){\makebox(0,0){\tiny \textsf{N}}}
  \put(53,22){\vector(-1,1){6}}
  \put(57,22){\vector(1,1){6}}
  \put(65,30){\makebox(0,0){\textbf{\underline{5}}}}
  \put(65,30){\circle{6}}
  \put(67,26){\makebox(0,0)[t]{\scriptsize \shortstack{\textbf{$s < 5$?} \\ (19/84)}}}
  \put(61,31.4){\makebox(0,0){\tiny \textsf{Y}}}
  \put(69,31.4){\makebox(0,0){\tiny \textsf{N}}}
  \put(63,32){\vector(-1,1){6}}
  \put(67,32){\vector(1,1){6}}
  \put(35,10){\makebox(0,0){\underline{\textbf{3}}}}
  \put(35,10){\circle{6}}
  \put(35,6){\makebox(0,0)[t]{\scriptsize \shortstack{\textbf{$s < 3$?} \\ (22/121)}}}
  \put(29,11){\makebox(0,0){\tiny \textsf{Y}}}
  \put(41,11){\makebox(0,0){\tiny \textsf{N}}}
  \put(32.5,11){\vector(-2,1){15}}
  \put(37.5,11){\vector(2,1){15}}
 \end{picture}
\end{center}
 \figurecaption{Example of a binary search tree with the sequential decisions
  required for coding data from a \mbox{6-symbol} alphabet using binary encoders.
  Leaf nodes represent data symbols, and the numbers above them represent their
  number of occurrences, $\tilde{P}(s)$. The binary information indicated by each question
  mark is coded with the probability estimate shown in parenthesis.}
                                                           \label{fgFDecTree}
\end{figure}

A binary-search tree as in Figure~\ref{fgFDecTree} can be automatically
generated using, for example, the bisection algorithm~\cite{Gill82,Rice83,Press93}. 
Algorithms \ref{alBSIntUpdt} and~\ref{alBSIntSel} show such implementations
for encoding, decoding, and overall probability estimation updates.
Note that they use the binary coding and decoding functions of
Algorithms \ref{alBNIntUpdt} and~\ref{alBNIntSel}, and
use only one vector, ${\bf\bar{C}}$, with dimension $M-1$, to store the
tree-branch occurrence counters. Each component $\bar{C}(m)$, $0 < m < M$,
contains the number of the number of times we had ``$s<m$?'' on tree node 
$\underline{m}$, and $\bar{C}(0)$ contains the total number of symbols 
coded. This vector is initialized with the number of leaf nodes reachable
from the left branch of the corresponding node.

The binary conditional probabilities estimates are computed during the
encoding and decoding directly from $\bar{\bf C}$.
For example, in the example of Figure~\ref{fgFDecTree}
we have
\[
  \bar{\bf C} = [\, 121 \; 4 \; 8 \; 22 \; 15 \; 19 \, ].
\]
In order to code the decision at node 
\underline{3} we use probability estimate 22/121, which is defined by 
counters $\bar{C}(0)=121$ and $\bar{C}(3)=22$. If we move to
node~\underline{1} then the next probability estimate is
$\bar{C}(1)/\bar{C}(3)=4/22$.
On the other hand, if we move node~\underline{4} we need to use
probability estimate $\bar{C}(4)/[\bar{C}(0)-\bar{C}(3)]=15/99$,
which is also readily computed from components of ${\bf\bar{C}}.$
See Algorithms \ref{alBSIntUpdt} and~\ref{alBSIntSel} for details.

\begin{Algorithm}{{Procedure Interval{\_}Update} $(s,b,l,M,{\bf\bar{C}})$}
1. set \{ \= $u \leftarrow 0;\;n \leftarrow M$; \COM{Initialize bisection search limits}
   \> $k \leftarrow \bar{C}(0)$; \COM{First divisor = symbol counter}
   \> $\bar{C}(0) \leftarrow \bar{C}(0) + 1$; \} \COM{Increment symbol counter}
\pushtabs
2. \= while $n - u > 1$ do \COM{Bisection search loop}
   \> 2.1. set $m \leftarrow \left\lfloor {(u + n) / 2} \right\rfloor $; \CON{Compute middle point}
   \> 2.2. \= if $s < m$ \CON{If symbol is smaller than middle}
   \> \> then \= set \{ \= $n \leftarrow m$; \COO{then update upper limit}
   \> \> \> \> \textsf{Binary{\_}Interval{\_}Update} $(0,\;b,\;l,\;\bar{C}(m),\;k)$; \COO{code symbol 0}
   \> \> \> \> $k \leftarrow \bar{C}(m)$; \COO{set next divisor}
   \> \> \> \> $\bar{C}(m) \leftarrow \bar{C}(m) + 1$; \} \COO{increment 0-symbol counter}
   \> \> else \> set \{ \> $u \leftarrow m$; \COO{else update lower limit}
   \> \> \> \> \textsf{Binary{\_}Interval{\_}Update} $(1,\;b,\;l,\;\bar{C}(m),\;k)$; \COO{code symbol 1}
   \> \> \> \> $k \leftarrow k - \bar{C}(m)$; \} \COO{set next divisor}
\poptabs 
3. return.
                                                     \label{alBSIntUpdt}
\end{Algorithm}

\begin{Algorithm}{{Function Interval{\_}Selection} $(v,b,l,M,{\bf \bar{C}})$}
1. set \{ \= $s \leftarrow 0; \quad n \leftarrow M$; \COM{Initialize bisection search bounds}
   \> $k \leftarrow \bar{C}(0)$; \COM{First divisor = symbol counter}
   \> $\bar{C}(0) \leftarrow \bar{C}(0) + 1$; \} \COM{Increment symbol counter}
\pushtabs
2. \= while $n - s > 1$ do \COM{Bisection search loop}
   \> 2.1. set $m \leftarrow \left\lfloor {(s + n) / 2} \right\rfloor $; \CON{Compute middle point}
   \> 2.2. \= if \textsf{Binary{\_}Interval{\_}Selection} $(v,\;b,\;l,\;\bar{C}(m),\;k) = 0$ \CON{If symbol is smaller than middle}
   \> \> then \= set \{ \= $n \leftarrow m$; \COO{then update upper limit}
   \> \> \> \> $k \leftarrow \bar{C}(m)$; \COO{set next divisor}
   \> \> \> \> $\bar{C}(m) \leftarrow \bar{C}(m) + 1$; \} \COO{increment 0-symbol counter}
   \> \> else \> set \{ \> $s \leftarrow m$; \COO{else update lower limit}
   \> \> \> \> $k \leftarrow k - \bar{C}(m)$; \} \COO{set next divisor}
\poptabs
3. return $s$.
                                                     \label{alBSIntSel}
\end{Algorithm}

Since we use bisection search in Algorithms \ref{alBSIntUpdt} and~\ref{alBSIntSel},
the number of times the binary encoding and decoding functions are called is between 
$\left\lfloor {\log_2 M} \right\rfloor$ and $\left\lceil {\log_2 M} \right\rceil$,
for all data symbols. Thus, by using binary-tree searches and binary arithmetic
coding we greatly reduce the \emph{worst-case} complexity required to update probability
estimates.

\begin{Example}
Figure~\ref{fgSDecTree} shows a second example of a binary-search tree
that can be used to code the \mbox{6-symbol} data of Example~\ref{exByteOut}.
Different search algorithms can be used to create different trees.
For example, we could have used a sequential search, composed of tests,
``$s < 5$?'', ``$s < 4$?'', ``$s < 3$?'', \ldots , ``$s < 1$?''

The trees created from bisection search minimize the maximum number of 
binary symbols to be coded, but not the average. The tree of Figure~\ref{fgSDecTree} was 
designed so that the most frequent symbols are reached with the smallest number 
of tests. Table~\ref{tbTreeSearch} shows the average number of symbols required for coding 
symbols from the source of Example~11 (considering probability estimates as 
the actual probabilities), for different trees created from different search 
methods.

Because we have the symbols sorted by increasing probability, the
performance of the tree defined by sequential search, starting from the 
most probable symbol, is quite good. The tree of Figure~\ref{fgSDecTree} is the one
that minimizes the average number of coded binary symbols. Below we explain 
how it is designed.
                                                      \label{exOptTree}
\end{Example}

%
\begin{figure}[tp]
\setlength{\unitlength}{1.5mm}
\begin{center}
 \begin{picture}(90,60)
%
%
  \thicklines
  \put(5,50){\makebox(0,0){\textbf{\color{red} 0}}}
  \put(5,50){\circle{6}}
  \put(5,55){\makebox(0,0){\small \textsf{\color{acblue} 4}}}
  \put(25,50){\makebox(0,0){\textbf{\color{red} 1}}}
  \put(25,50){\circle{6}}
  \put(25,55){\makebox(0,0){\small \textsf{\color{acblue} 8}}}
  \put(35,40){\makebox(0,0){\textbf{\color{red} 2}}}
  \put(35,40){\circle{6}}
  \put(35,45){\makebox(0,0){\small \textsf{\color{acblue} 10}}}
  \put(55,40){\makebox(0,0){\textbf{\color{red} 3}}}
  \put(55,40){\circle{6}}
  \put(55,45){\makebox(0,0){\small \textsf{1\color{acblue} 5}}}
  \put(75,40){\makebox(0,0){\textbf{\color{red} 4}}}
  \put(75,40){\circle{6}}
  \put(75,45){\makebox(0,0){\small \textsf{\color{acblue} 19}}}
  \put(85,20){\makebox(0,0){\textbf{\color{red} 5}}}
  \put(85,20){\circle{6}}
  \put(85,25){\makebox(0,0){\small \textsf{\color{acblue} 65}}}
  \put(15,40){\makebox(0,0){\underline{\textbf{1}}}}
  \put(15,40){\circle{6}}
  \put(13,36){\makebox(0,0)[t]{\scriptsize \shortstack{\textbf{$s < 1$?} \\ (4/12)}}}
  \put(11,41.4){\makebox(0,0){\tiny \textsf{Y}}}
  \put(19,41.4){\makebox(0,0){\tiny \textsf{N}}}
  \put(13,42){\vector(-1,1){6}}
  \put(17,42){\vector(1,1){6}}
  \put(25,30){\makebox(0,0){\textbf{\underline{2}}}}
  \put(25,30){\circle{6}}
  \put(25,26){\makebox(0,0)[t]{\scriptsize \shortstack{\textbf{$s < 2$?} \\ (12/22)}}}
  \put(21,31.4){\makebox(0,0){\tiny \textsf{Y}}}
  \put(29,31.4){\makebox(0,0){\tiny \textsf{N}}}
  \put(23,32){\vector(-1,1){6}}
  \put(27,32){\vector(1,1){6}}
  \put(45,20){\makebox(0,0){\textbf{\underline{3}}}}
  \put(45,20){\circle{6}}
  \put(45,16){\makebox(0,0)[t]{\scriptsize \shortstack{\textbf{$s < 3$?} \\ (22/56)}}}
  \put(39,21){\makebox(0,0){\tiny \textsf{Y}}}
  \put(51,21){\makebox(0,0){\tiny \textsf{N}}}
  \put(42.5,21){\vector(-2,1){15}}
  \put(47.5,21){\vector(2,1){15}}
  \put(65,30){\makebox(0,0){\underline{\textbf{4}}}}
  \put(65,30){\circle{6}}
  \put(65,26){\makebox(0,0)[t]{\scriptsize \shortstack{\textbf{$s < 4$?} \\ (15/34)}}}
  \put(61,31.4){\makebox(0,0){\tiny \textsf{Y}}}
  \put(69,31.4){\makebox(0,0){\tiny \textsf{N}}}
  \put(63,32){\vector(-1,1){6}}
  \put(67,32){\vector(1,1){6}}
  \put(65,10){\makebox(0,0){\underline{\textbf{5}}}}
  \put(65,10){\circle{6}}
  \put(65,6){\makebox(0,0)[t]{\scriptsize \shortstack{\textbf{$s < 5$?} \\ (56/121)}}}
  \put(59,11){\makebox(0,0){\tiny \textsf{Y}}}
  \put(71,11){\makebox(0,0){\tiny \textsf{N}}}
  \put(62.5,11){\vector(-2,1){15}}
  \put(67.5,11){\vector(2,1){15}}
 \end{picture}
\end{center}
 \figurecaption{Another example of a binary-search tree with for coding data
   from a \mbox{6-symbol} source. The number of symbol occurrences is the same
   as shown in Figure~\ref{fgFDecTree}. This tree has been designed to
   minimize the average number of binary symbols coded.}
                                                      \label{fgSDecTree}
\end{figure}

%

\begin{table}[tbp]
\begin{center}
\begin{tabular}{|c|c|c|c|c|c|c|c|} \hline
Data & Probability & \multicolumn{6}{|c|}{Number of binary symbols coded} \\ \cline{3-8}
Symbol & estimate & \multicolumn{2}{|c|}{Sequential search} & \multicolumn{2}{|c|}{Bisection search} & 
\multicolumn{2}{|c|}{Optimal search} \\ \cline{3-8}
$s$ & $p(s)$ & $N(s)$ & $p(s)N(s)$ & $N(s)$ & $p(s)N(s)$ & $N(s)$ & $p(s)N(s)$ \\ \hline
 0  & 0.033 & 6 & 0.198 & 2 & 0.066 & 4 & 0.132 \\ \hline
 1  & 0.066 & 5 & 0.331 & 3 & 0.198 & 4 & 0.264 \\ \hline
 2  & 0.083 & 4 & 0.331 & 3 & 0.248 & 3 & 0.248 \\ \hline
 3  & 0.124 & 3 & 0.372 & 2 & 0.248 & 3 & 0.372 \\ \hline
 4  & 0.157 & 2 & 0.314 & 3 & 0.471 & 3 & 0.471 \\ \hline
 5  & 0.537 & 1 & 0.537 & 3 & 1.612 & 1 & 0.537 \\ \hline
Sum & 1.000 & 21 & 2.083 & 16 & 2.843 & 18 & 2.025 \\ \hline
\end{tabular}
\end{center}
 \tablecaption{Number binary symbols coded using trees created from 
  different types of binary searches, applied to data source of
  Example~10. The trees corresponding to bisection and 
  optimal searches are shown in Figures \ref{fgFDecTree} and 
  \ref{fgSDecTree}, respectively.}
                                                      \label{tbTreeSearch}
\end{table}

Prefix coding~\cite{Gallager68,Jelinek68,McEliece84,Cover91,Sayood99,Salomon00}
is the process of coding information using decision trees as defined above.
The coding process we have shown above is identical, except 
that we call a binary arithmetic encoding/decoding function at each node. 
Thus, we can use all the known facts about prefix coding to analyze the 
computational complexity of binary arithmetic encoding/decoding, if we measure 
complexity by the number of coded binary symbols.

Since the optimal trees for prefix coding are created using the Huffman 
algorithm~\cite{Huffman52}, these trees are also optimal for binary 
arithmetic encoding/decoding~\cite{Langdon84}. Strictly speaking, if the data symbols 
are not sorted according to their probability, the optimal Huffman tree does 
not satisfy the requirements for binary-search trees, i.e., ``keys'' are not 
properly sorted, and we cannot define a node with a simple comparison of the 
type ``$s < m$?'' This problem is solved by storing the paths from the root node
to leaf nodes, i.e., the Huffman codewords.

\subsection{Tree-based Update of Cumulative Distributions}   \label{ssTreeUpdt}

In this section, we show that we can use binary-search trees
(Section~\ref{ssBinCod}) to efficiently combine computing the cumulative
distribution, updating it, encoding and decoding, without having to use
a binary arithmetic encoder. We present techniques similar to the 
methods proposed by Moffat~\cite{Moffat90} and Fenwick~\cite{Fenwick94}.
We start with an example of how to compute the cumulative distribution vector
${\bf \tilde{C}}$ from the statistics $\bar{\bf C}$ gathered while using the
binary search trees.

\begin{Example}
Let us consider the binary search tree shown in Figure~\ref{fgFDecTree}.
Let us assume that we had been using Algorithms \ref{alBSIntUpdt} and
\ref{alBSIntSel} to compute the number of symbols 
occurrences in the tree, ${\bf \bar{C}}$, and we want to compute the 
cumulative distribution ${\bf \tilde{C}}$ from
${\bf \bar{C}} = [\, 121 \; 4 \; 8 \; 22 \; 15 \; 19 \,]$.

From the tree structure we can find out that, except for the root node,
the counter at each node has 
the number of occurrences of all symbols found following the left branch, 
i.e.,
\[
 \begin{array}{rll}
  \bar{C}(0) \equiv & \tilde{P}(0) + \tilde{P}(1) + \tilde{P}(2) + \tilde{P}(3) + \tilde{P}(4) + \tilde{P}(5) & = 121 \\
  \bar{C}(1) = & \tilde{P}(0) & = 4 \\
  \bar{C}(2) = & \tilde{P}(1) & = 8 \\
  \bar{C}(3) = & \tilde{P}(0) + \tilde{P}(1) + \tilde{P}(2) & = 22 \\
  \bar{C}(4) = & \tilde{P}(3) & = 15 \\
  \bar{C}(5) = & \tilde{P}(4) & = 19 \\
 \end{array}
\]
where $\tilde{P}(s)$ is the number of time symbol $s$ has occurred. From these 
equations we can compute the vector with cumulative distributions as
\[
 \begin{array}{rlll}
  \tilde{C}(0) \equiv & 0 & \\
  \tilde{C}(1) = & \tilde{P}(0) & = \bar{C}(1) & = 4 \\
  \tilde{C}(2) = & \tilde{P}(0) + \tilde{P}(1) & = \bar{C}(1) + \bar{C}(2) & = 12 \\
  \tilde{C}(3) = & \tilde{P}(0) + \tilde{P}(1) + \tilde{P}(2) & = \bar{C}(3) & = 22 \\
  \tilde{C}(4) = & \tilde{P}(0) + \tilde{P}(1) + \tilde{P}(2) + \tilde{P}(3) & = \bar{C}(3) + \bar{C}(4) & = 37 \\
  \tilde{C}(5) = & \tilde{P}(0) + \tilde{P}(1) + \tilde{P}(2) + \tilde{P}(3) + \tilde{P}(4) & = \bar{C}(3) + \bar{C}(4) + \bar{C}(5) & = 56 \\
  \tilde{C}(6) = & \tilde{P}(0) + \tilde{P}(1) + \tilde{P}(2) + \tilde{P}(3) + \tilde{P}(4) + \tilde{P}(5) & = \bar{C}(0) & = 121 \\
 \end{array}
\]

We can do the same with the tree of Figure~\ref{fgSDecTree}, and find different sets of 
equations. In this case the counters are
\[
 \begin{array}{rll}
  \bar{C}(0) \equiv & \tilde{P}(0) + \tilde{P}(1) + \tilde{P}(2) + \tilde{P}(3) + \tilde{P}(4) + \tilde{P}(5) & = 121 \\
  \bar{C}(1) = & \tilde{P}(0) & = 4 \\
  \bar{C}(2) = & \tilde{P}(0) + \tilde{P}(1) & = 12 \\
  \bar{C}(3) = & \tilde{P}(0) + \tilde{P}(1) + \tilde{P}(2) & = 22 \\
  \bar{C}(4) = & \tilde{P}(3) & = 15 \\
  \bar{C}(5) = & \tilde{P}(0) + \tilde{P}(1) + \tilde{P}(2) + \tilde{P}(3) + \tilde{P}(4) & = 56 \\
\end{array}
\]
and their relationship with the cumulative distribution is given by
\[
 \begin{array}{rlll}
  \tilde{C}(0) \equiv & 0 & \\
  \tilde{C}(1) = & \tilde{P}(0) & = \bar{C}(1) & = 4 \\
  \tilde{C}(2) = & \tilde{P}(0) + \tilde{P}(1) & = \bar{C}(2) & = 12 \\
  \tilde{C}(3) = & \tilde{P}(0) + \tilde{P}(1) + \tilde{P}(2) & = \bar{C}(3) & = 22 \\
  \tilde{C}(4) = & \tilde{P}(0) + \tilde{P}(1) + \tilde{P}(2) + \tilde{P}(3) & = \bar{C}(3) + \bar{C}(4) & = 37 \\
  \tilde{C}(5) = & \tilde{P}(0) + \tilde{P}(1) + \tilde{P}(2) + \tilde{P}(3) + \tilde{P}(4) & = \bar{C}(5) & = 56 \\
  \tilde{C}(6) = & \tilde{P}(0) + \tilde{P}(1) + \tilde{P}(2) + \tilde{P}(3) + \tilde{P}(4) + \tilde{P}(5) & = \bar{C}(0) & = 121 \\
\end{array}
\]

                                                      \label{exTreeUpd}
\end{Example}

The equations that we obtain from any decision tree are linearly
independent, and thus it is always possible to 
compute the cumulative distribution ${\bf \tilde{C}}$ from the 
counters ${\bf \bar{C}}$. We show next how to use the tree structure 
to efficiently compute the components of ${\bf \tilde{C}}$ required 
for encoding symbol $s$, $\tilde{C}(s)$ and~$\tilde{C}(s + 1)$. 

In order to compute $\tilde{C}(s)$, when we move from the root of the tree, up 
to the leaf representing symbol~$s$, we simply add the value of $\bar{C}(n)$ for 
each node $\underline{n}$ that does not have its condition satisfied (i.e., we 
move up its right-side branch). For example, to compute $\tilde{C}(2)$ 
using the tree of Figure~\ref{fgFDecTree}, we start from node~\underline{3}, and move 
left to node~\underline{1}, right to node \underline{2}, and right to leaf~2.
Since we move along the right branch of nodes \underline{1} and 
\underline{2}, we conclude that $\tilde{C}(2)=\bar{C}(1)+\bar{C}(2)=12$.

The value of $\tilde{C}(s+1)$ can be computed together 
with $\tilde{C}(s)$: we just have to add the running sum for $\tilde{C}(s)$
and $\bar{C}(n)$ at the last left-side branch taken before reaching 
leaf $s$. For example, when computing $\tilde{C}(2)$,
the last left-side branch is taken at root node~$\underline{3}$, and thus
$\tilde{C}(3)$ is equal to the running sum (zero) plus $\bar{C}(3)=22$.

Algorithms \ref{alSSIntUpdt} and~\ref{alSSIntSel} show how to
combine all the techniques above to simultaneously compute and update
the cumulative distribution, and use the computed values for encoding
and decoding. They use a tree defined by bisection search, but it
is easy to modify them to other tree structures.
After Step~2 of Algorithm~\ref{alSSIntUpdt} we have $E=\tilde{C}(s)$ and
$F=\tilde{C}(s+1)$ computed and updated with a number of additions
proportional to $\log_2(M)$.

\begin{Algorithm}{{Procedure Interval{\_}Update} $(s,b,l,M,{\bf \bar{C}})$}
1. \= set \{ \= $u \leftarrow 0;\;n \leftarrow M$; \COM{Initialize bisection search limits}
   \> \> $E \leftarrow 0;\;F \leftarrow \bar{C}(0)$; \COM{Initialize cumulative sum bounds}
   \> \> $\gamma \leftarrow l / \bar{C}(0)$; \} \COM{Compute result of division}
\pushtabs
2. \= while $n - u > 1$ do \COM{Bisection search loop}
   \> 2.1. set $m \leftarrow \left\lfloor {(u + n) / 2} \right\rfloor$; \CON{Compute middle point}
   \> 2.2. \= if $s < m$ \CON{If symbol is smaller than middle}
   \> \> then \= set \{ \= $n \leftarrow m$; \COO{then update bisection upper limit}
   \> \> \> \> $F \leftarrow E + \bar{C}(m)$; \COO{set upper bound on cum. sum}
   \> \> \> \> $\bar{C}(m) \leftarrow \bar{C}(m) + 1$; \} \COO{Increment node occurrence counter}
   \> \> else \> set \{ \> $u \leftarrow m$; \COO{else update bisection lower limit}
   \> \> \> \> $E \leftarrow E + \bar{C}(m)$; \} \COO{increment lower bound on cum. sum}
\poptabs
3. if $s = M - 1$ \COM{Set interval end according to symbol}
   \> then \> set $y \leftarrow b + l$; \CON{Exact multiplication}
   \> else \> set $y \leftarrow b + \gamma \cdot F$; \CON{Base of next subinterval}
4. set \{ $b \leftarrow b + \gamma \cdot E$; \COM{Update interval base}
   \> \> $l \leftarrow y - b$; \COM{Update interval length as difference}
   \> \> $\bar{C}(0) \leftarrow \bar{C}(0) + 1$; \} \COM{Increment symbol counter}
5. return.
                                                     \label{alSSIntUpdt}
\end{Algorithm}

\begin{Algorithm}{{Function Interval{\_}Selection} $(v,b,l,M,{\bf \bar{C}})$}
1. set \{ \= $s \leftarrow 0;\;n \leftarrow M$; \COM{Initialize bisection search symbol limits}
   \> $x \leftarrow b; \; y \leftarrow b + l$; \COM{Set search interval bounds}
   \> $E \leftarrow 0$; \COM{Initialize cumulative sum}
   \> $\gamma \leftarrow l / \bar{C}(0)$; \} \COM{Compute result of division}
\pushtabs
2. \= while $n - s > 1$ do \COM{Bisection search loop}
   \> 2.1. \= set \{ \= $m \leftarrow \left\lfloor {(s + n) / 2} \right\rfloor$; \CON{Compute middle point}
   \> \> \> $z \leftarrow b + \gamma \cdot [E+\bar{C}(m)]$; \} \CON{Value at middle point}
   \> 2.2. if $z > v$ \CON{If symbol is smaller than middle}
   \> \> then \> set \{ \= $n \leftarrow m$; \COO{then update bisection upper limit}
   \> \> \> \> $y \leftarrow z;$  \COO{new interval end}
   \> \> \> \> $\bar{C}(m) \leftarrow \bar{C}(m) + 1$; \} \COO{increment node occurrence counter}
   \> \> else \> set \{ \> $s \leftarrow m$; \COO{else update bisection lower limit}
   \> \> \> \> $x \leftarrow z;$  \COO{new interval base}
   \> \> \> \> $E \leftarrow E + \bar{C}(m)$; \} \COO{increment lower bound on cum. sum}
\poptabs
3. set \{ $b \leftarrow x$; \COM{Update interval base}
   \> $l \leftarrow y - x$; \COM{Update interval length as difference}
   \> $\bar{C}(0) \leftarrow \bar{C}(0) + 1$; \} \COM{Increment symbol counter }
4. return $s$.
                                                     \label{alSSIntSel}
\end{Algorithm}

\subsection{Periodic Updates of the Cumulative Distribution}  \label{ssPerUpdt}

We have seen in Section~\ref{ssDirUpdt} that adaptive coding can
increase the arithmetic coding computational complexity significantly,
because of the effort to update the cumulative distributions.
In Sections \ref{ssBinCod} and~\ref{ssTreeUpdt} we present tree-based
updating techniques that reduce this complexity very significantly. 
However, adaptation can still be a substantial fraction of the
overall coding computational complexity, and it happens that there is
not much we can do if we insist on the
assumption that the probability model has to be updated immediately
after each encoded/decoded symbol, i.e., estimates are in the form given
by~(\ref{eqAdapCod3}), with a division by the factor $\tilde{C}(M)$.

However, with accurate and fast source modeling we can avoid having
probabilities changing so quickly that we need to refresh estimates on
a symbol-by-symbol basis. For example, an image encoder may use different
probabilities depending on a classification of the part being encoded,
which can be something like
``constant,'' ``smooth,'' ``edge,'' ``high-frequency pattern," etc.
With these techniques, we can assume that the source state may change
quickly, but the source properties (symbol probabilities) for each state
change slowly.

Under these assumptions, and unless the number of data symbols is very
large (thousands or millions~\cite{Moffat98}), a significantly more efficient form of adaptive
arithmetic coding updates only the vector with symbol occurrence counters
$({\bf \tilde{P}})$ after each symbol, and updates the distribution estimate
(${\bf c}$) periodically, or following some special events. For example, the
Q-coder updates its probability estimation only during renormalization~\cite{Pennebaker88a}.
For periodic updates, we define $R$ as number of symbol coded between updates
of ${\bf c}$. Typically, the period $R$ is a small multiple of $M$ (e.g., $R=4M$),
but to improve the accuracy while minimizing the computations, we can start
with frequent updates, and then decrease their frequency.

One immediate consequence of this approach is that while coding we can
use the simpler procedures of Section~\ref{ssFPImplem} and Appendix~A,
and can completely avoid the divisions in equations (\ref{eqAdapCod3})
to~(\ref{eqAdapCod5}). Furthermore,
if the period $R$ is large enough, then it is feasible to do many complex
tasks while updating ${\bf c}$, in order to increase the encoding/decoding
speed. For instance, in Section~\ref{ssSymSearch} we explain how to make
decoding faster by sorting the symbols according to their probability,
finding the optimal decoding sequence (Huffman tree), or computing the data
for fast table look-up decoding.

\section{Complexity Analysis}                                \label{ssComplexA}

Large computational complexity had always been a barrier to the adoption of 
arithmetic coding. In fact, for many years after arithmetic coding was 
invented, it was considered little more than a mathematical curiosity 
because the additions made it slow, multiplications made it very expensive, 
and divisions made it impractical. In this section, we analyze the 
complexity of arithmetic coding and explain how the technological advances 
that gives us fast arithmetic operations change the complexity analysis.

Many of the conclusions in this section are based on experimental tests
designed for analyzing the arithmetic coding complexity, which are described
in reference~\cite{Said04}. Another earlier experimental evaluation of
complexity is in reference~\cite{Moffat94}.

We have to observe that the \emph{relative} computational complexity of
different coding operations changed dramatically. Arithmetic operations
today are much more efficient, and not only due the great increase in the
CPU clock frequency. In the past, the ratio between clock cycles used by
some simple  operations (comparisons, bit shifts, table look-up) and arithmetic
operations (specially division) was quite large. Today, this ratio is much 
smaller~\cite{Texas00,IBM01,Intel01,Sun01}, invalidating previous assumptions for 
complexity reduction. For instance, a set of comparisons, bit shifts, and 
table look-up now takes significantly more time than a multiplication.

The speed of arithmetic coding needs to be measured by the rate of data
(true information) bits coming out of the encoder, or into the decoder 
(\textit{information throughput}). For example, binary coders can be very
simple and fast, but their throughput is limited to a fraction of bits per
CPU clock cycle. Arithmetic coders with larger alphabets, on the other
hand, process information in a significantly more efficient manner, and
can easily yield throughputs of many bits per CPU clock cycle~\cite{Said04}.

In this section we use the ``big-$O$'' notation of~\cite{Cormen90}, where
$O(\cdot)$ indicates asymptotic upper bounds on the computational complexity.

The main factors that influence complexity are
\begin{itemize}
 \item Interval renormalization and compressed data input and output.
 \item Symbol search.
 \item Statistics estimation (adaptive models only).
 \item Arithmetic operations.
\end{itemize}

In the next sections we analyze each of these in some detail. However, we will 
not consider special hardware implementations (ASICs) for three reasons:
(a)~it is definitely outside the scope of this text; (b)~our model also applies 
to some specialized hardware, like DSP chips; (c)~many optimization 
techniques for general CPUs also apply to efficient hardware.

\subsection{Interval Renormalization and Compressed Data Input and Output}

In integer arithmetic implementations the interval renormalization can be 
calculated with only bit shifts. However, when $D=2$, renormalization
occurs quite frequently, consuming many clock cycles. Using larger output
alphabets in the form $D=2^r$ reduces the frequency of the renormalization
by a factor of $r$ (see Example~\ref{exByteOut}), and thus may reduce the
number of clock cycles used for renormalization very significantly.
Floating-point implementations may need multiplication during renormalization,
but if $D$ is large enough then these operations do not add much to
the overall complexity.

Data input and output, which occurs during renormalization, also has 
significant impact on the encoding and decoding speed. Since computers
and communication systems work with groups of 8~bits (bytes), 
processing one bit at a time requires extra clock cycles to properly align
the data. Thus, there is substantial speedup when renormalization
produces bits that can be easily aligned in bytes (e.g., $D=2^4$, $D=2^8$,
or $D=2^{16}$). The case $D=256$ has been shown to be particularly
efficient~\cite{Schindler98,Wu99}.

\subsection{Symbol Search}                                  \label{ssSymSearch}

The computational complexity of the arithmetic decoder may be many times larger
than the encoder's because the decoder needs to find out in which subinterval
the current code value is, i.e., solve
\begin{equation}
  \hat{s}(v) = \left\{\,s:\; c(s)l \le v - b < c(s + 1)l \right\}, \label{eqSymSearch1}
\end{equation}
or
\begin{equation}
  \hat{s}(v) = \left\{\,s:\;c(s)\le\frac{v-b}{l}<c(s + 1) \right\}.  \label{eqSymSearch2}
\end{equation}
Mathematically these are equivalent, but we use~(\ref{eqSymSearch2}) to represent
the case in which the division $(v-b)/l$ is computed first
(e.g., Algorithm~\ref{alASCIntSel}).

This is a line-search problem~\cite{Gill82,Cormen90}, where we need to minimize both
the number of points tested and the computations used for determining these
points. The possible difference in complexity between the encoder and decoder grows
with the number of data symbols, $M$. We can see from Algorithms \ref{alBNIntUpdt}
and~\ref{alBNIntSel} that when $M=2$, the complexity is practically the
same for the encoder and decoder.

There are five main schemes for symbol search that are described next.

\paragraph{(a) Sequential search on sorted symbols} {\ } \par \nopagebreak

\noindent This search method, used in Algorithms \ref{alFPIntSel} and~\ref{alADCIntSel},
in the worst-case 
searches $M-1$ intervals to find the decoded symbol. We can try to
improve the average performance by sorting the symbols according to their 
probability. Assuming that the symbols are sorted with increasing 
probability, the average number of tests is
\begin{equation}
                                                     \label{ssComplex1}
  \bar{N}_s(\Omega) = \sum_{m=0}^{M-1} {p(m)(M-m)} = M -
    \sum_{m=0}^{M-1} {m\,p(m) = M-\bar{s}(\Omega)} \le M,
\end{equation}
where $\bar{s}(\Omega)$ is the average symbol number (after sorting) put 
out by data source $\Omega$. Thus, sequential search can be efficient only
if the symbol distribution is very skewed ($\bar{s}(\Omega) \approx M-1)$.

\paragraph{(b) Bisection search} {\ } \par \nopagebreak

\noindent With this type of search, shown in Algorithms \ref{alBSIntSel}, \ref{alSSIntSel}
and~\ref{alINTIntSel}, the number of tests is bounded by
\begin{equation}
                                                     \label{ssComplex2}
  \left\lfloor {\log_2(M)} \right\rfloor \le \bar{N}_b(\Omega) \le 
    \left\lceil {\log_2(M)} \right\rceil,
\end{equation}
independently of the probabilities of the data symbols. Note that the 
\emph{encoder} may also have to implement the bisection search when it is used
with binary coding (Section~\ref{ssBinCod}), or when using trees for
updating the probability estimates (Section~\ref{ssTreeUpdt}).

\paragraph{(c) Optimal tree-based search} {\ } \par \nopagebreak

\noindent We show in Section~\ref{ssTreeUpdt} how the process of encoding and decoding data from 
a \mbox{$M$-symbol} alphabet can be decomposed in a set of binary decisions, and in 
Example~\ref{exOptTree} we show its similarities to prefix coding, and an 
optimal decision tree designed with the Huffman algorithm. The interval 
search process during decoding is also defined by binary decisions, and the 
optimal set of decisions is also computed with the Huffman algorithm~\cite{Huffman52,Langdon84}. 
Thus, we can conclude that the average number of tests in such scheme is 
bounded as the average number of bits used to code the source with Huffman 
coding, which is given by~\cite{Gallagher78}
\begin{equation}
                                                     \label{ssComplex3}
  \max \{1,H(\Omega)\} \le \bar{N}_o (\Omega) \le H(\Omega) + 0.086 +
    \max_{m=0,1,\ldots,M-1} \{p(m)\}.
\end{equation}

Implementing the optimal binary-search tree requires some extra storage, 
corresponding to the data normally used for representing a Huffman code, and 
it is necessary to reorder the symbols according to probability (as in 
the example of Figure~\ref{fgSDecTree}), or use a modified definition of
the cumulative distribution.

With adaptive coding we have the problem that the optimal tree is defined 
from the symbol probabilities, which are unknown when encoding and decoding 
start. This problem can be solved by periodically redesigning the tree, 
together with the distribution updates (Section~\ref{ssPerUpdt}).

\paragraph{(d) Bisection search on sorted symbols} {\ } \par \nopagebreak

\noindent We can combine the simplicity of bisection with a technique that takes into 
account the symbol probabilities. When the data symbols are sorted with 
increasing probability, we can look for the symbol $m$ such that $c(m) \approx 0.5$,
and use it as the first symbol (instead of middle point) to divide the 
interval during bisection. If the distribution is nearly uniform, then this 
symbol should be near the middle and the performance is similar to standard 
bisection search. On the other hand, if the distribution is highly skewed, 
them $m \approx M-1$, meaning that the most probable symbols are tested 
first, reducing the average number of tests.

\begin{figure}[tp]
\begin{center}
  \includegraphics[scale=0.9]{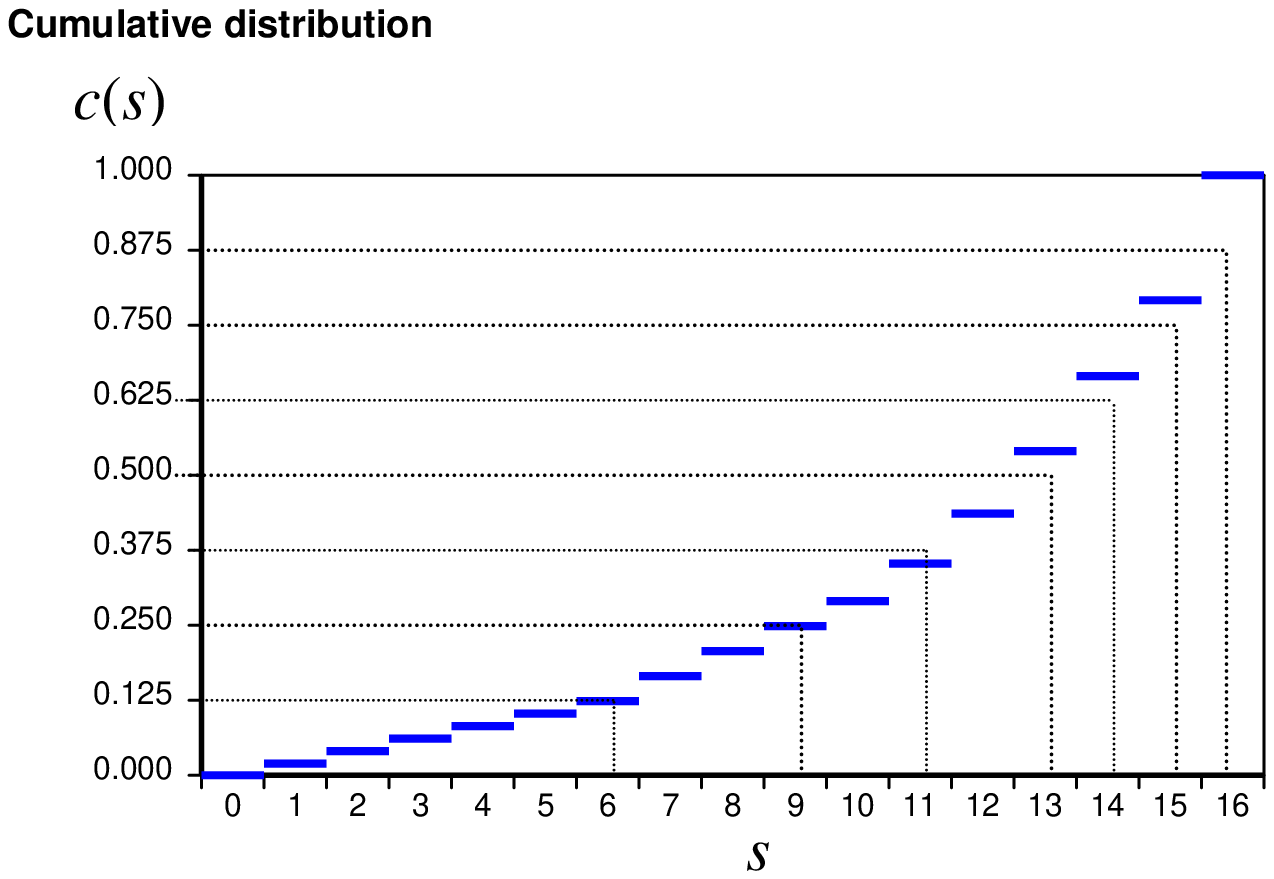}
\end{center}
 \figurecaption{Technique to find quasi-optimal points to initialize bisection
   search on sorted symbols.}
                                                          \label{fgCumDSearch}
\end{figure}

We can extend this technique to find also find the symbols with $c(m)$ near 
0.25 and 0.75, and then 0.125, 0.375, 0.625, and 0.825, and so on. Figure~\ref{fgCumDSearch} 
shows an example of a cumulative distribution, and the process of finding the 
symbols for initializing the bisection search. The corresponding first levels of the
binary-search tree are shown in Figure~\ref{fgCumDTree}. The decoder does not have to use 
all levels of that tree; it can use only the first level (and store one 
symbol), or only the first two levels (and store three symbols), or any 
desired number of levels. In Figure~\ref{fgCumDTree}, the complete binary-search tree is 
defined by applying standard bisection search following the tests shown in 
the figure.
If the encoder uses up to $V>1$ levels of the binary-search tree defined by 
the cumulative distribution, the average number of tests is near the optimal 
(\ref{ssComplex3}), and the worst case is not much larger than (\ref{ssComplex2}).

%

%
\begin{figure}[tp]
\setlength{\unitlength}{1.5mm}
\begin{center}
 \begin{picture}(70,35)
%
%
  \thicklines
  \put(5,25){\makebox(0,0){\underline{\textbf{6}}}}
  \put(5,25){\circle{6}}
  \put(3,21){\makebox(0,0)[t]{\scriptsize \textbf{$s < 6$?}}}
  \put(1,26.4){\makebox(0,0){\tiny \textsf{Y}}}
  \put(9,26.4){\makebox(0,0){\tiny \textsf{N}}}
  \put(3,27){\vector(-1,1){6}}
  \put(7,27){\vector(1,1){6}}
  \put(15,15){\makebox(0,0){\underline{\textbf{9}}}}
  \put(15,15){\circle{6}}
  \put(15,11){\makebox(0,0)[t]{\scriptsize \textbf{$s < 9$?}}}
  \put(11,16.4){\makebox(0,0){\tiny \textsf{Y}}}
  \put(19,16.4){\makebox(0,0){\tiny \textsf{N}}}
  \put(13,17){\vector(-1,1){6}}
  \put(17,17){\vector(1,1){6}}
  \put(25,25){\makebox(0,0){\textbf{\underline{11}}}}
  \put(25,25){\circle{6}}
  \put(27,21){\makebox(0,0)[t]{\scriptsize \textbf{$s < 11$?}}}
  \put(21,26.4){\makebox(0,0){\tiny \textsf{Y}}}
  \put(29,26.4){\makebox(0,0){\tiny \textsf{N}}}
  \put(23,27){\vector(-1,1){6}}
  \put(27,27){\vector(1,1){6}}
  \put(65,25){\makebox(0,0){\textbf{\color{red} 15}}}
  \put(65,25){\circle{6}}
  \put(55,15){\makebox(0,0){\underline{\textbf{15}}}}
  \put(55,15){\circle{6}}
  \put(55,11){\makebox(0,0)[t]{\scriptsize \textbf{$s < 15$?}}}
  \put(51,16.4){\makebox(0,0){\tiny \textsf{Y}}}
  \put(59,16.4){\makebox(0,0){\tiny \textsf{N}}}
  \put(53,17){\vector(-1,1){6}}
  \put(57,17){\vector(1,1){6}}
  \put(45,25){\makebox(0,0){\textbf{\underline{14}}}}
  \put(45,25){\circle{6}}
  \put(43,21){\makebox(0,0)[t]{\scriptsize \textbf{$s < 14$?}}}
  \put(41,26.4){\makebox(0,0){\tiny \textsf{Y}}}
  \put(49,26.4){\makebox(0,0){\tiny \textsf{N}}}
  \put(43,27){\vector(-1,1){6}}
  \put(47,27){\vector(1,1){6}}
  \put(35,5){\makebox(0,0){\underline{\textbf{13}}}}
  \put(35,5){\circle{6}}
  \put(35,1){\makebox(0,0)[t]{\scriptsize \textbf{$s < 13$?}}}
  \put(29,6){\makebox(0,0){\tiny \textsf{Y}}}
  \put(41,6){\makebox(0,0){\tiny \textsf{N}}}
  \put(32.5,6){\vector(-2,1){15}}
  \put(37.5,6){\vector(2,1){15}}
 \end{picture}
\end{center}
 \figurecaption{Decision process corresponding to the quasi-optimal binary-search
   initialization of Figure~\ref{fgCumDSearch}.}
                                                      \label{fgCumDTree}
\end{figure}

\paragraph{(e) Bisection search starting from table look-up} {\ } \par \nopagebreak

\noindent The interval search methods that use~(\ref{eqSymSearch2}) require one
division, but its advantage is that it is equivalent to rescaling the current
interval to unit length before search. This means that from a quantized version
of the fraction $(v-b)/l$, like
\begin{equation}
  E(v, b, l) = \Floor{\frac{K_t(v-b)}{l}},
\end{equation}
we can know better initial values for the bisection search, which can be stored
in tables with $K_t$ elements, enabling significantly faster decoding.

The table entries are
\begin{eqnarray}
  \hat{s}_{\min}(E) & = & \left\{\,s:\;c(s)\le\frac{E}{K_t}<c(s + 1) \right\}, \\
  \hat{s}_{\max}(E) & = & \left\{\,s:\;c(s) < \frac{E+1}{K_t} \le c(s + 1) \right\}.
\end{eqnarray}
Note that $\hat{s}_{\max}(E) <= \hat{s}_{\min}(E+1)$, so one table is enough
for correct decoding.

\begin{Example}
Let us consider again the source of Example~\ref{exFirstEnc}, with
${\bf c} = [\,0\;\;0.2\;\;0.7\;\;0.9\;\;1\,]$. Table~\ref{tbTableDecd} shows
how each value of $E(v, b, l)$ corresponds to an interval of possible
values of $(v-b)/l$. By analyzing ${\bf c}$ and these intervals we can
identify the range of possible decoded symbols, which correspond to
$\hat{s}_{\min}(E)$ and $\hat{s}_{\max}(E)$.
For example, if while decoding we have $E(v, b, l) = 0$,
then we know that $(v-b)/l < 0.125$, and consequently $\hat{s}(v) = 0$.
On the other hand, if $E(v, b, l) = 5$, then $0.625 \leq (v-b)/l < 0.75$,
which means that we can only say that $1 \leq \hat{s}(v) \leq 2$, and one
more test is necessary for finishing decoding.
                                                            \label{exTableDecd}
\end{Example}

%

\begin{table}[tbp]
\begin{center}
  \begin{tabular}{|c|c|c|c|} \hline
  $E(v, b, l)$ & $(v-b)/l$ & $\hat{s}_{\min}(E)$ & $\hat{s}_{\max}(E)$ \\ \hline \hline
  0 & [0.000, 0.125) & 0 & 0 \\ \hline
  1 & [0.125, 0.250) & 0 & 1 \\ \hline
  2 & [0.250, 0.375) & 1 & 1 \\ \hline
  3 & [0.375, 0.500) & 1 & 1 \\ \hline
  4 & [0.500, 0.625) & 1 & 1 \\ \hline
  5 & [0.625, 0.750) & 1 & 2 \\ \hline
  6 & [0.750, 0.875) & 2 & 2 \\ \hline
  7 & [0.875, 1.000) & 2 & 3 \\ \hline
\end{tabular}
\end{center}
 \tablecaption{Values of $(v-b)/l$ quantized to 8 integer values to allow fast
   decoding using table look-up. The minimum and maximum possible values
   of the decoded symbol (table entries) were computed using
   ${\bf c} = [\,0\;\;0.2\;\;0.7\;\;0.9\;\;1\,]$.}
                                                            \label{tbTableDecd}
\end{table}

Note that, while the number of tests required by the original bisection
search does not depend on the probabilities, when we use $\hat{s}_{\min}(E)$
and $\hat{s}_{\max}(E)$ to initialize the search the initial interval
depends on the probabilities, and the most probable symbols are found with
a smaller number of tests.
For instance, we can see in Table~\ref{tbTableDecd} that most of the table
entries correspond to the most probable symbols.

The main constraint for table look-up decoding is that it is practical only for
static or periodically updated models (Section~\ref{ssPerUpdt}). The overhead
of computing the table each time the model is updated can be small
because table look-up is only for initializing the bisection search,
and even small tables (e.g., 16~entries) can make it significantly more
efficient.

\subsection{Cumulative Distribution Estimation}          \label{ssCumEstComplx}

Adaptive arithmetic coding requires an extra effort to update the
cumulative distribution estimates after coding each symbol, or in
a periodical manner.
Section~\ref{ssAdapCod} covers all the important aspects of how adaptive
encoders efficiently update the cumulative distributions ${\bf c}$. We
analyze two main strategies, depending on whether the cumulative distribution
updating is done together or independently of the symbol search
(Section~\ref{ssSymSearch}).

\paragraph{(a) Combined updating, coding and decoding} {\ } \par \nopagebreak

\noindent In this case, the most efficient implementations use binary-search
trees, and updating is applied to vector $\bar{\bf C}$, instead of
${\bf \tilde{C}}$. Binary coding (Section~\ref{ssBinCod}) and
tree-based updating (Section~\ref{ssTreeUpdt}) have the same
asymptotic complexity, depending on the tree being used. In addition,
by comparing Algorithms \ref{alBSIntUpdt} and~\ref{alBSIntSel} and
Algorithms \ref{alSSIntUpdt} and~\ref{alSSIntSel}, we can see that
the encoder and the decoder have very similar computational complexity.

The worst-case effort is minimized when we use bisection search, resulting
in a number of operations per coded symbol
proportional to $\log_2(M)$ (equation~(\ref{ssComplex2})).
Huffman trees, that optimize the average performance, require an
average number of operations proportional to $\bar{N}_o(\Omega)$, defined in
equation~(\ref{ssComplex3}).

\paragraph{(b) Independent updating} {\ } \par \nopagebreak

\noindent When source $\Omega$ has low entropy, some symbols should occur
much more frequently, and it may be efficient to update~${\bf \tilde{C}}$
directly, if the symbols are sorted by increasing probabilities.
Implementations by Witten \emph{et al.} \cite{Witten87,Moffat98}
use this strategy. However, when all symbols are equally probable,
the effort for sorting and updating ${\bf \tilde{C}}$ is on average
proportional to $M/2$, and in the worst-case proportional to $M-1$.

As explained in Section~\ref{ssPerUpdt}, with periodic updates of ${\bf \tilde{C}}$
we can recompute and sort symbols with reasonable worst-case
complexity. We assume that the updating period is $R$ coded symbols.
One choice is keep the data symbols not sorted, and to update ${\bf \tilde{C}}$
using~(\ref{eqAdapCod3}), which requires $O(M)$ operations per update, and an
average complexity of $O(M/R)$ operations per coded symbol. Decoding can
use bisection for symbol search. If we choose, for example, $R = M$
then we have $O(1)$ operations per update, which is quite reasonable.

Sorting the symbols according to their probability during each update
can be done with $O(M\log(M)/R)$ operations per symbol, in the worst-case~\cite{Cormen90}.
Since symbols are normally already sorted from previous updating passes,
insertion sort~\cite{Cormen90} typically can be done with an average of
$O(M/R)$ operations. When the symbols are sorted, we can use the more
efficient symbol search of Section~\ref{ssSymSearch}(d). Choosing $R=M$
results in $O(log(M))$ operations per symbol.

With periodic updating we can both sort symbols, and find the optimal
search strategy (Huffman tree), with a reasonable complexity of
$O(M\log(M)/R)$ operations per symbol~\cite{Cormen90}. However, even though the
asymptotic complexity of finding the Huffman code is the same as
simply sorting, it requires considerable more effort.

Table~\ref{tbComplex} presents a summary of the results above. Note that 
for the complexity analysis of optimal searches, we can use $O(H(\Omega)+1)$,
instead of the more complex and tighter bounds given by~(\ref{ssComplex3}).
The last columns in Table~\ref{tbComplex} indicates the need to use
divisions in the form~(\ref{eqAdapCod3}). Changing the cumulative
distribution every symbol requires using different divisors
for each coded symbol. Periodic updating, on the other hand, allows us
to compute the inverse of divisor once, and use it for coding several ($R$)
symbols.

%

\begin{table}[tbp]
\begin{center}
  \begin{tabular}{|l|c|c|c|c|} \hline
  \textbf{Searching \& updating} & \textbf{Symbol} & \textbf{Decoder} & \textbf{Distribution} & \textbf{Divisions} \\
  \textbf{methods} & \textbf{encoding} & \textbf{interval search} & \textbf{updating} & \\ \hline \hline
  Fixed code & $O(1)$ & $O(H(\Omega)+1)$ & NA & NA \\
   optimal search & & & &  \\ \hline
  Sequential search & $O(1)$ & $O(M)$ & $O(M)$ & O(1) \\
   on sorted symbols & & & &  \\ \hline
  Combined updating and & 
    \multicolumn{3}{|c|}{$O(\log(M))$} & O(1) \\
   coding, bisection tree & \multicolumn{3}{|c|}{} &  \\ \hline
  Combined updating and & 
    \multicolumn{3}{|c|}{$O(H(\Omega)+1)$} & O(1) \\
   coding, optimal tree & \multicolumn{3}{|c|}{} & \\ \hline
  Periodic update, &
    $O(1)$ & $O(\log(M))$ & $O(M/R)$ & O(1/R) \\
   bisection search & & & &  \\ \hline
  Periodic update, & 
    $O(1)$ & $O(H(\Omega)+1)$ & $O(M\log(M)/R)$ & O(1/R) \\
   optimal search & & & &  \\ \hline
\end{tabular}
\end{center}
 \tablecaption{Computational complexity of symbol search and adaptation (cumulative
   distribution updating) for fixed and different adaptive arithmetic coding
   techniques. Typically, $R=4M$ minimizes the complexity without degrading
   performance.}
                                                      \label{tbComplex}
\end{table}

\subsection{Arithmetic Operations}

Presently, additions and multiplications are not much slower than other 
operations, such as bit shifts and comparison. Divisions, on the other hand,
have a much longer latency (number of clock cycles required to perform the
operation), and cannot be streamlined with other operations (like special
circuitry for multiply-add)~\cite{Texas00,IBM01,Intel01,Sun01}.

First, let us consider the arithmetic required by fixed coders, and
adaptive coders with periodic updating.
The arithmetic coding recursion in the forms~(\ref{eqAritCode3})
and~(\ref{eqPropAM4}) require two multiplications and, respectively,
one and three additions. Thus, except for processor with very slow
multiplication, encoding requirements are quite reasonable. Decoding
is more complex due to the effort to find the correct decoding interval.
If we use~(\ref{eqAritCode10}) for interval search, we have one extra
division, plus several comparisons (Section~\ref{ssSymSearch}). The
multiplication-only version~(\ref{eqPropAM2}) requires several multiplications
and comparisons, but with reasonably efficient symbol search this
is faster than~(\ref{eqAritCode10}), and eliminates the need to
define division when multiplications are approximated.

Adaptive coding can be considerably slower because of the divisions
in (\ref{eqAdapCod4}) and~(\ref{eqAdapCod5}). They can add up to two divisions
per interval updating. In Algorithms \ref{alADCIntUpdt}, \ref{alADCIntSel}
and~\ref {alASCIntSel}, we show how to avoid one of the divisions.
Furthermore, updating the cumulative distribution may require a significant
number of additions. Note that having only periodic updates of the
cumulative distribution significantly reduces this adaptation overhead,
because all these divisions and additions are not required for every
coded symbol. For example, a single division, for the computation of
$1/\tilde{C}(M)$, plus a number of multiplications and additions
proportional to $M$, may be needed per update. If the update occurs
every $M$ coded symbols, then the average number of divisions per
coded symbol is proportional to $1/M$, and the average number of
multiplications and additions are constants.

\section{Further Reading}

We presented the coding method that is now commonly associated with the
name arithmetic coding, but the reader should be aware that other types
of arithmetic coding had been proposed~\cite{Pasco76,Rissanen76}.
Standard implementations of arithmetic coding had been defined in some
international standards~\cite{ITU92,Pennebaker92,ITU93,Taubman02}.
Several techniques for arithmetic coding complexity reduction not
covered here are in the references~\cite{Langdon78p,Langdon82,Langdon84,Witten87,Pennebaker88a,Pennebaker88b,Howard92a,Howard92b,Fenwick94,Moffat98,Schindler98}.
We did not mention the fact that errors in arithmetic coded streams
commonly lead to catastrophic error propagation, and thus error-resilient
arithmetic coding~\cite{Sodagar00} is very important in some applications.

%
%

\appendix
\chapter{Integer Arithmetic Implementation}
\fancyhead[RE]{\sc Integer Arithmetic Implementation}
\fancyhead[LO]{\sc Appendix}

The following algorithms show the required adaptations in the algorithms in 
Section~\ref{ssFPImplem} for use with integer arithmetic, and with a $D$-symbol output 
alphabet. Typically $D$ is small power of two, like 2, 4, 16, or~256. As 
explained in Section~\ref{ssIntArit}, we assume that all numbers are represented as 
integers, but here we define $B = D^Pb$, $L = D^Pl$, and $C(s) = D^Pc(s)$.
In addition, we assume multiplications computed with $2P$ digits and
results truncated to $P$ digits. Renormalization~(\ref{eqFixPrec5}) sets
$L > D^{P-1}$, and thus the minimum probability allowed is defined by
\begin{equation}
                                                         \label{eqAppendix1}
  \left\lfloor [C(s+1) - C(s)]\; D^{-P}\,L \right\rfloor \geq 1 \Rightarrow C(s+1) - C(s) \geq D,
\end{equation}
i.e., $p(s) \geq D^{1-P}$.

\begin{AAlgorithm}{{Function Arithmetic{\_}Encoder} $(N,S,M,{\bf C},{\bf d})$}
1. set \{ \= $B \leftarrow 0; \quad L \leftarrow D^P-1$; \COM{Initialize interval}
   \> $t \leftarrow 0$; \} \COM{and symbol counter}
\pushtabs
2. \= for $k=1$ to $N$ do \COM{Encode N data symbols}
   \> 2.1. \textsf{Interval{\_}Update} $(s_k,M,B,L,{\bf C})$; \CON{Update interval according to symbol}
   \> 2.2. \= if $L < D^{P - 1}$ then \CON{If interval is small enough}
   \> \> \textsf{Encoder{\_}Renormalization} $(B,L,t,{\bf d})$; \COO{then renormalize interval}
\poptabs
3. \textsf{Code{\_}Value{\_}Selection} $(B,L,t,{\bf d})$; \COM{Choose final code value}
4. return $t$. \COM{Return number of code symbols}
                                                     \label{alINTEnc}
\end{AAlgorithm}

Algorithm~\ref{alINTEnc}, for integer arithmetic, is almost identical to
Algorithm~\ref{alFPEnc}, except for the initialization of $L$, and the decision
for renormalization.
The arithmetic operations that update the interval base may overflow the 
integer registers. To make clear that this is acceptable, we define the 
results modulo $D^P$, as in Algorithm~\ref{alINTIntUpdt}.
The results of the multiplications are multiplied by 
$D^{-P}$ and truncated, meaning that the least significant bits are 
discarded. Overflow is here detected by a reduction in the base value. For that 
reason, we implement carry propagation together with the interval 
update in Algorithm~\ref{alINTIntUpdt}.

\begin{AAlgorithm}{{Procedure Interval{\_}Update} $(s,M,B,L,{\bf C})$}
1. \= if $s=M-1$; \COM{If $s$ is last symbol set first product equal}
   \> then \= set $Y \leftarrow L$; \CON{to current interval length}
   \> else \> set $Y \leftarrow \left\lfloor {[L \cdot C(s+1)] \cdot D^{-P}} \right\rfloor $; \CON{or else equal to computed value}
2. \> set \{ \= $A \leftarrow B$; \COM{Save current base}
   \> \> $X \leftarrow \left\lfloor {[L \cdot C(s)] \cdot D^{-P}} \right\rfloor $; \COM{Compute second product}
   \> \> $B \leftarrow \left( {B + X } \right)\bmod D^P$; \COM{Update interval base}
   \> \> $L \leftarrow Y - X$; \} \COM{Update interval length}
3. if $A > B$ then \textsf{Propagate{\_}Carry} $(t, {\bf d})$; \COM{Propagate carry if overflow}
4. return.
                                                     \label{alINTIntUpdt}
\end{AAlgorithm}

\begin{AAlgorithm}{{Procedure Encoder{\_}Renormalization} $(B,L,t,{\bf d})$}
1. \= while $L < D^{P - 1}$ do \COM{Renormalization loop}
   \> 1.1. set \{ \= $t \leftarrow t + 1$; \CON{Increment symbol counter}
   \> \> $d(t) \leftarrow \left\lfloor {B \cdot D^{1 - P}} \right\rfloor$; \CON{Output symbol}
   \> \> $L \leftarrow \left( {D \cdot L} \right)\bmod D^P$; \CON{Update interval length}
   \> \> $B \leftarrow \left( {D \cdot B} \right)\bmod D^P$; \} \CON{Update interval base}
2. return.
                                                     \label{alINTENormal}
\end{AAlgorithm}

\begin{AAlgorithm}{{Procedure Propagate{\_}Carry} $(t,{\bf d})$}
1. set $n \leftarrow t$; \COM{Initialize pointer to last outstanding symbol}
2. \= while $d(n) = D - 1$ do \COM{While carry propagation}
   \> 2.1. set \{ \= $d(n) \leftarrow 0$; \CON{Set outstanding symbol to zero}
   \> \> $n \leftarrow n - 1$; \} \CON{Move to previous symbol}
3. set $d(n) \leftarrow d(n) + 1$; \COM{Increment first outstanding symbol}
4. return.
                                                     \label{alINTPrCarry}
\end{AAlgorithm}

Note that in the renormalization of Algorithm~\ref{alINTENormal}, we
assume that $D$ is a power of two, and all 
multiplications and divisions are actually implemented with bit shifts.
The carry propagation with a \mbox{$D$-symbol} output, shown in
Algorithm~\ref{alINTPrCarry}, is very similar to Algorithm~\ref{alFPPrCarry}.

In Algorithm~\ref{alFPValSelc} the code value selection is made to
minimize the number of bits, assuming that the decoder pads the buffer
${\bf d}$ with sufficient zeros. This inconvenience can be avoided by
simply adding a proper extra symbol at the end of the compressed data.
Algorithm~\ref{alINTValSelc} shows the required modifications. It shifts
the interval base by a small amount, and resets the interval length,
so that when procedure \textsf{Encoder{\_}Renormalization} is called, it
adds the proper two last output symbols to buffer ${\bf d}$. This way,
correct decoding does not dependent on the value of the symbols that are read by
the decoder (depending on register size $P$) past the last compressed
data symbols.

\begin{AAlgorithm}{{Procedure Code{\_}Value{\_}Selection} $(B,L,t,{\bf d})$}
1. set \{ \= $A \leftarrow B$; \COM{Save current base}
   \> $B \leftarrow \left( B + D^{P-1}/2 \right) \bmod D^P$; \COM{Increase interval base}
   \> $L \leftarrow D^{P-2}-1$; \} \COM{Set new interval length}
2. if $A > B$ then \textsf{Propagate{\_}Carry} $(t, {\bf d})$; \COM{Propagate carry if overflow}
3. \textsf{Encoder{\_}Renormalization} $(B,L,t,{\bf d})$ \COM{Output two symbols}
4. return.
                                                     \label{alINTValSelc}
\end{AAlgorithm}

With integer arithmetic the decoder does not have to simultaneously update 
the interval base and code value (see Algorithm~\ref{alFPDNormal}). Since decoding is always 
based on the difference $v-b$, we define $V= D^P(v-b)$ and use only $V$ 
and $L$ while decoding. The only other required change is that we must subtract 
from $V$ the numbers that would have been added to $B$.

\begin{AAlgorithm}{{Procedure Arithmetic{\_}Decoder} $(N,M,{\bf C},{\bf d})$}
1. \= set \{ \= $L \leftarrow D^P - 1$; \COM{Initialize interval length}
   \> \> $V = \sum\nolimits_{n = 1}^P {D^{P-n}d(n)} $; \COM{Read $P$ digits of code value}
   \> \> $t \leftarrow P$; \} \COM{Initialize symbol counter}
2. for $k$ = 1 to $N$ do \COM{Decoding loop}
   \> 2.1. set $s_k = $ \textsf{Interval{\_}Selection} $(V,\;L,\;M,\;{\rm {\bf C}})$; \CON{Decode symbol and update interval}
   \> 2.2. if $L < D^{P - 1}$ then \CON{If interval is small enough}
   \> \> \textsf{Decoder{\_}Renormalization} $(V,\;L,\;t,\;{\bf d})$; \COO{renormalize interval}
4. return.
                                                     \label{alINTDec}
\end{AAlgorithm}

\begin{AAlgorithm}{{Function Interval{\_}Selection} $(V,L,M,{\bf C})$}
1. set \{ \= $s \leftarrow 0; \quad n \leftarrow M$; \COM{Initialize bisection search limits}
   \> $X \leftarrow 0; \quad Y \leftarrow L$; \COM{Initialize bisection search interval}
\pushtabs
2. \= while $n - s > 1$ do \COM{Binary search loop}
   \> 2.1. \= set \{ \= $m \leftarrow \left\lfloor {(s + n) / 2} \right\rfloor $; \CON{Compute middle point}
   \> \> \> $Z \leftarrow \left\lfloor {[L \cdot C(m)] \cdot D^{-P}} \right\rfloor$; \} \CON{Compute value at middle point}
\pushtabs
2. \= while \kill
   \> 2.2. \= if $Z > V$ \CON{If new value larger than code value}
   \> \> then \= set \{ \= $n \leftarrow m$; \COO{then update upper limit}
   \> \> \> \> $Y \leftarrow Z$; \} \\
   \> \> else \> set \{ $s \leftarrow m$; \COO{else update lower limit}
   \> \> \> \> $X \leftarrow Z$; \} \\
\poptabs \poptabs
3. set \{ $V \leftarrow V - X$; \COM{Update code value}
   \> $L \leftarrow Y - X$; \} \COM{Update interval length as difference}
4. return $s$.
                                                   \label{alINTIntSel}
\end{AAlgorithm}

\begin{AAlgorithm}{{Procedure Decoder{\_}Renormalization} $(V,L,t,{\bf d})$}
1. \= while $L < D^{P - 1}$ do \COM{Renormalization loop}
   \> 1.1. set \{ \= $t \leftarrow t + 1$; \CON{Increment symbol counter}
   \>        \> $V \leftarrow \left( {D \cdot V} \right) \bmod D^P + d(t)$; \CON{Update code value}
   \>        \> $L \leftarrow \left( {D \cdot L} \right) \bmod D^P$; \} \CON{Update interval length}
2. return.
                                                 \label{alINTDNormal}
\end{AAlgorithm}

In Algorithm~\ref{alFPIntSel} we used sequential search for interval selection during 
decoding, which in the worst case requires testing $M-1$ intervals. In 
Algorithm~\ref{alINTIntSel} we use bisection~\cite{Gill82,Rice83,Press93} for
solving~(\ref{eqAritCode10}), which requires testing at most
$\left\lceil {\log_2(M)} \right\rceil$ intervals (see Sections \ref{ssBinCod}
and ~\ref{ssSymSearch}).
Finally, the decoder renormalization is shown in Algorithm~\ref{alINTDNormal}.

%
%

\newpage
\fancyhead[RE]{\sc Bibliography}
\fancyhead[LO]{\sc Bibliography}
\begin{small}

\end{small}


\end{document}